\documentclass[11pt]{article}
\usepackage{fullpage}
\usepackage{amsmath}
\usepackage{color}
\usepackage{amsfonts}
\usepackage{amssymb}
\usepackage{graphicx}
\usepackage{graphics}
\usepackage{subfig}
\usepackage{psfrag}
\usepackage{epsfig,latexsym,verbatim}
\usepackage{epstopdf}
\usepackage{rotating}
\usepackage{lscape}
\usepackage{geometry}
\usepackage{mathrsfs}
\usepackage{float}
\usepackage{algorithm,algorithmic}
\usepackage[colon]{natbib}
\usepackage{breakcites}
\usepackage{authblk}
\usepackage{enumitem}

\graphicspath{{fig/}}
\allowdisplaybreaks[3]
\def\1{{\boldsymbol 1}}
\def\0{{\boldsymbol 0}}

\def\bbeta{\boldsymbol \beta}
\def\hbbeta{\widehat{\boldsymbol \beta}}

\def\bdelta{\boldsymbol \delta}

\def\bata{\boldsymbol \eta}
\def\hbata{\widehat{\bata}}

\def\bupsilon{\boldsymbol \upsilon}

\def\real{\mathop{{\rm I}\kern-.2em\hbox{\rm R}}\nolimits}

\numberwithin{equation}{section}

\newtheorem{theorem}{Theorem}

\newtheorem{corollary}{Corollary}
\newtheorem{proposition}{Proposition}

\newtheorem{remark}{Remark}

\begin{document}

\begin{center}
{\LARGE \textbf{Exploration of heterogeneous treatment effects via concave
fusion}\let\thefootnote\relax\footnotetext{%
The research of Ma is supported in part by the U.S. NSF grant DMS-1712558.}}%
\footnote{\baselineskip=10pt}

\vskip 1cm Shujie Ma\\[0pt]
Department of Statistics, University of California at Riverside, Riverside,
CA 92521\\[0pt]
shujie.ma@ucr.edu\\[0pt]
\hskip 5mm\\[0pt]
Jian Huang\\[0pt]
Department of Statistics and Actuarial Science, University of Iowa, Iowa
City, IA 52242\\[0pt]
jian-huang@uiowa.edu \\[0pt]
\hskip 5mm\\[0pt]
Zhiwei Zhang\\[0pt]
Department of Statistics, University of California at Riverside, Riverside,
CA 92521\\[0pt]
zhiwei.zhang@ucr.edu \\[0pt]
\hskip 5mm\\[0pt]
Mingming Liu\\[0pt]
Department of Statistics, University of California at Riverside, Riverside,
CA 92521\\[0pt]
mliu034@ucr.edu \\[0pt]
\hskip 5mm\\[0pt]

{\Large \textbf{Abstract}}
\end{center}

Understanding treatment heterogeneity is essential to the development of
precision medicine, which seeks to tailor medical treatments to subgroups of
patients with similar characteristics. One of the challenges of achieving
this goal is that we usually do not have \textit{a priori} knowledge of the
grouping information of patients with respect to treatment effect. To
address this problem, we consider a heterogeneous regression model which
allows the coefficients for treatment variables to be subject-dependent with
unknown grouping information. We develop a concave fusion penalized method
for estimating the grouping structure and the subgroup-specific treatment
effects, and derive an alternating direction method of multipliers algorithm
for its implementation. We also study the theoretical properties of the
proposed method and show that under suitable conditions there exists a local
minimizer that equals the oracle least squares estimator based on \textit{a
priori} knowledge of the true grouping information with high probability.
This provides theoretical support for making statistical inference about the
subgroup-specific treatment effects using the proposed method. The proposed
method is illustrated in simulation studies and illustrated with real data
from an AIDS Clinical Trials Group Study.

\baselineskip=12pt

\baselineskip=12pt

\vfill\noindent \underline{\textbf{Key Words}}: Fusiongram; Oracle property;
Penalized least squares; 
Subgroup analysis; Treatment heterogeneity

\medskip \noindent \underline{\textbf{Short title}}: Heterogeneous treatment
effects

\clearpage\pagebreak\newpage \pagenumbering{arabic} \newlength{\gnat} %
\setlength{\gnat}{22pt} \baselineskip=\gnat

\section{Introduction\label{SEC:Intro}}

Treatment effects are often heterogeneous, that is, the same treatment can
have different effects on different patients %
\citep{Sorensen:1996,KravitzDB:2004}. For instance, molecularly targeted
cancer drugs are only effective for patients with tumors expressing targets %
\citep{Simon:2010}, and the relative efficacy of antiretroviral drugs for
treating human immunodeficiency virus infection sometimes depends on
baseline viral load and CD4 count \citep{Zhang.Nie.Soon.Liu:2014}.
Understanding the heterogeneity of treatment effects (HTE) is critical to
the eventual success of precision medicine, which seeks to tailor medical
treatments to individual patients.

Possible HTE is usually assessed in a subgroup analysis %
\citep{Gail.Simon:1985,Russek.Simon:1998,KravitzDB:2004,Rothwell:2005,Lagakos:2006}
or, more generally, a regression analysis relating the outcome of interest
to treatment and a collection of baseline covariates \citep{cai.tian.wong:2011}.
Such a regression model can incorporate HTE as interactions between
treatment and baseline covariates, and can be used to estimate
covariate-specific treatment effects indirectly. Alternatively, a
covariate-specific treatment effect model can be specified and estimated
directly without relying on a regression model for the outcome %
\citep{Tian.Alizadeh:2014,Zhang.Qu:2017}. There is also a growing literature on HTE estimation using machine learning methods \citep[e.g.][]{Su.Meneses.McNees.Johnson:2011,Imai.Ratkovic:2013,Su.Pena.Liu.Levine:2018}. All of these methods are based on
observed baseline covariates; they do not address possible HTE due to
unmeasured covariates.

The collection of observed baseline covariates is often limited, and thus
may be insufficient for characterizing the true HTE across individual
patients. The true HTE structure is not empirically identifiable unless all
effect modifiers are measured, but can be explored under appropriate
assumptions. \cite{Zhang.Wang.Nie.Soon:2013,Zhang.Nie.Soon.Liu:2014} use
random effect models to conduct sensitivity analyses concerning the joint
distribution of two potential outcomes (for two different treatments). \cite%
{ShenHe:2015} propose a two-group logistic-normal mixture model for the true
HTE and develop a model-based procedure for testing the equivalence of the
two groups. The model of \cite{ShenHe:2015} includes a normality assumption
for the outcome and a logistic regression model relating the latent group
membership to a collection of observed covariates.

In this article, we propose a general latent class model for exploring the
true HTE. Different from the aforementioned existing models, our model deals
with an unspecified number of latent groups, without assuming normality for
the outcome or a particular relationship between the latent group membership
and the observed covariates. In our model, we assume that the treatment
coefficients are subject-specific and belong to different groups with
unknown grouping information. The subgroups, therefore, can be driven by
observed covariates, unobserved covariates or an arbitrary combination of
both types of covariates. We recover the latent subgroups and estimate the
model using concave fusion penalization, an unsupervised machine learning
method, that applies a concave penalty, such as the smoothly clipped
absolute deviations penalty \citep[SCAD,][]{fan.li:2001} or the minimax
concave penalty \citep[MCP,][]{zhang:2010}, to pairwise differences of
coefficients for treatment effects. The fusion penalized approach has been
proposed for clustering analysis of grouping data objects %
\citep{chi.lange:2014} and grouping means of clinical outcomes with random
errors \citep{Ma.Huang:2015}, but not for exploring HTE. \cite{ma.huang:2017}
briefly mentioned potential applications of the fusion penalized approach to
the estimation of subject-specific coefficient models without providing
theoretical justifications and numerical studies.

We consider the present paper as the first work which applies the fusion
learning approach to the investigation of HTE through subject-specific
treatment coefficients and provides a theoretical analysis of the resulting
estimator for the proposed general latent class model with subject-specific
treatment effects. The consistency and asymptotic distributional properties
of the estimators follow from the fact that, under appropriate conditions,
the oracle least squares estimator based on \textit{a priori} knowledge of
the true group membership is a local minimizer of the objective function
with high probability. Moreover, we derive the conditions on the number of
subgroups and the number of covariates compared with the sample size as well
as the lower bound of the minimum difference of coefficient values between
the subgroups in order to identify the true subgroups of treatment effects.
Computationally, we apply an alternating direction method of multipliers
(ADMM) algorithm \citep{Boyd:2011} for implementing the proposed approach.
This algorithm has been used widely in convex optimization problems. We
derive the convergence properties of the ADMM in the present setting.
Another contribution is that using the proposed fusion penalized estimator,
we further propose a bootstrapping procedure to test homogeneity of the
treatment effects for confirmatory analysis.

The rest of this article is organized as follows. Section \ref{SEC:subgroup}
introduces the proposed latent class model. Section \ref{SEC:estimation}
describes the concave fusion penalization method for model estimation.
In Section \ref{SEC:theory} we establish the theoretical properties of the
proposed estimator. Section \ref{SEC:bootstrap} introduces a bootstrapping
procedure for testing homogeneity. In Section \ref{SEC:examples} we evaluate
the finite-sample properties of the proposed method via simulation studies,
and apply the method to AIDS Clinical Trials Group Study 175. Concluding
remarks are given in Section \ref{SEC:Discussion}. The computational
algorithm and the technical proofs are provided in the Appendix.


\section{Model}


\label{SEC:subgroup}Suppose the data consist of $(y_{i},{\boldsymbol{z}}_{i},%
{\boldsymbol{x}}_{i}),i=1,\ldots ,n$, where $y_{i}$ is an outcome variable, $%
{\boldsymbol{z}}_{i}$ a $q$-vector of patient characteristics related to the
outcome, and ${\boldsymbol{x}}_{i}$ a $p$-vector of treatment or exposure
characteristics. The vector ${\boldsymbol{x}}_{i}$ may include discrete
components (e.g., indicators of treatment groups or exposure status) and/or
continuous components (e.g., dose of drug or radiation). Our goal is to
understand how the causal effect of ${\boldsymbol{x}}_{i}$ may vary across
individual subjects. Let $y_i(\boldsymbol{x})$ denote the potential outcome for the $i$th subject under treatment $\boldsymbol{x}$ \citep{Rubin:1974}. We assume that treatment assignment is
strongly ignorable
\citep{Rosenbaum:1983} in the sense that
\begin{equation}
\boldsymbol{x}_i\perp y_i(\boldsymbol{x})|\boldsymbol{z}_i,
\label{ignorability}
\end{equation}
where $\perp$ denotes independence and $\boldsymbol{x}$ may be any treatment that could conceivably be given to the $i$th subject. From this assumption it follows that
$$
E\{y_i(\boldsymbol{x})|\boldsymbol{z}_i\}= E\{y_i(\boldsymbol{x})|\boldsymbol{z}_i,\boldsymbol{x}_i=\boldsymbol{x}\}
= E\{y_i|\boldsymbol{z}_i,\boldsymbol{x}_i=\boldsymbol{x}\},
$$
which is empirically identified for any possibly treatment $\boldsymbol{x}$ for the $i^{\text{th}}$ subject. Assumption \eqref{ignorability} is trivially true in a randomized clinical
trial where ${\boldsymbol{x}}_{i}$ is independent of all baseline variables;
in this case, adjustment for ${\boldsymbol{z}}_{i}$ is not strictly
necessary but may improve efficiency.

Suppose the data generation mechanism can be described by the following
linear regression model:
\begin{equation}
y_{i}={\boldsymbol{z}}_{i}^{\text{T}}\boldsymbol{\eta }+{\boldsymbol{x}}%
_{i}^{\text{T}}\boldsymbol{\beta }_{i}+\varepsilon _{i},i=1,\ldots ,n,
\label{Mod1}
\end{equation}%
where $\boldsymbol{\eta }$ and $\boldsymbol{\beta }_{i}$ are unknown
regression coefficients and the $\varepsilon _{i}$'s are i.i.d. random
errors with $E(\varepsilon _{i})=0$ and $\hbox{Var}(\varepsilon _{i})=\sigma
^{2}$. We assume that the first entry in each ${\boldsymbol{x}}_{i}$ is $1$
so the intercept is included in $\boldsymbol{\beta }_{i}$. Under this model
and the strongly ignorable treatment assignment assumption, the
subject-specific regression coefficient $\boldsymbol{\beta }_{i}$ represents
the causal effect of ${\boldsymbol{x}}_{i}$ on the $i^{\text{th}}$ subject,
and the distribution of $\boldsymbol{\beta }_{i}$ across $i$ represents the
true HTE. With more unknown parameters than observations, model \eqref{Mod1}
is not estimable without additional assumptions.

A simple way to constrain model \eqref{Mod1} and achieve identification is
to assume that the $\boldsymbol{\beta }_{i}$ are all equal. This leads to
the following model:
\begin{equation}
y_{i}={\boldsymbol{z}}_{i}^{\text{T}}\boldsymbol{\eta }+{\boldsymbol{x}}%
_{i}^{\text{T}}\boldsymbol{\beta }+\varepsilon _{i},i=1,\ldots ,n,
\label{Mod0}
\end{equation}%
which is commonly used in practice. Model \eqref{Mod0} is quite restrictive
as it rules out any HTE with respect to observed and unobserved covariates.
The model can be expanded by including interaction terms between some or all
components of ${\boldsymbol{z}}_{i}$ and ${\boldsymbol{x}}_{i}$, thereby
allowing $\boldsymbol{\beta }_{i}$ to be a specified function of ${%
\boldsymbol{z}}_{i}$. The expanded model can accommodate a particular form
of observed HTE, but it does not account for any latent HTE. Moreover, even
if the HTE is only driven by the observed covariates ${\boldsymbol{z}}_{i}$,
it is unclear in what specific form ${\boldsymbol{z}}_{i}$ causes the HTE.
For instance, the common assumption that $\boldsymbol{\beta }_{i}=%
\boldsymbol{\beta +}\boldsymbol{\Gamma }{\boldsymbol{t}}_{i}$, where ${%
\boldsymbol{t}}_{i}$ is a subset of ${\boldsymbol{z}}_{i}$, and $\boldsymbol{%
\beta }$ and $\boldsymbol{\Gamma }$ are coefficients implies a linear
interaction structure which may be too stringent in practice. When many
baseline variables are present, there can be many interaction terms. This
increases the chance of false positive results in interaction tests and may
lead to overstated and misleading conclusions \citep{Wang.Stephen:2007}.

Another approach is to treat $\boldsymbol{\beta }_{i}$ as a subject-specific
random vector following a specified conditional distribution given ${%
\boldsymbol{z}}_{i}$. A prominent example in this category is the common
linear mixed model where $\boldsymbol{\beta }_{i}$ is assumed to follow a
normal distribution, independently of ${\boldsymbol{z}}_{i}$. Such a model
may have identifiability issues in the present setting, where each subject
contributes only one observation. Another example in this category, which
achieves identifiability, is the two-group logistic-normal mixture model of
\cite{ShenHe:2015}. In the present notation, their model can be expressed as
\begin{equation}
y_{i}={\boldsymbol{x}}_{i}^{\text{T}}\boldsymbol{\beta }_{i}+\varepsilon _{i}%
\text{, where }\boldsymbol{\beta }_{i}=\boldsymbol{\alpha }_{1}+\boldsymbol{%
\alpha }_{2}w_{i}\text{ and }w_{i}=I({\boldsymbol{z}}_{i}^{\text{T}}%
\boldsymbol{\gamma }+\epsilon >0),  \label{MOD:mix}
\end{equation}%
where $\epsilon $ is an error distributed by the standard logistic
distribution, so that $P(w_{i}=1|{\boldsymbol{z}}_{i})=\exp ({\boldsymbol{z}}%
_{i}^{\text{T}}\boldsymbol{\gamma })/(1+\exp ({\boldsymbol{z}}_{i}^{\text{T}}%
\boldsymbol{\gamma }))$. This model assumes that the $\boldsymbol{\beta }%
_{i} $ in \eqref{Mod1} arise from a finite mixture model with two possible
values ($\boldsymbol{\alpha }_{1}$ and $\boldsymbol{\alpha }_{1}+\boldsymbol{%
\alpha }_{2}$), that $\varepsilon _{i}$ is normally distributed, and that
the latent group membership is related to observed covariates through a
logistic regression model. Under these assumptions, model \eqref{MOD:mix}
can be estimated using a standard EM algorithm, and \cite{ShenHe:2015}
further develop an EM test for the null hypothesis $\boldsymbol{\alpha }%
_{2}=0$, which indicates the absence of (observed or latent) HTE. This
approach, as a model-based way to estimate the true HTE, is limited by the
two-group assumption and the modeling assumptions.

Here we consider a more flexible latent class model and propose a machine
learning approach to recovering the subgroups. We assume that the $%
\boldsymbol{\beta }_{i}$ in \eqref{Mod1} arise from a mixture model with an
unspecified number of groups. Specifically, let $\mathcal{G}=(\mathcal{G}%
_{1},\ldots ,\mathcal{G}_{K})$ be a mutually exclusive partition of $%
\{1,\ldots ,n\}$. Suppose $\boldsymbol{\beta }_{i}=\boldsymbol{\alpha }_{k}$
for all $i\in \mathcal{G}_{k}$, where $\boldsymbol{\alpha }_{k}$ is the
common value for the $\boldsymbol{\beta }_{i}$'s in group $\mathcal{G}_{k}$.
In other words, we assume that
\begin{equation*}
\boldsymbol{\beta }_{i}=\boldsymbol{\alpha }_{1}w_{i1}+\boldsymbol{\alpha }%
_{2}w_{i2}+\cdots +\boldsymbol{\alpha }_{K}w_{iK},
\end{equation*}%
where $\boldsymbol{\alpha }_{k}=(\alpha _{k1},\ldots ,\alpha _{kp})^{\text{T}%
}$, and $w_{ik}\in (0,1)$ is the (latent) indicator for the $k^{\text{th}}$
group $\mathcal{G}_{k}$, i.e., $w_{ik}=1$ for $i\in \mathcal{G}_{k\text{ }}$
and $w_{ik}=0$ otherwise. Substituting this into model (\ref{Mod1}) yields
the latent class model:
\begin{equation}
y_{i}={\boldsymbol{z}}_{i}^{\text{T}}\boldsymbol{\eta }+{\boldsymbol{x}}%
_{i}^{\text{T}}(\boldsymbol{\alpha }_{1}w_{i1}+\boldsymbol{\alpha }%
_{2}w_{i2}+\cdots +\boldsymbol{\alpha }_{K}w_{iK})+\varepsilon _{i},
\label{EQ:model}
\end{equation}%
where the number of subgroups $K$ is unknown and the group indicators $%
w_{ik} $ are unobservable. Model (\ref{EQ:model}) includes the case of no
HTE as a special case (i.e., $K=1$ and hence $w_{i1}\equiv 1$). Of note, we
do not parameterize the distribution of $\varepsilon _{i}$, nor do we
specify how the $w_{ik}$'s may be related to ${\boldsymbol{z}}_{i}$. Thus,
our model is considerably more flexible than the model of \cite{ShenHe:2015}%
. 

Although the subgroups in model \eqref{EQ:model} are not (fully)
ascertainable using observed covariates, an exploratory analysis based on
model \eqref{EQ:model} can provide unique insights into the true HTE and
helpful guidance for future research. For example, if the results indicate
that a new treatment only benefits a small (and unidentified) proportion of
the patient population, that finding might motivate scientists to collect
more covariate data (e.g., biomarkers) in future studies and search for
predictors of treatment benefit. Conversely, if there is no indication of
clinically important HTE, that information would support a decision to
(re)direct limited resources toward other, more promising areas of research.

\section{Estimation}


\label{SEC:estimation}

To identify the subgroups of the heterogeneous treatment effects, we first
need to estimate model (\ref{EQ:model}), i.e., we need to estimate the
number of subgroups $K$, the coefficients $\boldsymbol{\eta }$ and $%
\boldsymbol{\alpha }$, and the group membership $w_{ik}$ for each
observation. Estimating model (\ref{EQ:model}) is the same as estimation of
model (\ref{Mod1}), since they are the same models with different notations.
We propose a concave fusion method for model estimation described below.

For any vector $\mathbf{a}$, denote its $L_{2}$ norm by $\Vert \mathbf{a}%
\Vert =(\sum |a_{i}|^{2})^{1/2}$. Consider the criterion
\begin{equation}
Q_{n}(\boldsymbol{\eta },\boldsymbol{\beta })=\frac{1}{2}\sum%
\nolimits_{i=1}^{n}(y_{i}-{\boldsymbol{z}}_{i}^{\text{T}}\boldsymbol{\eta }-{%
\boldsymbol{x}}_{i}^{\text{T}}\boldsymbol{\beta }_{i})^{2}+\sum\nolimits_{1%
\leq i<j\leq n}p(\Vert \boldsymbol{\beta }_{i}-\boldsymbol{\beta }_{j}\Vert
,\lambda ),  \label{EQ:objective}
\end{equation}%
where $\boldsymbol{\beta }=(\boldsymbol{\beta }_{1}^{\text{T}},\ldots ,%
\boldsymbol{\beta }_{n}^{\text{T}})^{\text{T}}$, and $p(\cdot ,\lambda )$ is
a penalty function with a tuning parameter $\lambda \geq 0$. For a given $%
\lambda >0$, let 
\begin{equation}
(\widehat{\boldsymbol{\eta }}(\lambda ),\widehat{\boldsymbol{\beta }}%
(\lambda ))=\mathop{\rm argmin}_{\boldsymbol{\eta }\in \mathop{{\rm
I}\kern-.2em\hbox{\rm R}}\nolimits^{q},\,\boldsymbol{\beta }\in \mathop{{\rm
I}\kern-.2em\hbox{\rm R}}\nolimits^{np}}Q_{n}(\boldsymbol{\eta },\boldsymbol{%
\beta };\lambda ).  \label{Min1}
\end{equation}

We use sparsity-inducing penalties (to be discussed later) in %
\eqref{EQ:objective}. For a sufficiently large $\lambda $, the penalty
shrinks some of $\Vert \boldsymbol{\beta }_{i}-\boldsymbol{\beta }_{j}\Vert $
to zero. We partition the treatment effects into subgroups according to the
unique values of $\widehat{\boldsymbol{\beta }}$. Specifically, let $%
\widehat{{\lambda }}$ be the value of the tuning parameter on the path
selected based on a data-driven procedure such as the BIC. For simplicity,
write $(\widehat{\boldsymbol{\eta }},\widehat{\boldsymbol{\beta }})\equiv (%
\widehat{\boldsymbol{\eta }}(\widehat{\lambda }),\widehat{\boldsymbol{\beta }%
}(\widehat{\lambda }))$. Let $\{\widehat{\boldsymbol{\alpha }}_{1},\ldots ,%
\widehat{\boldsymbol{\alpha }}_{\widehat{K}}\}$ be the distinct values of $%
\widehat{\boldsymbol{\beta }}$, where $\widehat{K}$ is the number of the
distinct values. These are the estimates of subgroup-specific treatment
effects. The samples can then be divided into subgroups accordingly. Denote
the set of indices of the $k$th subgroup by $\widehat{\mathcal{G}}_{k}=\{i:%
\widehat{\boldsymbol{\beta }}_{i}=\widehat{\boldsymbol{\alpha }}_{k},1\leq
i\leq n\},1\leq k\leq \widehat{K}$. Accordingly, we have $\widehat{w}_{ik}=1$%
, if $i\in \widehat{\mathcal{G}}_{k}$ and $\widehat{w}_{ik}=0$, otherwise.


A popular sparsity-inducing penalty is the $L_{1}$ or lasso penalty with $%
p_{\gamma }(t,\lambda )=\lambda |t|$ \citep{Tibshirani:1996}, but this
penalty tends to produce too many subgroups \citep{Ma.Huang:2015}. Thus, we
focus on two concave penalty functions: the smoothly clipped absolute
deviation penalty \citep[SCAD,][]{fan.li:2001} and the minimax concave
penalty \citep[MCP,][]{zhang:2010}. The SCAD penalty is
\begin{equation*}
p_{\gamma }(t,\lambda )=\lambda \int_{0}^{|t|}\min \{1,(\gamma -x/\lambda
)_{+}/(\gamma -1)\}dx.
\end{equation*}%
The MCP has the form
\begin{equation*}
p_{\gamma }(t,\lambda )=\lambda \int_{0}^{|t|}(1-x/(\gamma \lambda ))_{+}dx.
\end{equation*}%
These penalties lead to nearly unbiased estimators of the parameters due to
the fact that their derivatives equal to zero at large (in magnitude) values
of the parameter estimates. Moreover, they are more aggressive in enforcing
a sparser solution. Thus, they are better suited for the current problem,
where the number of subgroups may be expected to be much smaller than the
sample size.


We compute $(\widehat{\boldsymbol{\eta }}(\lambda ),\widehat{\boldsymbol{%
\beta }}(\lambda ))$ given in (\ref{Min1}) for $\lambda $ in a given
interval $[\lambda _{\min },\lambda _{\max }]$, where $\lambda _{\max }$ is
the value that forces a constant $\widehat{\boldsymbol{\beta }}$ solution,
i.e., $\widehat{\boldsymbol{\beta }}_{j}(\lambda _{\max })=\widehat{%
\boldsymbol{\beta }}_{k}(\lambda _{\max }),1\leq j<k\leq n$; $\lambda _{\min
}$ is a small positive number.
We are particularly interested in the path $\{\widehat{\boldsymbol{\beta }}%
(\lambda ):\lambda \in \lbrack \lambda _{\min },\lambda _{\max }]\}$. The
ADMM algorithm for computing the solution path on a grid of $\lambda $
values is described in detail in Section A.1 of the Appendix. We also derive
the convergence property of the ADMM algorithm with concave penalties. The
result is given in Proposition A.1 of the Appendix.

\begin{figure}[tbp]
\caption{\textit{Solution paths for }$(\protect\widehat{\protect\beta }_{21}(%
\protect\lambda ),\ldots ,\protect\widehat{\protect\beta }_{2n}(\protect%
\lambda ))$\textit{\ against $\protect\lambda $ with $n=200$ for data from
Example 1 in Section \protect\ref{SEC:examples}.}}
\label{FIG:pathEX1}\centering
$%
\begin{array}{ccc}
\includegraphics[width=4.5cm,height=4.5cm]{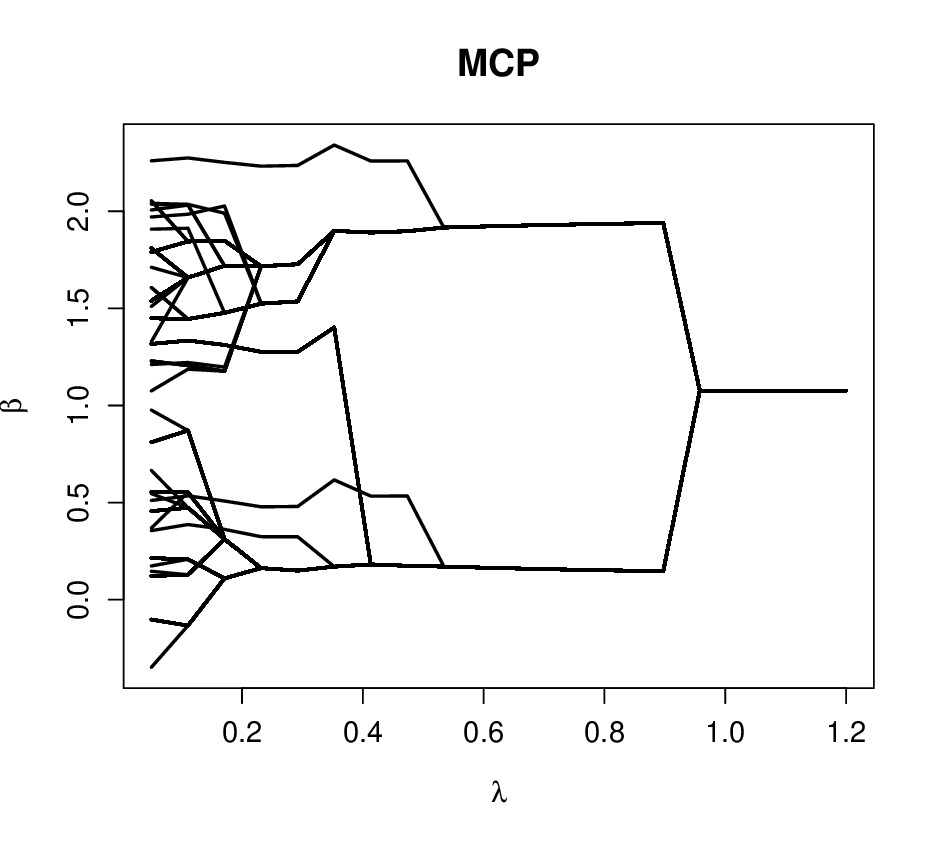} & %
\includegraphics[width=4.5cm,height=4.5cm]{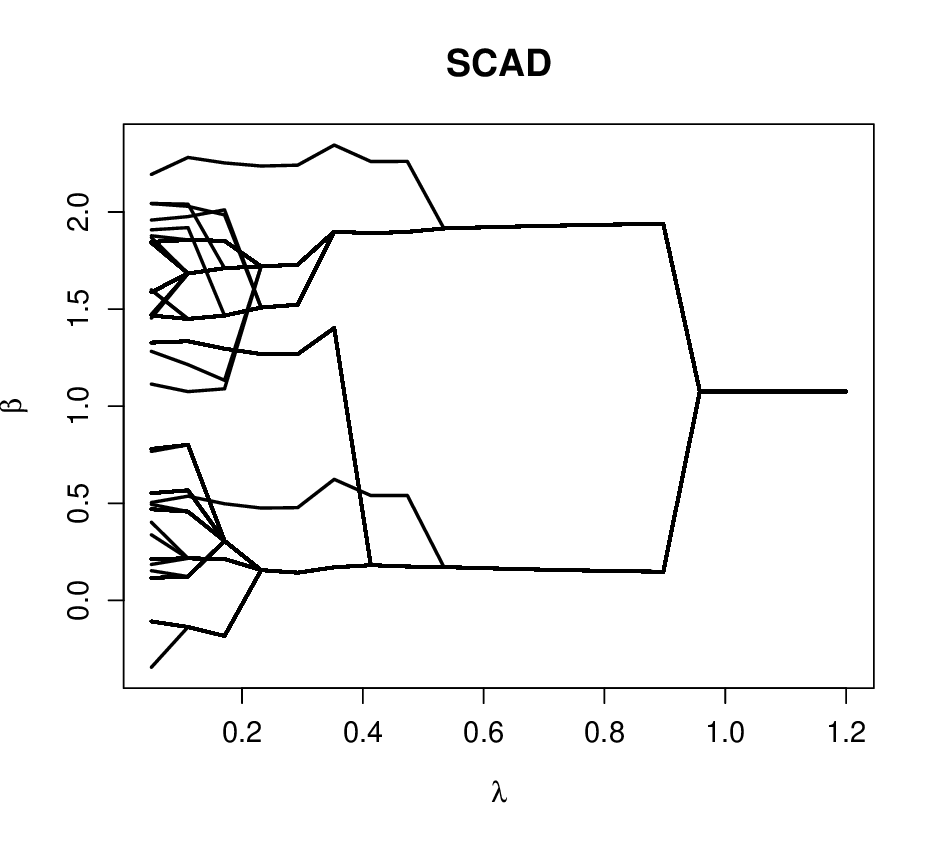} & %
\includegraphics[width=4.5cm,height=4.5cm]{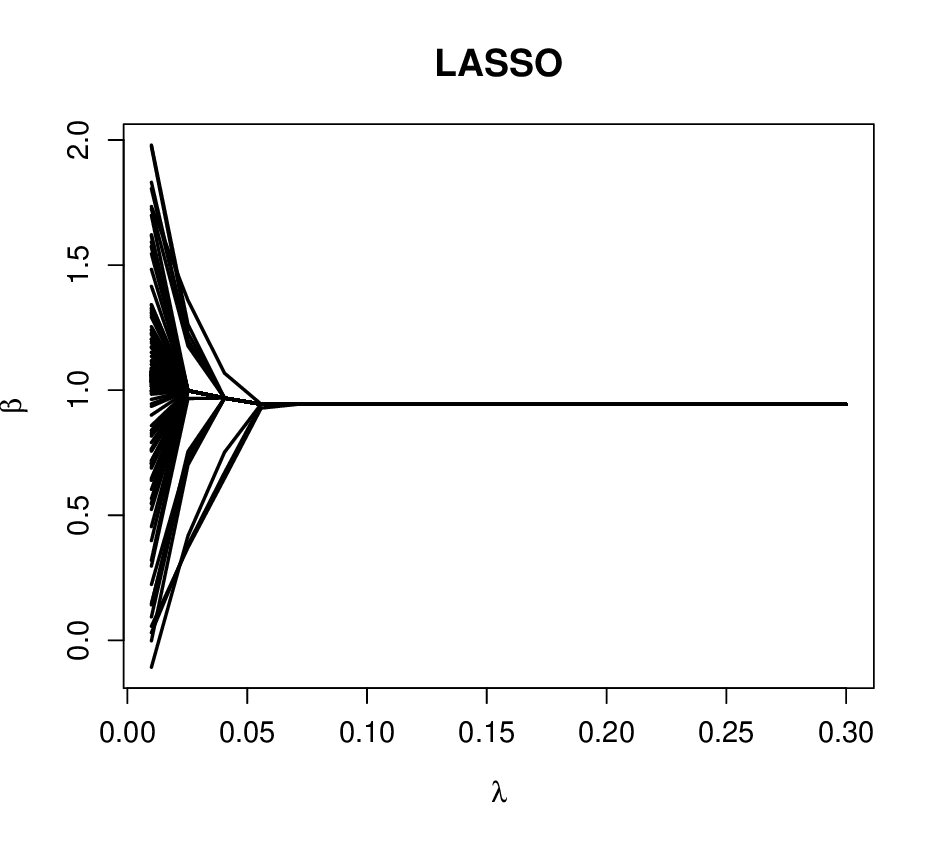}%
\end{array}
$%
\end{figure}

Figure \ref{FIG:pathEX1} illustrates the solution path for the estimates of
the treatment coefficients \newline
$(\widehat{\beta }_{21}(\lambda ),\ldots ,\widehat{\beta }_{2n}(\lambda ))$
against $\lambda $ using MCP, SCAD and lasso penalties for simulated data in
Example 1 of Section \ref{SEC:examples}, which has two subgroups with
`treatment effects' 0 and $2$, respectively. The path is calculated using a
\textquotedblleft bottom up\textquotedblright\ approach starting from $%
\lambda _{\min }$. It looks similar to the dendrogram for agglomerative
hierarchical clustering. However, unlike the clustering algorithms which
form the clusters based on a direct measure of dissimilarity, the fusion of
the coefficients is based on solving the optimization problems along the
solution path. We shall refer to the solution path $\{\widehat{\boldsymbol{%
\beta }}(\lambda ),\lambda \in \lbrack \lambda _{\min },\lambda _{\max }]\}$
as a \textit{fusiongram}.

In Figure \ref{FIG:pathEX1}, the fusiongrams for SCAD and MCP look similar.
They both include a segment containing nearly unbiased estimates of the
treatment effects. When the $\lambda $ value reaches around 0.6, the
estimates of $(\beta _{21},\ldots ,\beta _{2n})$ merge into two values that
are close to the true values $0$ and $2$. When the $\lambda $ value exceeds
1.0, the estimates shrink 
to one value. For the lasso, the estimates of $(\beta _{21},\ldots ,\beta
_{2n})$ merge to one value quickly at $\lambda =0.05$ due to the
overshrinkage of the $L_{1}$ penalty.

\section{Theoretical properties\label{SEC:theory}}

In this section, we study the theoretical properties of the proposed
estimator. Specifically, we provide sufficient conditions under which there
exists a local minimizer of the objective function equal to the oracle least
squares estimator with \textit{a priori} knowledge of the true groups with
high probability. We also derive the lower bound of the minimum difference
of coefficients between subgroups in order to be able to estimate the
subgroup-specific treatment effects.

\subsection{Notation and conditions}

Let $\widetilde{\mathbf{W}}\mathbf{=\{}w_{ik}\}$ be an $n\times K$ matrix
with $w_{ik}=1$ for $i\in \mathcal{G}_{k\text{ }}$ and $w_{ik}=0$ otherwise.
Let $\mathbf{W=}\widetilde{\mathbf{W}}\mathbf{\otimes I}_{p}$.
Let 
$\mathcal{M}_{\mathcal{G}}=\{\boldsymbol{\beta }\in
\mathop{{\rm
I}\kern-.2em\hbox{\rm R}}\nolimits^{np}:\boldsymbol{\beta }_{i}=\boldsymbol{%
\beta }_{j}\text{, for any }i,j\in \mathcal{G}_{k},1\leq k\leq K\}.$
For each $\boldsymbol{\beta }\in \mathcal{M}_{\mathcal{G}}$, it can be
written as $\boldsymbol{\beta }=\mathbf{W}\boldsymbol{\alpha }$, where $%
\boldsymbol{\alpha }=(\boldsymbol{\alpha }_{1}^{\text{T}},\ldots ,%
\boldsymbol{\alpha }_{K}^{\text{T}})^{\text{T}}$ and $\boldsymbol{\alpha }%
_{k}$ is a $p\times 1$ vector of the $k$th subgroup-specific parameter for $%
k=1,\ldots ,K$. Simple calculation shows
\begin{equation*}
\mathbf{W}^{\text{T}}\mathbf{W}=\text{diag}(\left\vert \mathcal{G}%
_{1}\right\vert ,\ldots ,\left\vert \mathcal{G}_{K}\right\vert )\mathbf{%
\otimes I}_{p},
\end{equation*}%
where $\left\vert \mathcal{G}_{k}\right\vert $ denotes the number of
elements in $\mathcal{G}_{k}$. Denote the minimum and maximum group sizes by
$\left\vert \mathcal{G}_{\min }\right\vert \mathcal{=}\min_{1\leq k\leq
K}\left\vert \mathcal{G}_{k}\right\vert $ and $\left\vert \mathcal{G}_{\max
}\right\vert \mathcal{=}\max_{1\leq k\leq K}\left\vert \mathcal{G}%
_{k}\right\vert $, respectively. For any positive numbers $a_{n}$ and $b_{n}$%
, let $a_{n}\gg b_{n}$ denote $a_{n}^{-1}b_{n}=o(1)$. For any vector $%
\mathbf{\boldsymbol{\zeta }}=\left( \zeta _{1},\ldots ,\zeta _{s}\right) ^{%
\text{T}}\in \mathop{{\rm I}\kern-.2em\hbox{\rm R}}\nolimits^{s}$, let $%
\left\Vert \mathbf{\boldsymbol{\zeta }}\right\Vert _{\infty }=\max_{1\leq
l\leq s}\left\vert \zeta _{l}\right\vert .$ For any symmetric matrix $%
\mathbf{A}_{s\times s}$, denote its $L_{2}$ norm by $\left\Vert \mathbf{A}%
\right\Vert =\max_{\mathbf{\boldsymbol{\zeta }\in }R^{s}\mathbf{,\Vert
\boldsymbol{\zeta }\Vert =}1}\left\Vert \mathbf{A\boldsymbol{\zeta }}%
\right\Vert $, and let $\lambda _{\min }(\mathbf{A)}$ and $\lambda _{\max }(%
\mathbf{A)}$ be the smallest and largest eigenvalues of $\mathbf{A}$,
respectively. For any matrix $\mathbf{A=}\left( A_{ij}\right)
_{i=1,j=1}^{s,t}$, denote $\left\Vert \mathbf{A}\right\Vert _{\infty
}=\max_{1\leq i\leq s}\sum\nolimits_{j=1}^{t}\left\vert A_{ij}\right\vert $.
Let $\mathbf{y=(}y_{1},\ldots ,y_{n})^{\text{T}}$, $\mathbf{Z}=({\boldsymbol{%
z}}_{1},\ldots ,{\boldsymbol{z}}_{n})^{\text{T}}$, and $\mathbf{X=}$diag$({%
\boldsymbol{x}}_{1}^{\text{T}},\ldots ,{\boldsymbol{x}}_{n}^{\text{T}})$.
Denote $\widetilde{\mathbf{X}}=\mathbf{XW}$ and ${\mathbf{U}}=(\mathbf{Z},%
\mathbf{XW)}$. Finally, denote the scaled penalty function by
\begin{equation*}
\rho (t)=\lambda ^{-1}p_{\gamma }(t,\lambda ).
\end{equation*}

We make the following basic assumptions.

\begin{enumerate}
\item[(C1)] The function $\rho(t)$ is symmetric, non-decreasing and concave
on $[0,\infty )$. It is constant for $t\geq a\lambda $ for some constant $%
a>0 $, and $\rho (0)=0$. In addition, $\rho ^{\prime}(t)$ exists and is
continuous except for a finite number values of $t$ and $\rho
^{\prime}(0+)=1 $.

\item[(C2)] The noise vector $\boldsymbol{\varepsilon =(}\varepsilon
_{1},\ldots ,\varepsilon _{n})^{\text{T}}$ has sub-Gaussian tails such that $%
P(|\mathbf{a}^{\text{T}}\boldsymbol{\varepsilon }|>\mathbf{\|a\|}x)\leq
2\exp (-c_{1}x^{2})$ for any vector $\mathbf{a}\in
\mathop{{\rm
I}\kern-.2em\hbox{\rm R}}\nolimits^{n}$ and $x>0$, where $0<c_{1}<\infty $.

\item[(C3)] 
Assume $\sum_{i=1}^{n}z_{il}^{2}=n$ for $1\leq l\leq q$, and $%
\sum\nolimits_{i=1}^{n}x_{ij}^{2}1\{i\in \mathcal{G}_{k\text{ }%
}\}=\left\vert \mathcal{G}_{k}\right\vert $ for $1\leq j\leq p$, \newline
$\lambda _{\min }({\mathbf{U}}^{\text{T}}{\mathbf{U}})\geq C_{1}\left\vert
\mathcal{G}_{\min }\right\vert $, $\sup_{i}\Vert {\boldsymbol{x}}_{i}\Vert
\leq C_{2}\sqrt{p}$ and $\sup_{i}\Vert {\boldsymbol{z}}_{i}\Vert \leq C_{3}%
\sqrt{q}$ for some constants $0<C_{1}<\infty $, $0<C_{2}<\infty $ and $%
0<C_{3}<\infty $.
\end{enumerate}

Conditions (C1) and (C2) are common assumptions in penalized regression in
high-dimensional settings. The concave penalties such as MCP and SCAD
satisfy (C1). In the literature, it is commonly assumed that the smallest
eigenvalue of the transpose of the design matrix multiplied by the design
matrix is bounded by $C_{1}n$, which may not hold for ${\mathbf{U}}^{\text{T}%
}{\mathbf{U}}$. By some calculation and $\widetilde{\mathbf{X}}=\mathbf{XW}$%
, we have
\begin{equation*}
\widetilde{\mathbf{X}}^{\text{T}}\widetilde{\mathbf{X}}=\text{diag}%
(\sum\nolimits_{i\in \mathcal{G}_{k\text{ }}}\boldsymbol{x}_{i}\boldsymbol{x}%
_{i}^{\text{T}},k=1,\ldots ,K).
\end{equation*}%
By assuming that $\lambda _{\min }(\sum\nolimits_{i\in \mathcal{G}_{k\text{ }%
}}\boldsymbol{x}_{i}\boldsymbol{x}_{i}^{\text{T}})\mathbf{\geq }c\left\vert
\mathcal{G}_{k}\right\vert $ for some constant $0<c<\infty $, we have $%
\lambda _{\min }(\widetilde{\mathbf{X}}^{\text{T}}\widetilde{\mathbf{X}}%
)\geq c\left\vert \mathcal{G}_{\min }\right\vert $. If $\mathbf{Z}^{\text{T}}%
\widetilde{\mathbf{X}}=0$ and $\lambda _{\min }(\mathbf{Z}^{\text{T}}\mathbf{%
Z)\geq }Cn$, we have
\begin{equation*}
\lambda _{\min }({\mathbf{U}}^{\text{T}}{\mathbf{U}})=\min \{\lambda _{\min
}(\mathbf{Z}^{\text{T}}\mathbf{Z),}\lambda _{\min }(\widetilde{\mathbf{X}}^{%
\text{T}}\widetilde{\mathbf{X}})\}\geq \min (c\left\vert \mathcal{G}_{\min
}\right\vert ,Cn\mathbf{),}
\end{equation*}%
and $\left\vert \mathcal{G}_{\min }\right\vert \leq n/K$. Therefore, we let
the smallest eigenvalue in Condition (C3) be bounded below by $%
C_{1}\left\vert \mathcal{G}_{\min }\right\vert $.

\subsection{Heterogeneous model}

In this section, we study the theoretical properties of the proposed
estimator under the heterogeneous model in which there are at least two
subgroups, that is, $K\geq 2$. If the underlying groups $\mathcal{G}%
_{1},\ldots ,\mathcal{G}_{K}$ were known, the oracle estimator of $(%
\boldsymbol{\eta },\boldsymbol{\beta })$ would be
\begin{equation}
(\widehat{\boldsymbol{\eta }}^{or},\hat{\boldsymbol{\beta }}^{or})=%
\mathop{\rm argmin}_{\boldsymbol{\eta }\in \mathop{{\rm
I}\kern-.2em\hbox{\rm R}}\nolimits^{q},\,\boldsymbol{\beta }\in \mathcal{M}_{%
\mathcal{G}}}\frac{1}{2}\Vert \mathbf{y}-\mathbf{Z}\boldsymbol{\eta }-%
\mathbf{X}\boldsymbol{\beta }\Vert ^{2}.  \label{EQ:oracle}
\end{equation}%
%
%
%
%
%
%
%
%
%
%
%
%
%
%
%
%
%
%
%
%
%
%
%
%
%
%
%
%
%
%
%
Since $\boldsymbol{\beta }=\widetilde{\mathbf{W}}\boldsymbol{\alpha }$, the
oracle estimators for the common coefficient $\boldsymbol{\alpha }$ and the
coefficients $\boldsymbol{\eta }$ are 
\begin{equation*}
(\widehat{\boldsymbol{\eta }}^{or},\widehat{\boldsymbol{\alpha }}^{or})=%
\mathop{\rm argmin}_{\boldsymbol{\eta }{\in }\mathop{{\rm
I}\kern-.2em\hbox{\rm R}}\nolimits^{q},\,\boldsymbol{\alpha }\mathbf{\in }%
\mathop{{\rm I}\kern-.2em\hbox{\rm R}}\nolimits^{Kp}}\frac{1}{2}\Vert
\mathbf{y}-\mathbf{Z}\boldsymbol{\eta }-\mathbf{\widetilde{\mathbf{X}}}%
\boldsymbol{\alpha }\Vert ^{2}=({\mathbf{U}}^{\text{T}}{\mathbf{U}})^{-1}{%
\mathbf{U}}^{\text{T}}\mathbf{y.}
\end{equation*}%
%
%
%
%
%
%
%
%
%
%
%
%
Let $\boldsymbol{\alpha }_{k}^{0}$ be the true common coefficient vector for
group $\mathcal{G}_{k}$, $k=1,\ldots ,K$ and $\boldsymbol{\alpha }^{0}=((%
\boldsymbol{\alpha }_{k}^{0})^{\text{T}},k=1,\ldots ,K)^{\text{T}}$. Of
course, oracle estimators are not real estimators; they are theoretical
constructions useful for stating the properties of the proposed estimators.

\begin{theorem}
\label{THM:norm} Suppose 
$\left\vert \mathcal{G}_{\min }\right\vert \gg \sqrt{(q+Kp)n\log n}$.
Then under Conditions (C1)-(C3), we have with probability at least $%
1-2(Kp+q)n^{-1}$,%
\begin{equation}
\left\Vert ((\widehat{\boldsymbol{\eta }}^{or}-\boldsymbol{\eta }^{0})^{%
\text{T}},(\widehat{\boldsymbol{\alpha }}^{or}-\boldsymbol{\alpha }^{0})^{%
\text{T}})^{\text{T}}\right\Vert \leq \phi _{n},  \label{EQ:supnormmubeta}
\end{equation}%
and
\begin{equation*}
\left\Vert \widehat{\boldsymbol{\beta }}^{or}-\boldsymbol{\beta }%
^{0}\right\Vert \leq \sqrt{\left\vert \mathcal{G}_{\max }\right\vert }\phi
_{n},\text{ }\sup_{i}\left\Vert \widehat{\boldsymbol{\beta }}_{i}^{or}-%
\boldsymbol{\beta }_{i}^{0}\right\Vert \leq \phi _{n},
\end{equation*}%
where
\begin{equation}
\phi _{n}=c_{1}^{-1/2}C_{1}^{-1}\sqrt{q+Kp}\left\vert \mathcal{G}_{\min
}\right\vert ^{-1}\sqrt{n\log n}.  \label{EQ:phn}
\end{equation}%
Moreover, for any vector $\mathbf{a}_{n}\in
\mathop{{\rm
I}\kern-.2em\hbox{\rm R}}\nolimits^{q+Kp}$ with $||\mathbf{a}_{n}||=1$, we
have as $n\rightarrow \infty $,
\begin{equation}
\sigma _{n}(\mathbf{a}_{n})^{-1}\mathbf{a}_{n}^{\text{T}}((\widehat{%
\boldsymbol{\eta }}^{or}-\boldsymbol{\eta }^{0})^{\text{T}},(\widehat{%
\boldsymbol{\alpha }}^{or}-\boldsymbol{\alpha }^{0})^{\text{T}})^{\text{T}%
}\rightarrow _{D}N(0,1\mathbf{),}  \label{EQ:normal}
\end{equation}%
where
\begin{equation}
\sigma _{n}(\mathbf{a}_{n})=\sigma \left[ \mathbf{a}_{n}^{\text{T}}({\mathbf{%
U}}^{\text{T}}{\mathbf{U}})^{-1}\mathbf{a}_{n}\right] ^{1/2}.
\label{EQ:sign}
\end{equation}
\end{theorem}

\begin{remark}
\label{Remark4} 
Since $\left\vert \mathcal{G}_{\min }\right\vert \leq n/K$, by the condition
$\left\vert \mathcal{G}_{\min }\right\vert \gg \sqrt{(q+Kp)n\log n}$, then $%
q $, $K$ and $p$ must satisfy $K\sqrt{q+Kp}=o\{\sqrt{n(\log n)^{-1}}\}$.
\end{remark}

\begin{remark}
\label{Remark4b} By letting $\left\vert \mathcal{G}_{\min }\right\vert
=\delta n/K$ for some constant $0<\delta \leq 1$, the bound (\ref%
{EQ:supnormmubeta}) is $\phi_n=c_{1}^{-1/2}C_{1}^{-1}\delta ^{-1}K\sqrt{q+Kp}%
\sqrt{\log n/n}$. Moreover, if $q$, $K$ and $p$ are fixed quantities, then $%
\phi_n=C^{\ast }\sqrt{\log n/n}$ for some constant $0<C^{\ast }<\infty$.
\end{remark}

Let
\begin{equation*}
b_{n}=\min_{i\in \mathcal{G}_{k},j\in \mathcal{G}_{k^{\prime }},k\neq
k^{\prime }}\Vert \boldsymbol{\beta }_{i}^{0}-\boldsymbol{\beta }%
_{j}^{0}\Vert =\min_{k\neq k^{\prime }}\Vert \boldsymbol{\alpha }_{k}^{0}-%
\boldsymbol{\alpha }_{k^{\prime }}^{0}\Vert
\end{equation*}%
be the minimal difference of the common values between two groups.

\begin{theorem}
\label{THM:selection} Suppose the conditions in Theorem \ref{THM:norm} hold.
If $b_{n}>a\lambda $ and $\lambda \gg \phi _{n}$, for some constant $a>0$,
where $\phi _{n}$ is given in (\ref{EQ:phn}), then there exists a local
minimizer $(\widehat{\boldsymbol{\eta }}(\lambda ),\widehat{\boldsymbol{%
\beta }}(\lambda ))$ of the objective function $Q_{n}(\boldsymbol{\eta },%
\boldsymbol{\beta }{;\lambda })$ given in (\ref{EQ:objective}) satisfying
\begin{equation*}
P\left( (\widehat{\boldsymbol{\eta }}(\lambda ),\widehat{\boldsymbol{\beta }}%
(\lambda ))=(\widehat{\boldsymbol{\eta }}^{or},\widehat{\boldsymbol{\beta }}%
^{or})\right) \rightarrow 1.
\end{equation*}
\end{theorem}

\begin{remark}
\label{Remark5} 
Theorem \ref{THM:selection} shows that the oracle estimator $(\widehat{%
\boldsymbol{\eta }}^{or},\widehat{\boldsymbol{\beta }}^{or})$ is a local
minimizer of the objective function with a high probability, and thus the
true groups can be recovered with the estimated common value for group $k$
given as $\widehat{\boldsymbol{\alpha }}_{k}(\lambda )=\widehat{\boldsymbol{%
\beta }}_{i}^{or}$ for $i\in \mathcal{G}_{k}$. This result holds given that $%
b_{n}\gg \phi _{n}$. As discussed in Remark \ref{Remark4b}, when $K$, $p$
and $q$ are finite and fixed numbers and $\left\vert \mathcal{G}_{\min
}\right\vert =\delta n/K$ for some constant $0<\delta \leq 1$, $b_{n}\gg
C^{\ast }\sqrt{\log n/n}$ for some constant $0<C^{\ast }<\infty $.
\end{remark}

Let $\widehat{\boldsymbol{\alpha }}(\lambda )=(\widehat{\boldsymbol{\alpha }}%
_{1}(\lambda )^{\text{T}},\ldots ,\widehat{\boldsymbol{\alpha }}_{K}(\lambda
)^{\text{T}})^{\text{T}}$ be the estimated treatment effects such that $%
\widehat{\boldsymbol{\alpha }}_{k}(\lambda )=\widehat{\boldsymbol{\beta }}%
_{i}(\lambda )$ for $i\in \mathcal{G}_{k}$, where $k=1,\ldots ,K$, and $%
\widehat{\boldsymbol{\beta }}(\lambda )=\{\widehat{\boldsymbol{\beta }}%
_{i}(\lambda )^{\text{T}},1\leq i\leq n\}^{\text{T}}$ is the local minimizer
given in Theorem \ref{THM:selection}. Based on the results in Theorems \ref%
{THM:norm} and \ref{THM:selection}, we obtain the asymptotic distribution of
$(\widehat{\boldsymbol{\eta }}(\lambda )^{\text{T}},\widehat{\boldsymbol{%
\alpha }}(\lambda )^{\text{T}})^{\text{T}}$ given in the following corollary.

\begin{corollary}
\label{COR:distribution} Under the conditions in Theorem \ref{THM:selection}%
, we have for any $\mathbf{a}_{n}\in \mathop{{\rm I}\kern-.2em\hbox{\rm R}}%
\nolimits^{q+Kp}$ with $||\mathbf{a}_{n}||=1$, as $n\rightarrow \infty $,
\begin{equation*}
\sigma _{n}(\mathbf{a}_{n})^{-1}\mathbf{a}_{n}^{\text{T}}((\widehat{%
\boldsymbol{\eta }}(\lambda )-\boldsymbol{\eta }^{0})^{\text{T}},(\widehat{%
\boldsymbol{\alpha }}(\lambda )-\boldsymbol{\alpha }^{0})^{\text{T}})^{\text{%
T}}\rightarrow _{D}N(0,1\mathbf{),}
\end{equation*}%
with $\sigma _{n}(\mathbf{a}_{n})$ given in (\ref{EQ:sign}). As a result, we
have for any vectors $\mathbf{a}_{n1}\in
\mathop{{\rm I}\kern-.2em\hbox{\rm
R}}\nolimits^{q}$ with $||\mathbf{a}_{n1}||=1$ and $\mathbf{a}_{n2}\in
\mathop{{\rm I}\kern-.2em\hbox{\rm
R}}\nolimits^{Kp}$ $||\mathbf{a}_{n2}||=1$, as $n\rightarrow \infty $,
\begin{equation*}
\sigma _{n1}^{-1}(\mathbf{a}_{n1})\mathbf{a}_{n1}^{\text{T}}(\widehat{%
\boldsymbol{\eta }}(\lambda )-\boldsymbol{\eta }^{0})\rightarrow _{D}N(0,1)%
\text{ and }\sigma _{n2}^{-1}(\mathbf{a}_{n2})\mathbf{a}_{n2}^{\text{T}}(%
\widehat{\boldsymbol{\alpha }}(\lambda )-\boldsymbol{\alpha }%
^{0})\rightarrow _{D}N(0,1),
\end{equation*}%
where
\begin{eqnarray*}
\sigma _{n1}(\mathbf{a}_{n1}) &=&\sigma \left[ \mathbf{a}_{n1}^{\text{T}}[{%
\mathbf{Z}}^{\text{T}}{\mathbf{Z}}-{\mathbf{Z}}^{\text{T}}\widetilde{\mathbf{%
X}}(\widetilde{\mathbf{X}}^{\text{T}}\widetilde{\mathbf{X}})^{-1}\widetilde{%
\mathbf{X}}^{\text{T}}{\mathbf{Z}}]^{-1}\mathbf{a}_{n1}\right] ^{1/2}, \\
\sigma _{n2}(\mathbf{a}_{n2}) &=&\sigma \left[ \mathbf{a}_{n2}^{\text{T}}[%
\widetilde{\mathbf{X}}^{\text{T}}\widetilde{\mathbf{X}}-\widetilde{\mathbf{X}%
}^{\text{T}}{\mathbf{Z}}({\mathbf{Z}}^{\text{T}}{\mathbf{Z}})^{-1}{\mathbf{Z}%
}^{\text{T}}\widetilde{\mathbf{X}}]^{-1}\mathbf{a}_{n2}\right] ^{1/2}.
\end{eqnarray*}
\end{corollary}

\begin{remark}
\label{Remark6star} 
{From the oracle property in Theorem \ref{THM:selection}, we have that $P(%
\widehat{K}=K)\rightarrow 1$, $\widehat{K}$ is the estimated number of
subgroups. Moreover, since }$\widehat{\boldsymbol{\beta }}^{or}=\widetilde{%
\mathbf{W}}\boldsymbol{\alpha }^{or}$ and $\widehat{\boldsymbol{\beta }}%
(\lambda )=\widehat{\widetilde{\mathbf{W}}}\widehat{\boldsymbol{\alpha }}%
(\lambda )$, then {$P(\widehat{\widetilde{\mathbf{W}}}=\widetilde{\mathbf{W}}%
)\rightarrow 1$, where }$\widehat{\widetilde{\mathbf{W}}}=\mathbf{\{}%
\widehat{w}_{ik}\}$ with $\widehat{w}_{ik}=1$ for $i\in \widehat{\mathcal{G}}%
_{k\text{ }}$ and $w_{ik}=0$ otherwise. Hence, the subgroup memberships can
be recovered with a high probability.
\end{remark}

\begin{remark}
\label{Remark6} 
The asymptotic distribution of the penalized estimators provides a
theoretical justification for further conducting statistical inference about
heterogeneity. By the results in Corollary \ref{COR:distribution}, for given
$\mathbf{a}_{n1}\in \mathop{{\rm I}\kern-.2em\hbox{\rm R}}\nolimits^{q}$ and
$\mathbf{a}_{n2}\in \mathop{{\rm I}\kern-.2em\hbox{\rm R}}\nolimits^{Kp}$, $%
100(1-\alpha )\%$ confidence intervals for $\mathbf{a}_{n1}^{\text{T}}%
\boldsymbol{\eta }^{0}$ and $\mathbf{a}_{n2}^{\text{T}}\mathbf{\boldsymbol{%
\alpha }}^{0}$ are given by
\begin{equation*}
\mathbf{a}_{n1}^{\text{T}}\widehat{\boldsymbol{\eta }}(\lambda )\pm
z_{\alpha /2}\widehat{\sigma }_{n1}(\mathbf{a}_{n1})\ \text{ and }\ \mathbf{a%
}_{n2}^{\text{T}}\widehat{\boldsymbol{\alpha }}(\lambda )\pm z_{\alpha /2}%
\widehat{\sigma }_{n2}(\mathbf{a}_{n2}),
\end{equation*}%
respectively, where $z_{\alpha /2}$ is the $(1-\alpha /2)100$ percentile of
the standard normal, and $\widehat{\sigma }_{n1}(\mathbf{a}_{n1})$ and $%
\widehat{\sigma }_{n2}(\mathbf{a}_{n2})$ are estimates of $\sigma _{n1}(%
\mathbf{a}_{n1})$ and $\sigma _{n2}(\mathbf{a}_{n2})$ with $\sigma ^{2}$
estimated by
\begin{equation*}
\widehat{\sigma }^{2}=(n-q-\widehat{K}p)^{-1}\sum\nolimits_{i=1}^{n}(y_{i}-%
\boldsymbol{z}_{i}^{\text{T}}\widehat{\boldsymbol{\eta }}-\boldsymbol{x}%
_{i}^{\text{T}}\widehat{\boldsymbol{\beta }}_{i})^{2}.
\end{equation*}
\end{remark}

\subsection{Homogeneous model}

When the true model is the homogeneous model given as $y_{i}={\boldsymbol{z}}%
_{i}^{\text{T}}\boldsymbol{\eta }+{\boldsymbol{x}}_{i}^{\text{T}}\boldsymbol{%
\alpha }+\varepsilon _{i},i=1,\ldots ,n$, we have $\boldsymbol{\beta }%
_{1}=\cdots =\boldsymbol{\beta }_{n}=\boldsymbol{\alpha }$ and $K=1$. The
penalized estimator $(\widehat{\boldsymbol{\eta }}(\lambda ),\widehat{%
\boldsymbol{\beta }}(\lambda ))$ of $(\boldsymbol{\eta },\boldsymbol{\beta }%
) $, where $\boldsymbol{\beta }=(\boldsymbol{\beta }_{1}^{\text{T}},\ldots ,%
\boldsymbol{\beta }_{n}^{\text{T}})^{\text{T}}$, also has the oracle
property given as follows. We define the oracle estimator for $(\boldsymbol{%
\eta },\boldsymbol{\alpha })$ as%
\begin{eqnarray*}
(\widehat{\boldsymbol{\eta }}^{or},\widehat{\boldsymbol{\alpha }}^{or}) &=&%
\mathop{\rm argmin}_{\boldsymbol{\eta }{\in }\mathop{{\rm
I}\kern-.2em\hbox{\rm R}}\nolimits^{q},\,\boldsymbol{\alpha }\mathbf{\in }%
\mathop{{\rm I}\kern-.2em\hbox{\rm R}}\nolimits^{p}}\frac{1}{2}\Vert \mathbf{%
y}-\mathbf{Z}\boldsymbol{\eta }-\mathbf{x}\boldsymbol{\alpha }\Vert ^{2} \\
&=&({\mathbf{U}}^{\ast \text{T}}{\mathbf{U}}^{\ast })^{-1}{\mathbf{U}}^{\ast
\text{T}}\mathbf{y},
\end{eqnarray*}%
where $\mathbf{x=}({\boldsymbol{x}}_{1},\ldots ,{\boldsymbol{x}}_{n})^{\text{%
T}}$ and ${\mathbf{U}}^{\ast }=(\mathbf{Z,x)}$. Let $\widehat{\boldsymbol{%
\beta }}^{or}=(\widehat{\boldsymbol{\beta }}_{1}^{or\text{T}},\ldots ,%
\widehat{\boldsymbol{\beta }}_{n}^{or\text{T}})^{\text{T}}$, where $\widehat{%
\boldsymbol{\beta }}_{i}^{or}=\widehat{\boldsymbol{\alpha }}^{or}$ for all $%
i $. Let $\boldsymbol{\eta }^{0}$ and $\boldsymbol{\alpha }^{0}$ be the true
coefficient vectors. We introduce the following condition.

\begin{enumerate}
\item[(C3$^{\ast }$)]
Assume $\sum_{i=1}^{n}z_{il}^{2}=n$ for $1\leq l\leq q$, and $%
\sum\nolimits_{i=1}^{n}x_{ij}^{2}=n$ for $1\leq j\leq p$, $\lambda _{\min }({%
\mathbf{U}}^{\ast \text{T}}{\mathbf{U}}^{\ast })\geq C_{1}n$, $\sup_{i}\Vert
\mathbf{x}_{i}\Vert \leq C_{2}\sqrt{p}$ and $\sup_{i}\Vert \boldsymbol{z}%
_{i}\Vert \leq C_{3}\sqrt{q}$ for some constants $0<C_{1}<\infty $, $%
0<C_{2}<\infty $ and $0<C_{3}<\infty $.
\end{enumerate}

\begin{theorem}
\label{THM:normhomo} Suppose Conditions (C1), (C2)\ and (C3$^{\ast }$)\
hold. If $p=o(n(\log n)^{-1})$ and $q=o(n(\log n)^{-1})$, the oracle
estimator has the property that with probability at least $1-2(p+q)n^{-1}$,
\begin{eqnarray}
\left\Vert ((\widehat{\boldsymbol{\eta }}^{or}-\boldsymbol{\eta }^{0})^{%
\text{T}},(\widehat{\boldsymbol{\alpha }}^{or}-\boldsymbol{\alpha }^{0})^{%
\text{T}})^{\text{T}}\right\Vert &\leq &\phi _{n},  \notag \\
\sup_{i}\left\Vert \widehat{\boldsymbol{\beta }}_{i}^{or}-\boldsymbol{\beta }%
_{i}^{0}\right\Vert &\leq &\phi _{n},  \label{EQ:supnormbetahomo}
\end{eqnarray}%
where
\begin{equation*}
\phi _{n}=c_{1}^{-1/2}C_{1}^{-1}\sqrt{q+p}\sqrt{n^{-1}\log n},
\end{equation*}%
in which $c_{1}$ and $C_{1}$ are given in Conditions (C2) and (C3$^{\ast }$%
), respectively, and for any vector $\mathbf{a}_{n}\in
\mathop{{\rm
I}\kern-.2em\hbox{\rm R}}\nolimits^{q+p}$ with $||\mathbf{a}_{n}||=1$, as $%
n\rightarrow \infty $,
\begin{equation}
\sigma _{n}(\mathbf{a}_{n})^{-1}\mathbf{a}_{n}^{\text{T}}((\widehat{%
\boldsymbol{\eta }}^{or}-\boldsymbol{\eta }^{0})^{\text{T}},(\widehat{%
\boldsymbol{\alpha }}^{or}-\boldsymbol{\alpha }^{0})^{\text{T}})^{\text{T}%
}\rightarrow N(0,1\mathbf{),}  \label{EQ:normhomo}
\end{equation}%
where
\begin{equation*}
\sigma _{n}(\mathbf{a}_{n})=\sigma \left[ \mathbf{a}_{n}^{\text{T}}({\mathbf{%
U}}^{\ast \text{T}}{\mathbf{U}}^{\ast })^{-1}\mathbf{a}_{n}\right] ^{1/2}.
\end{equation*}%
Moreover, if $\lambda \gg \phi _{n}$, then there exists a local minimizer $(%
\widehat{\boldsymbol{\eta }}(\lambda ),\widehat{\boldsymbol{\beta }}(\lambda
))$ of the objective function $Q_{n}(\boldsymbol{\eta },\boldsymbol{\beta }{%
;\lambda })$ given in (\ref{EQ:objective}) satisfying
\begin{equation}
P\left( (\widehat{\boldsymbol{\eta }}(\lambda ),\widehat{\boldsymbol{\beta }}%
(\lambda ))=(\widehat{\boldsymbol{\eta }}^{or},\widehat{\boldsymbol{\beta }}%
^{or})\right) \rightarrow 1.  \label{EQ:penhomo}
\end{equation}
\end{theorem}

\begin{remark}
\label{Remark7} By Theorem \ref{THM:normhomo}, the local minimizer $\widehat{%
\boldsymbol{\beta }}_{i}(\lambda )=\widehat{\boldsymbol{\alpha }}(\lambda )=%
\widehat{\boldsymbol{\alpha }}^{or}$ for all $i$ with probability
approaching 1. Then, we have for any vectors $\mathbf{a}_{n1}\in
\mathop{{\rm I}\kern-.2em\hbox{\rm
R}}\nolimits^{q}$ with $||\mathbf{a}_{n1}||=1$ and $\mathbf{a}_{n2}\in
\mathop{{\rm I}\kern-.2em\hbox{\rm
R}}\nolimits^{p}$ with $||\mathbf{a}_{n2}||=1$, as $n\rightarrow \infty $,
\begin{equation*}
\sigma _{n1}^{-1}(\mathbf{a}_{n1})\mathbf{a}_{n1}^{\text{T}}(\widehat{%
\boldsymbol{\eta }}(\lambda )-\boldsymbol{\eta }^{0})\rightarrow _{D}N(0,1)%
\text{ and }\sigma _{n2}^{-1}(\mathbf{a}_{n2})\mathbf{a}_{n2}^{\text{T}}(%
\widehat{\boldsymbol{\alpha }}(\lambda )-\boldsymbol{\alpha }%
^{0})\rightarrow _{D}N(0,1),
\end{equation*}%
where
\begin{eqnarray*}
\sigma _{n1}(\mathbf{a}_{n1}) &=&\sigma \left[ \mathbf{a}_{n1}^{\text{T}}[{%
\mathbf{Z}}^{\text{T}}{\mathbf{Z}}-{\mathbf{Z}}^{\text{T}}\mathbf{x}(\mathbf{%
x}^{\text{T}}\mathbf{x})^{-1}\mathbf{x}^{\text{T}}{\mathbf{Z}}]^{-1}\mathbf{a%
}_{n1}\right] ^{1/2}, \\
\sigma _{n2}(\mathbf{a}_{n2}) &=&\sigma \left[ \mathbf{a}_{n2}^{\text{T}}[%
\mathbf{x}^{\text{T}}\mathbf{x}-\mathbf{x}^{\text{T}}{\mathbf{Z}}({\mathbf{Z}%
}^{\text{T}}{\mathbf{Z}})^{-1}{\mathbf{Z}}^{\text{T}}\mathbf{x}]^{-1}\mathbf{%
a}_{n2}\right] ^{1/2}.
\end{eqnarray*}
\end{remark}

\section{Testing of a homogeneous model\label{SEC:bootstrap}}

Next, we propose a residual bootstrapping procedure to test homogeneity of
the treatment effects, i.e., to test whether the model is the homogeneous
model, given as $y_{i}={\boldsymbol{z}}_{i}^{\text{T}}\boldsymbol{\eta }+{%
\boldsymbol{x}}_{i}^{\text{T}}\boldsymbol{\alpha }+\varepsilon
_{i},i=1,\ldots ,n$. We consider this model as the reduced model, and the
full model is given in (\ref{Mod1}). We estimate the reduced model by OLS
and obtain the resulting estimators as $\widehat{\boldsymbol{\eta }}^{\text{R%
}}$ and $\widehat{\boldsymbol{\alpha }}^{\text{R}}$. We then estimate the
full model by our proposed method and denote the resulting estimators as $%
\widehat{\boldsymbol{\eta }}^{\text{F}}$ and $\widehat{\boldsymbol{\beta }}%
_{i}^{\text{F}}$. Let the fitted values be $\widehat{\mu }_{i}^{\text{R}}={%
\boldsymbol{z}}_{i}^{\text{T}}\widehat{\boldsymbol{\eta }}^{\text{R}}+{%
\boldsymbol{x}}_{i}^{\text{T}}\widehat{\boldsymbol{\alpha }}^{\text{R}}$ and
$\widehat{\mu }_{i}^{\text{F}}={\boldsymbol{z}}_{i}^{\text{T}}\widehat{%
\boldsymbol{\eta }}^{\text{F}}+{\boldsymbol{x}}_{i}^{\text{T}}\widehat{%
\boldsymbol{\beta }}_{i}^{\text{F}}$ for the reduced and full models,
respectively. Borrowing the idea from \cite{Hardle.Mammen:1993}, we use the
integrated squared deviation between $\widehat{\mu }_{i}^{\text{R}}$ and $%
\widehat{\mu }_{i}^{\text{F}}$ as the test statistic, which would be $%
\mathcal{T}_{n}=\sum\nolimits_{i=1}^{n}(\widehat{\mu }_{i}^{\text{F}}-%
\widehat{\mu }_{i}^{\text{R}})^{2}/n$. Let the residuals be $\widehat{%
\epsilon }_{i}=y_{i}-\widehat{\mu }_{i}^{\text{F}}$ for $i=1,\ldots ,n$. We
obtain a randomly resampled residual $\widehat{\epsilon }_{i}^{\ast }$ and
then create synthetic response variables $y_{i}^{\ast }=\widehat{\mu }_{i}^{%
\text{R}}+\widehat{\epsilon }_{i}^{\ast }$. Using $\left( {\boldsymbol{z}}%
_{i},{\boldsymbol{x}}_{i},y_{i}^{\ast }\right) $ as bootstrap observations,
we obtain the fitted values $\widehat{\mu }_{i}^{\ast \text{R}}$ and $%
\widehat{\mu }_{i}^{\ast \text{F}}$, respectively, by refitting the reduced
and full models, and then creat the bootstrapped version of the test
statistic, denoted as $\mathcal{T}_{n}^{\ast }=\sum\nolimits_{i=1}^{n}(%
\widehat{\mu }_{i}^{\ast \text{F}}-\widehat{\mu }_{i}^{\ast \text{R}})^{2}/n$%
. Using the Monte Carlo simulations to approximate the conditional
distribution $\mathcal{L}^{\ast }(\mathcal{T}_{n}^{\ast })=\mathcal{L(T}%
_{n}^{\ast }|({\boldsymbol{z}}_{i},{\boldsymbol{x}}_{i})_{i=1}^{n})$, we
obtain the $(1-\alpha )^{\text{th}}$ quantile $\widehat{t}_{\alpha }$ and
reject the hypothesis of homogeneity if $\mathcal{T}_{n}>\widehat{t}_{\alpha
}$ at the significance level $\alpha $. Alternatively, we can obtain the $P$%
-value $p_{\text{v}}$ by finding the $(1-p_{\text{v}})^{\text{th}}$ quantile
$\widehat{t}_{p_{\text{v}}}$ which satisfies $\widehat{t}_{p_{\text{v}}}=%
\mathcal{T}_{n}$.

\section{Numerical studies}

\label{SEC:examples}

\subsection{Simulation studies}


We use the modified Bayes Information Criterion (BIC) %
\citep{wang.li.tsai:2007} 
for high-dimensional data settings to select the tuning parameter by
minimizing%
\begin{equation}
\text{BIC}(\lambda )=\log [\sum\nolimits_{i=1}^{n}(y_{i}-{\boldsymbol{z}}%
_{i}^{\text{T}}\widehat{\boldsymbol{\eta }}(\lambda )-{\boldsymbol{x}}_{i}^{%
\text{T}}\widehat{\boldsymbol{\beta }}_{i}(\lambda ))^{2}/n]+C_{n}\frac{\log
n}{n}(\widehat{K}(\lambda )p+q),  \label{EQ:BIC}
\end{equation}%
where $C_{n}$ is a positive number which can depend on $n$. When $C_{n}=1$,
the modified BIC reduces to the traditional BIC \citep{schwarz:1978}.
Following \cite{lee.noh.park:2014}, we use $C_{n}=\log (np+q)$.
We select $\lambda $ by minimizing the modified BIC.

One important evaluation criterion for clustering methods is their ability
to reconstruct the true underlying cluster structure. We, therefore, use the
Rand Index measure \citep{Rand:1971} to evaluate the accuracy of the
clustering results. The Rand Index is viewed as a measure of the percentage
of correct decisions made by an algorithm. It is computed by using the
formula:%
\begin{equation}
\text{RI}=\frac{\text{TP}+\text{TN}}{\text{TP}+\text{FP}+\text{FN}+\text{TN}}%
,  \label{EQ:RI}
\end{equation}%
where a true positive (TP) decision assigns two observations from the same
ground truth group to the same cluster, a true negative (TN) decision
assigns two observations from different groups to different clusters, a
false positive (FP) decision assigns two observations from different groups
to the same cluster, and a false negative (FN) decision assigns two
observations from the same group to different clusters. The Rand Index lies
between 0 and 1. Higher values of the Rand Index indicate better performance
of the algorithm.

\textbf{Example 1} (Two subgroups). We simulate data from the heterogeneous
model with two groups:
\begin{equation}
y_{i}=\boldsymbol{z}_{i}^{\text{T}}\boldsymbol{\eta }+{\boldsymbol{x}}_{i}^{%
\text{T}}\boldsymbol{\beta }_{i}+\varepsilon _{i},i=1,\ldots ,n\text{, with }%
\boldsymbol{\beta }_{i}=\boldsymbol{\alpha }_{1}w_{i1}+\boldsymbol{\alpha }%
_{2}w_{i2},  \label{EQ:DGP1}
\end{equation}%
where $\boldsymbol{z}_{i}=(z_{i1},z_{i2},z_{i3})^{\text{T}}\sim \mathcal{N}(%
\mathbf{0},\boldsymbol{\Sigma })$, in which $\boldsymbol{\Sigma }=\{\sigma
_{jj^{\prime }}\}$, $\sigma _{jj}=1$ and $\sigma _{jj^{\prime }}=0.3$ for $%
j\neq j^{\prime }$, and ${\boldsymbol{x}}_{i}=(1,x_{i})^{\text{T}}$, in
which $x_{i}$ is simulated from centered and standardized binomial with
probability $0.7$ for one outcome. We simulate the error terms $\varepsilon
_{i}$ from independent $N(0,0.5^{2})$. Let $\boldsymbol{\eta }=(1,1,1)^{%
\text{T}}$, $\boldsymbol{\alpha }_{1}=(2,2)^{\text{T}}$ and $\boldsymbol{%
\alpha }_{2}=(0,0)^{\text{T}}$. Moreover, let $%
w_{i1}=I(z_{i1}^{2}+u_{i}-1<0) $ and $w_{i2}=1-w_{i1}$, where $u_{i}\sim
\mathcal{N}(0,1)$.

\begin{table}[tbph]
\caption{The sample mean, median and standard deviation (s.d.) of $\protect%
\widehat{K}$, the Rand Index (RI) value and the percentage (per) of $\protect%
\widehat{K}$ equaling to the true number of subgroups by MCP and SCAD based
on 500 replications with $n=200,400$ in Example 1.}
\label{TAB:KhatEX1}
\begin{center}
\begin{tabular*}{0.95\textwidth}{l|@{\extracolsep{\fill}}ccccc|ccccc}
\hline
& \multicolumn{5}{c|}{$n=200$} & \multicolumn{5}{c}{$n=400$} \\ \hline
& mean & median & s.d. & RI & per & mean & median & s.d. & RI & per \\ \hline
MCP & 2.100 & 2.000 & 0.302 & 0.799 & 0.890 & 2.080 & 2.000 & 0.272 & 0.826
& 0.920 \\
SCAD & 2.100 & 2.000 & 0.303 & 0.799 & 0.890 & 2.080 & 2.000 & 0.273 & 0.825
& 0.920 \\ \hline
\end{tabular*}%
\end{center}
\end{table}

We select the $\lambda $ value by minimizing the modified BIC given in (\ref%
{EQ:BIC}). Table \ref{TAB:KhatEX1} reports the sample mean, median and
standard deviation (s.d.) of the estimated number of groups $\widehat{K}$,
the average value of Rand Index (RI) defined in (\ref{EQ:RI}) for measuring
clustering accuracy, and the percentage (per) of $\widehat{K}$ equaling to
the true number of subgroups by the MCP and SCAD methods based on $500$
simulation realizations with $n=200,400$. 
The median of $\widehat{K}$ is 2 which is the true number of subgroups for
all cases. As $n$ increases, the mean gets closer to 2 and the standard
deviation becomes smaller. Moreover, the Rand Index (RI) value and the
percentage of correctly selecting the number of subgroups become closer to 1
as $n$ increases.

\begin{table}[tbph]
\caption{The sample mean, median and asymptotic standard error (ASE) of the
estimators $\protect\widehat{\boldsymbol{\protect\alpha }}_{1}$ and $\protect%
\widehat{\boldsymbol{\protect\alpha }}_{2}$ by MCP and SCAD and oracle
estimators $\protect\widehat{\boldsymbol{\protect\alpha }}_{1}^{or}$ and $%
\protect\widehat{\boldsymbol{\protect\alpha }}_{2}^{or}$ based on 500
replications with $n=200,400$ in Example 1.}
\label{TAB:alphaEX1}
\begin{center}
\begin{tabular*}{0.95\textwidth}{l|c|@{\extracolsep{\fill}}rrr|rrr}
\hline
&  & \multicolumn{3}{c|}{$n=200$} & \multicolumn{3}{c}{$n=400$} \\ \hline
&  & mean & median & ASE & mean & median & ASE \\ \hline
$\widehat{\alpha }_{11}$ & MCP & 2.016 & 2.019 & 0.046 & 2.012 & 2.014 &
0.034 \\
& SCAD & 2.016 & 2.018 & 0.046 & 2.012 & 2.014 & 0.034 \\
$\widehat{\alpha }_{11}^{or}$ &  & 1.996 & 1.997 & 0.047 & 2.009 & 2.009 &
0.033 \\ \hline
$\widehat{\alpha }_{12}$ & MCP & 1.927 & 1.929 & 0.045 & 1.947 & 1.945 &
0.034 \\
& SCAD & 1.927 & 1.929 & 0.045 & 1.947 & 1.946 & 0.034 \\
$\widehat{\alpha }_{12}^{or}$ &  & 1.990 & 1.989 & 0.046 & 1.999 & 1.999 &
0.034 \\ \hline
$\widehat{\alpha }_{21}$ & MCP & 0.016 & 0.016 & 0.053 & 0.006 & 0.010 &
0.038 \\
& SCAD & 0.016 & 0.016 & 0.053 & 0.006 & 0.010 & 0.038 \\
$\widehat{\alpha }_{21}^{or}$ &  & 0.006 & 0.010 & 0.055 & 0.009 & 0.009 &
0.039 \\ \hline
$\widehat{\alpha }_{22}$ & MCP & 0.087 & 0.086 & 0.054 & 0.086 & 0.084 &
0.038 \\
& SCAD & 0.087 & 0.086 & 0.054 & 0.086 & 0.084 & 0.038 \\
$\widehat{\alpha }_{22}^{or}$ &  & -0.011 & -0.014 & 0.055 & 0.001 & 0.001 &
0.039 \\ \hline
\end{tabular*}%
\end{center}
\end{table}

To further study the estimation accuracy and evaluate 
the asymptotic properties stated in Section \ref{SEC:theory}, Table \ref%
{TAB:alphaEX1} presents the sample mean, median and asymptotic standard
error (ASE) obtained according to Corollary \ref{COR:distribution} of the
estimators $\widehat{\boldsymbol{\alpha }}_{1}=(\widehat{\alpha }_{11},%
\widehat{\alpha }_{12})^{\text{T}}$ and $\widehat{\boldsymbol{\alpha }}_{2}=(%
\widehat{\alpha }_{21},\widehat{\alpha }_{22})^{\text{T}}$ by the MCP and
SCAD methods and oracle estimators $\widehat{\boldsymbol{\alpha }}_{1}^{or}=(%
\widehat{\alpha }_{11}^{or},\widehat{\alpha }_{12}^{or})^{\text{T}}$ and $%
\widehat{\boldsymbol{\alpha }}_{2}^{or}=(\widehat{\alpha }_{21}^{or},%
\widehat{\alpha }_{22}^{or})^{\text{T}}$ based on 500 replications with $%
n=200$ and $400$. 
The medians and means of $\widehat{\boldsymbol{\alpha }}_{1}$ and $\widehat{%
\boldsymbol{\alpha }}_{2}$ are close to the true values 2 and 0 for all
cases. Moreover, the asymptotic standard errors of the penalized estimators $%
\widehat{\boldsymbol{\alpha }}_{1}$ and $\widehat{\boldsymbol{\alpha }}_{2}$
are close to those of the oracle estimators $\widehat{\boldsymbol{\alpha }}%
_{1}^{or}$ and $\widehat{\boldsymbol{\alpha }}_{2}^{or}$. This
supports the oracle property established in Theorem \ref{THM:selection}.

\begin{figure}[h]
\caption{\textit{The boxplots of the MSEs of $\protect\widehat{\boldsymbol{%
\protect\eta }}$ using MCP and SCAD, respectively, with $n=200$ (white)\ and
$n=400$ (grey) in Example 1.}}
\label{FIG:etaEX1}
\begin{center}
\vspace{-0.6cm} $\includegraphics[width=3in]{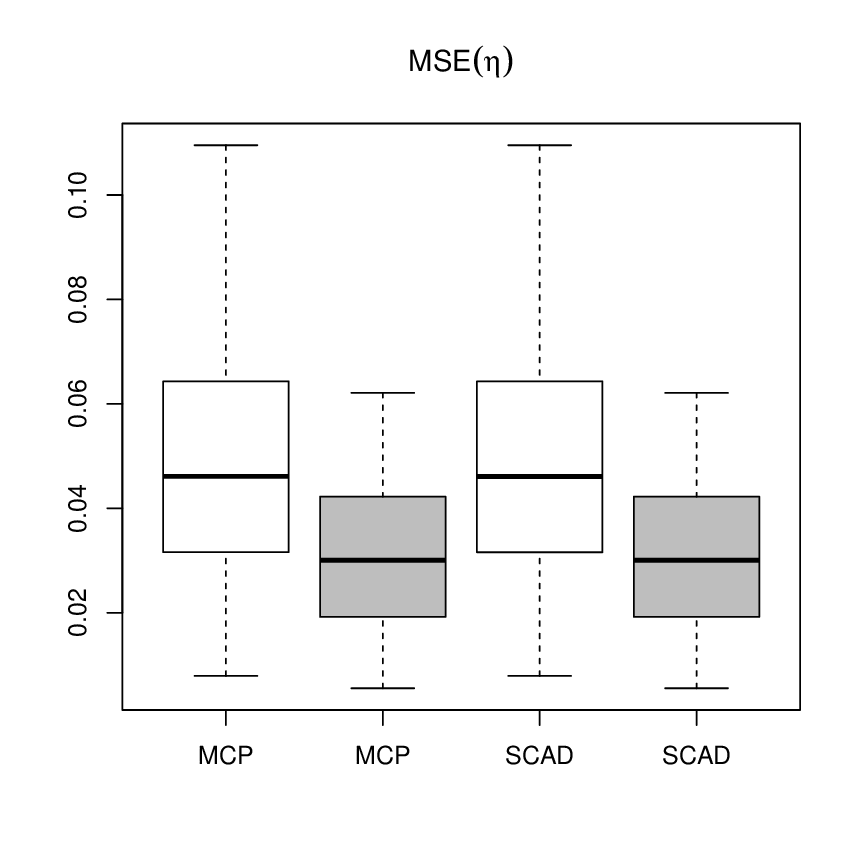} $%
\end{center}
\end{figure}

Next, we calculate the mean squared error (MSE) of the estimates $\widehat{%
\boldsymbol{\eta }}$ by using the formula $\Vert \widehat{\boldsymbol{\eta }}%
-\boldsymbol{\eta }\Vert /\sqrt{q}$. Figure \ref{FIG:etaEX1} depicts the
boxplots of the MSEs of $\widehat{\boldsymbol{\eta }}$ by the MCP and SCAD,
respectively, at $n=200$ (white)\ and $n=400$ (grey). The MCP\ and SCAD
result in similar MSEs of $\widehat{\boldsymbol{\eta }}$. The MSE values
decrease as $n$ increases for both MCP and SCAD.

Lastly, we fit the two-group logistic-normal mixture model given in (\ref%
{MOD:mix}) using the R package \textquotedblleft flexmix\textquotedblright ,
and obtain the average values of RI, which are $0.516$ and $0.518$, for $%
n=200$ and $400$, respectively, based on 500 simulation realizations. We see
that our method leads to higher RI values than the structured mixture
modelling approach, even though our method does not need to specify the true
number of subgroups beforehand. Table \ref{Table:mixture} reports the sample
mean, median and empirical standard deviation (ESD) of the estimators $%
\widehat{\boldsymbol{\alpha }}_{1}$ and $\widehat{\boldsymbol{\alpha }}_{2}$
by using the clustering result from the mixture modelling approach to fit
regressions of the two groups. Clearly, the estimated values from our method
given in Table \ref{TAB:alphaEX1} are closer to the true values $\boldsymbol{%
\alpha }_{1}$ and $\boldsymbol{\alpha }_{2}$ than those obtained from the
mixture modelling approach given in Table \ref{Table:mixture}. {\color{black} The mixture modelling method requires the indicator function which is $w_{i}=I({\boldsymbol{z}}_{i}^{\text{T}}%
\boldsymbol{\gamma }+\epsilon >0)$ given in (\ref{MOD:mix}) to be a parametric linear function of $\boldsymbol{z}_{i}$, so that it fits a mis-specified model for this example. As a result, it leads to biased estimates of $\boldsymbol{\alpha }_{1}$ and $\boldsymbol{\alpha }_{2}$.}
\begin{table}[tbph]
\caption{The sample mean, median and empirical standard deviation (ESD) of
the estimators $\protect\widehat{\boldsymbol{\protect\alpha }}_{1}$ and $%
\protect\widehat{\boldsymbol{\protect\alpha }}_{2}$ by the mixture modelling
approach based on 500 replications in Example 1.}
\label{Table:mixture}
\begin{center}
\begin{tabular*}{0.95\textwidth}{c|@{\extracolsep{\fill}}ccc|ccc}
\hline
& \multicolumn{3}{c|}{$n=200$} & \multicolumn{3}{c}{$n=400$} \\ \hline
& mean & median & ESD & mean & median & ESD \\ \hline
$\widehat{\alpha }_{11}$ & 1.834 & 1.809 & 0.266 & 1.819 & 1.794 & 0.179 \\
$\widehat{\alpha }_{12}$ & 1.254 & 1.262 & 0.137 & 1.263 & 1.266 & 0.090 \\
$\widehat{\alpha }_{21}$ & 0.809 & 0.842 & 0.419 & 0.786 & 0.796 & 0.343 \\
$\widehat{\alpha }_{22}$ & 0.949 & 0.977 & 0.210 & 0.940 & 0.957 & 0.151 \\
\hline
\end{tabular*}%
\end{center}
\end{table}

\textbf{Example 2} (Three subgroups). We simulate data from the
heterogeneous model with three groups:
\begin{equation}
y_{i}={\boldsymbol{z}}_{i}^{\text{T}}\boldsymbol{\eta }+{\boldsymbol{x}}%
_{i}^{\text{T}}\boldsymbol{\beta }_{i}\mathbf{+}\varepsilon _{i},i=1,\ldots
,n\text{, with }\boldsymbol{\beta }_{i}=\boldsymbol{\alpha }_{1}w_{i1}+%
\boldsymbol{\alpha }_{2}w_{i2}+\boldsymbol{\alpha }_{3}w_{i3},
\label{EQ:DGP2}
\end{equation}%
where ${\boldsymbol{z}}_{i}$, $\varepsilon _{i}$ and $\boldsymbol{\eta }$
are simulated in the same way as in Example 1. Let ${\boldsymbol{x}}%
_{i}=(1,x_{i})^{\text{T}}$, where $x_{i}$ is generated from centered and
standardized binomial with probability $0.5$ for one outcome. Let $%
\boldsymbol{\alpha }_{1}=(-c,-c)^{\text{T}}$, $\boldsymbol{\alpha }%
_{2}=(0,0)^{\text{T}}$ and $\boldsymbol{\alpha }_{2}=(c,c)^{\text{T}}$.
Moreover, let $w_{i1}=I(|z_{i1}+z_{i2}+z_{i3}|<0.9)$, $%
w_{i2}=I(z_{i1}+z_{i2}+z_{i3}\geq 0.9)$ and $w_{i3}=1-w_{i1}-w_{i2}$. Let $%
n=200$.

\begin{figure}[tbp]
\caption{\textit{Fusiongram for }$(\protect\widehat{\protect\beta }_{21}(%
\protect\lambda),\ldots ,\protect\widehat{\protect\beta }_{2n}(\protect%
\lambda)$\textit{, the second component in }$\protect\widehat{\mathit{%
\mathbf{\boldsymbol{\protect\beta }}}}$\textit{$_{i}$}$(\protect\lambda )$%
\textit{'s, against $\protect\lambda$ with $c=2$ in Example 2.}}
\label{FIG:pathEX2}{\normalsize \vspace{-0.3cm} }
\par
\begin{center}
$%
\begin{array}{ccc}
\includegraphics[width=4.5cm]{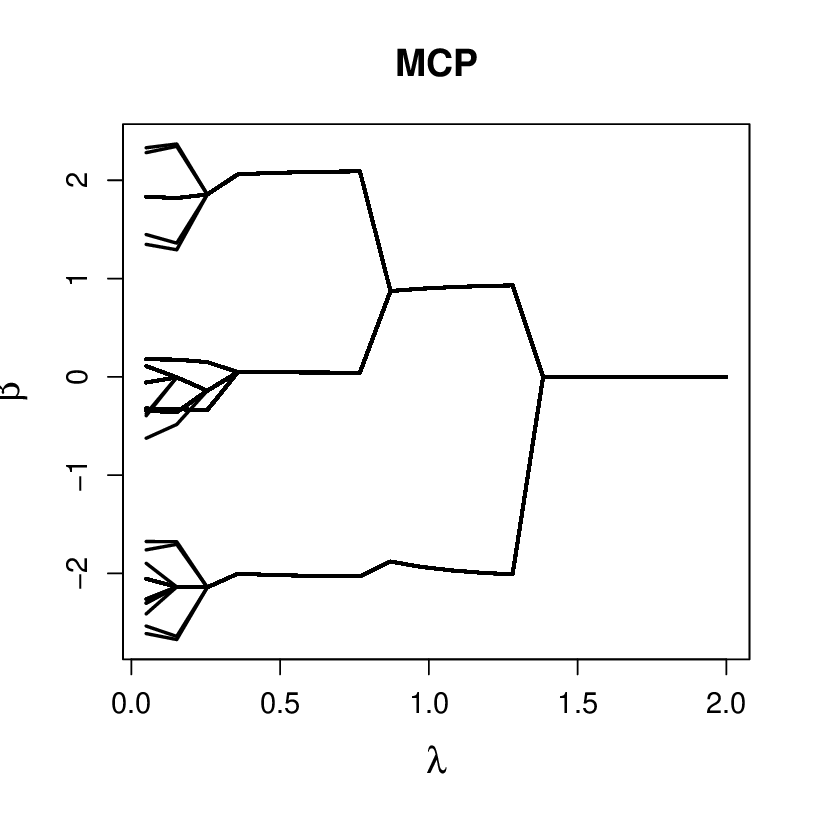} & %
\includegraphics[width=4.5cm]{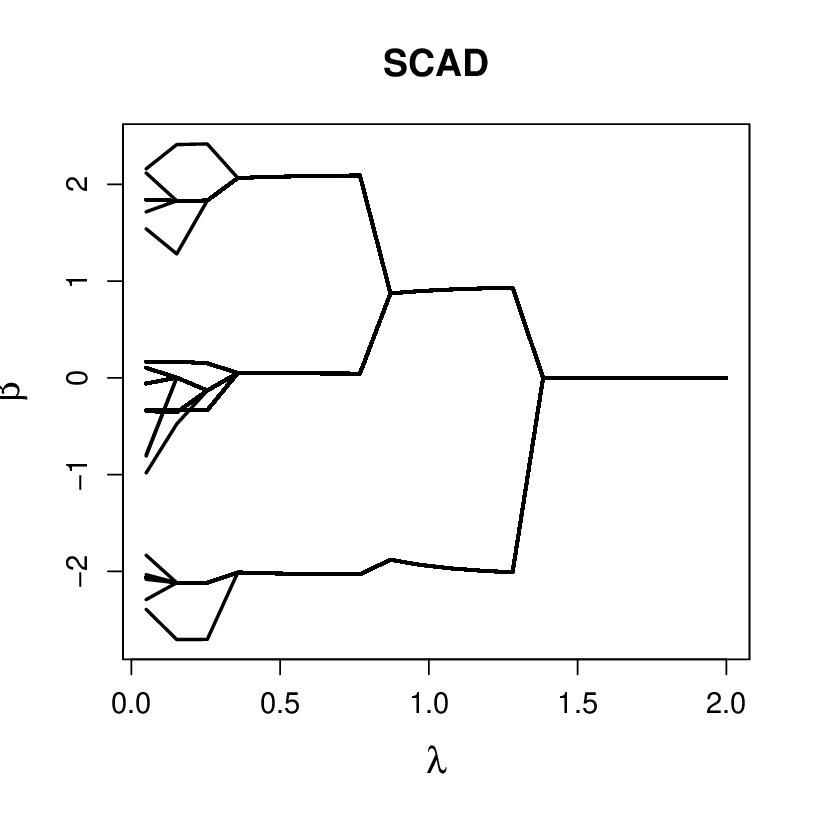} & %
\includegraphics[width=4.5cm]{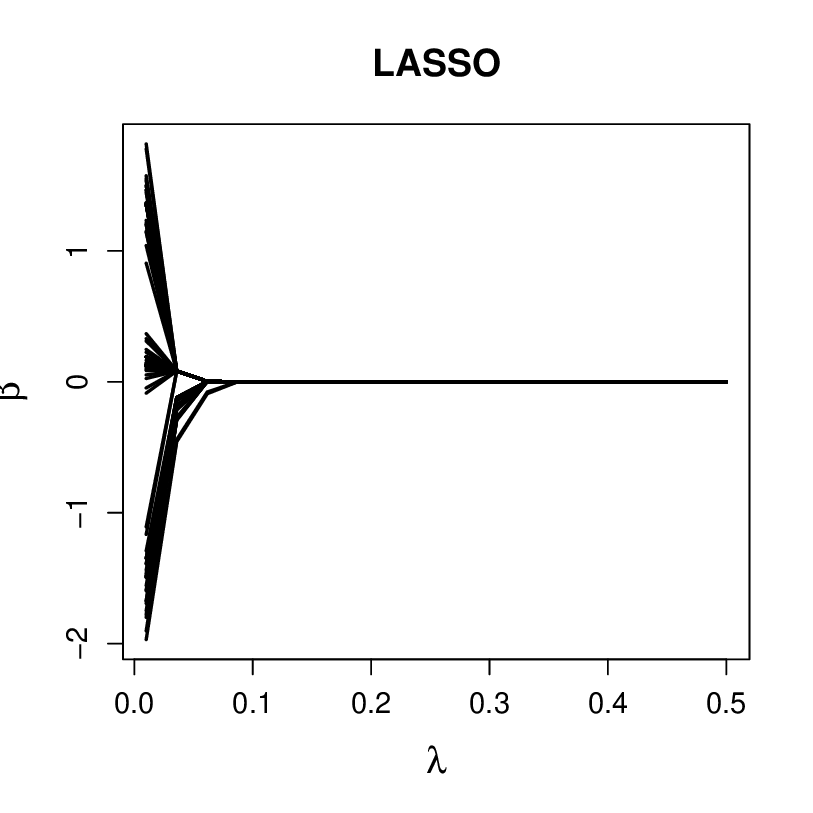}%
\end{array}%
$%
\end{center}
\end{figure}

Figure \ref{FIG:pathEX2} displays the fusiongram for $(\widehat{\beta }%
_{21}(\lambda),\ldots ,\widehat{\beta }_{2n}(\lambda))$, the elements of the
second component in $\widehat{\boldsymbol{\beta }}_{i}(\lambda)$'s, against $%
\lambda $ values with $c=2$. 
The fusiongrams for SCAD and MCP look similar. They generate three subgroups
for $\lambda $ in a certain interval, and the estimates of the treatment
effects are nearly unbiased on this segment. For the LASSO, the estimates
merge to a single value quickly due to the overshrinkage of the $L_{1}$
penalty.

\begin{table}[tbph]
\caption{The sample mean, median and standard deviation (s.d.) of $\protect%
\widehat{K}$, the Rand Index (RI) value and the percentage (per) of $\protect%
\widehat{K}$ equaling to the true number of subgroups by MCP and SCAD based
on 500 replications for $c=2$, $3$ and $4$ in Example 2.}
\label{TAB:KhatEX2}
\begin{center}
\begin{tabular*}{0.95\textwidth}{l|@{\extracolsep{\fill}}cc|cc|cc}
\hline
& \multicolumn{2}{c|}{$c=2$} & \multicolumn{2}{c|}{$c=3$} &
\multicolumn{2}{c}{$c=4$} \\ \hline
& MCP & SCAD & MCP & SCAD & MCP & SCAD \\ \hline
mean & 3.450 & 3.470 & 3.400 & 3.410 & 3.350 & 3.370 \\
median & 3.000 & 3.000 & 3.000 & 3.000 & 3.000 & 3.000 \\
s.d. & 0.775 & 0.765 & 0.769 & 0.758 & 0.748 & 0.726 \\
RI & 0.651 & 0.651 & 0.658 & 0.657 & 0.672 & 0.672 \\
per & 0.630 & 0.600 & 0.680 & 0.660 & 0.800 & 0.760 \\ \hline
\end{tabular*}%
\end{center}
\end{table}

We next conduct the simulations by selecting $\lambda $ via minimizing the
modified BIC given in (\ref{EQ:BIC}). Table \ref{TAB:KhatEX2} reports the
sample mean, median and standard deviation (s.d.) of the estimated number of
groups $\widehat{K}$, the Rand Index (RI) defined in (\ref{EQ:RI}) for
measuring clustering accuracy, and the percentage (per) of $\widehat{K}$
equaling to the true number of subgroups by the MCP and SCAD methods based
on $500$ simulation realizations for $c=2$, $3$ and $4$. We observe that the
median values of $\widehat{K}$ are 3, which is the true number of subgroups,
for all cases. Moreover, as the $c$ value becomes larger, the Rand Index
(RI) value and the percentage of correctly selecting the number of subgroups
increase.

\textbf{Example 3} (No treatment heterogeneity). We generate data from a
model with homogeneous treatment effects given by $y_{i}={\boldsymbol{z}}%
_{i}^{\text{T}}\boldsymbol{\eta }+{\boldsymbol{x}}_{i}^{\text{T}}\boldsymbol{%
\beta }\mathbf{+}\varepsilon _{i},i=1,\ldots ,n$, where $\boldsymbol{z}_{i}$%
, ${\boldsymbol{x}}_{i}$, $\varepsilon _{i}$ and $\boldsymbol{\eta }$ are
simulated in the same way as in Example 1. Set $\boldsymbol{\beta }=(\beta
_{1},\beta _{2})^{\text{T}}=(2,2)$ and $n=200$. We use our proposed
penalized estimation method to fit the model. The sample mean of the
estimated number of groups $\widehat{K}$ is 1.19 and 1.18, respectively, for
the MCP and SCAD methods, and the sample median is 1 for both methods based
on 500 replications.

\begin{table}[tbph]
\caption{The empirical bias (Bias) for the estimates of $\boldsymbol{\protect%
\beta }$ and $\boldsymbol{\protect\eta }$, and the average asymptotic
standard error (ASE) calculated according to Corollary \protect\ref%
{COR:distribution} and the empirical standard error (ESE) for the MCP\ and
SCAD methods and oracle estimator (ORACLE) in Example 3.}
\label{TAB:biasEX3}
\begin{center}
{\normalsize
\begin{tabular*}{0.95\textwidth}{l|c|@{\extracolsep{\fill}}rrrrr}
\hline
&  & $\beta _{1}$ & $\beta _{2}$ & $\eta _{1}$ & $\eta _{2}$ & $\eta _{3}$
\\ \hline
& Bias & $-0.005$ & $0.003$ & $-0.002$ & 0.007 & 0.003 \\
MCP & ASE & 0.035 & 0.037 & 0.036 & 0.037 & 0.037 \\
& ESE & 0.034 & 0.036 & 0.041 & 0.038 & 0.041 \\ \hline
& Bias & $-0.004$ & $0.002$ & $-0.001$ & 0.007 & 0.003 \\
SCAD & ASE & 0.035 & 0.036 & 0.036 & 0.037 & 0.037 \\
& ESE & 0.034 & 0.036 & 0.040 & 0.037 & 0.041 \\ \hline
& Bias & $-0.004$ & $0.002$ & $-0.001$ & 0.006 & 0.004 \\
ORACLE & ASE & 0.036 & 0.036 & 0.037 & 0.038 & 0.038 \\
& ESE & 0.036 & 0.037 & 0.039 & 0.034 & 0.039 \\ \hline
\end{tabular*}
}
\end{center}
\end{table}

To evaluate the asymptotic normality established in Corollary \ref%
{COR:distribution}, Table \ref{TAB:biasEX3} lists the empirical bias (Bias)
for the estimates of $\boldsymbol{\beta }$ and $\boldsymbol{\eta }$,
the average asymptotic standard error (ASE) calculated according to
Corollary \ref{COR:distribution}, the empirical standard error (ESE) for the
MCP, SCAD as well as the oracle estimator (ORACLE). The bias, ASE and ESE
for the estimates of $\boldsymbol{\beta }$ by the MCP and SCAD are
calculated based on the replications with the estimated number of groups
equal to one. For other cases, they are calculated based on the 500
replications. 
The biases are small for all cases. 
The ESE and ASE are similar for both MCP and SCAD, and they are also close
to the corresponding values for the oracle estimator. These results indicate
that the proposed method works well for the homogeneous model.

\subsection{Empirical example\label{SEC:applications}}

We apply our method to the AIDS Clinical Trials Group Study 175.
This study was a randomized clinical trial to compare zidovudine with other
three therapies including zidovudine and didanosine, zidovudine and
zalcitabine, and didanosine in adults infected with the human
immunodeficiency virus type I. We use the log-transformed values of the CD4
counts at 20$\pm $5 weeks as the responses $y_{i}$ %
\citep{Tsiatis.Davidian.Zhang.Lu:2007}. For illustration of our method, we
use didanosine as the treatment variable and use a binary variable $x_{i}$
for this treatment. Thus, the coefficient of $x_{i}$ represents the
difference in the treatment effect between the two therapies: didanosine and
zidovudine. We randomly select $500$ patients from the study to consist of
our dataset. Moreover, we include 5 baseline covariates in the model, which
are age (years), weight (kg), Karnofsky score, log-transformed CD8 counts at
baseline, and gender ($0=$female, $1=$male).

\begin{figure}[tbp]
\caption{\textit{The kernel density plot of the residuals (a) by fitting the
homogeneous linear model (\protect\ref{Mod0}) and (b) by fitting the
heterogeneous linear model (\protect\ref{Mod1}).}}
\label{FIG:density}\centering
$%
\begin{array}{cc}
(a) & (b) \vspace{-1cm} \\
\includegraphics[width=8cm,height=8cm]{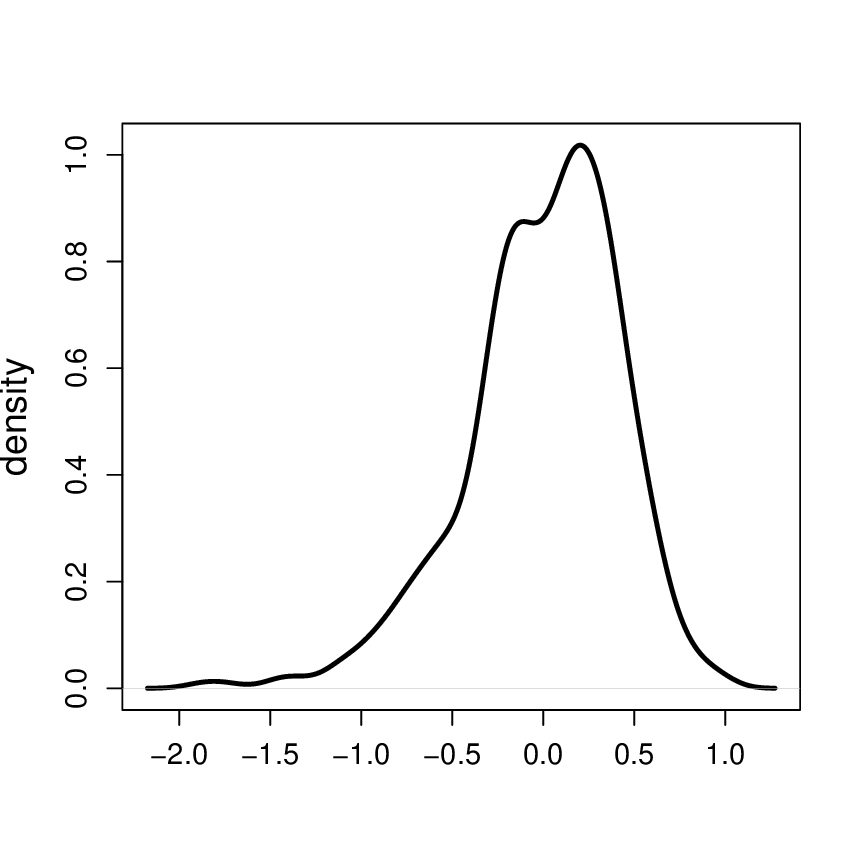} & %
\includegraphics[width=8cm,height=7.8cm]{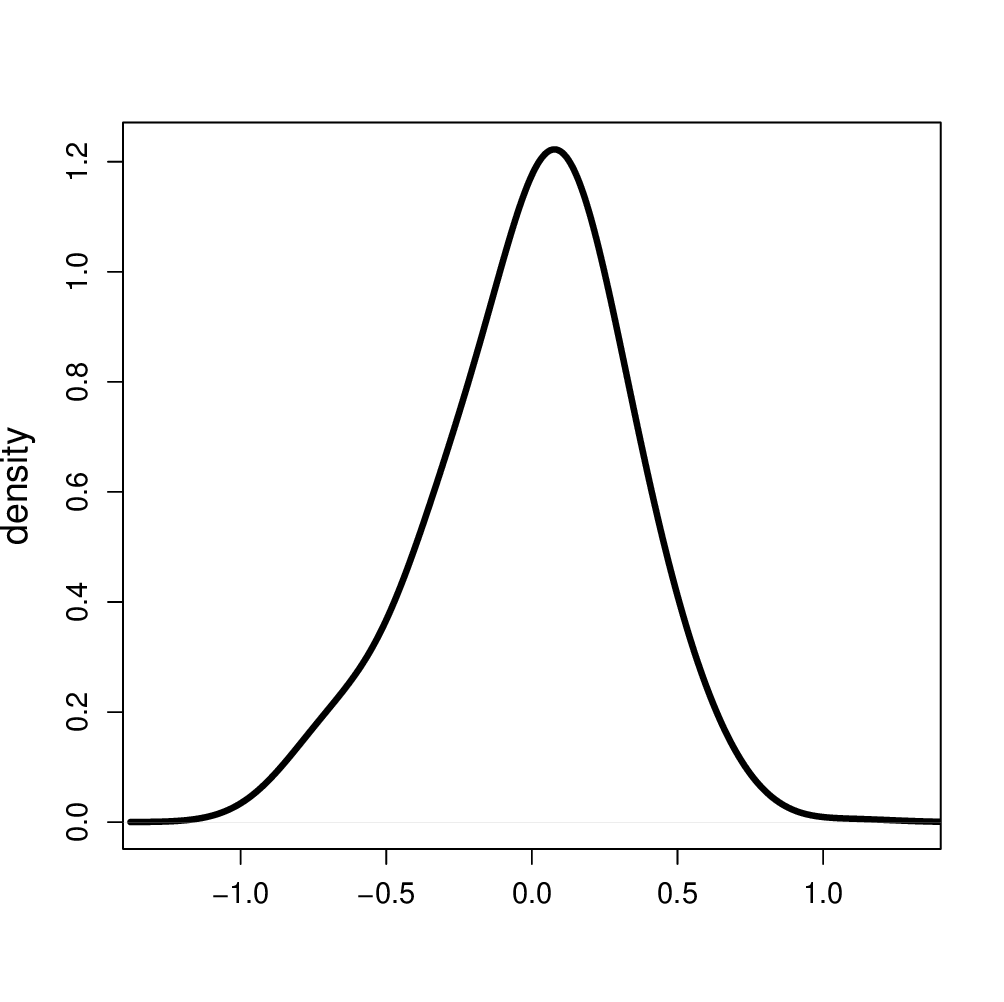}%
\end{array}
$%
\end{figure}

To see possible heterogeneity in treatment effects, we first fit the
homogeneous linear model (\ref{Mod0}) by using $y_{i}$ as the response and
the 5 baseline covariates and the treatment variable as predictors and
obtain the residuals, so that the effects of the baseline covariates are
controlled. In Figure \ref{FIG:density} (a), it shows the kernel density
plot of the residuals. We can see that the distribution has multiple modes
for these patients, which indicates possible heterogeneous treatment
effects. 

\begin{figure}[tbp]
\caption{\textit{Fusiongram for }$\protect\widehat{\mathit{\boldsymbol{%
\protect\beta }}}$\textit{$_{2}(\protect\lambda)=$}$\protect\widehat{\protect%
\beta }_{21}(\protect\lambda),\ldots ,\protect\widehat{\protect\beta }_{2n}(%
\protect\lambda)$ against $\protect\lambda$ \textit{.}}
\label{FIG:pathreal}\centering
$\includegraphics[scale=0.4]{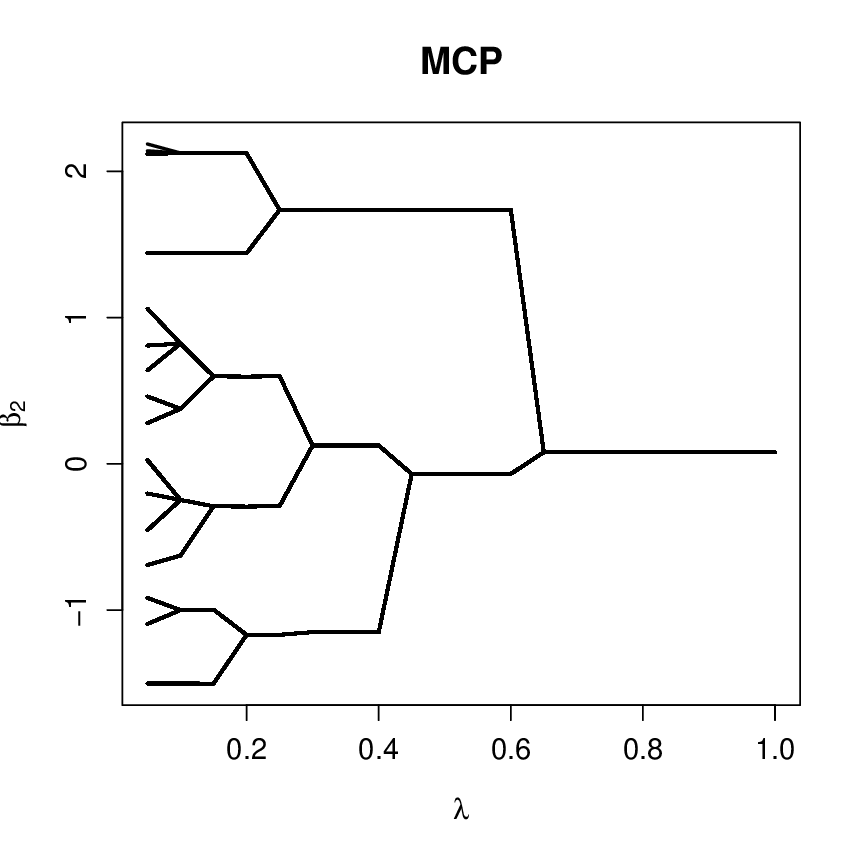} $%
\end{figure}

Next, we fit the heterogeneous model $y_{i}=\boldsymbol{z}_{i}^{\text{T}}%
\boldsymbol{\eta }+{\boldsymbol{x}}_{i}^{\text{T}}\boldsymbol{\beta }_{i}%
\mathbf{+}\varepsilon _{i},i=1,\ldots ,n$, where $\boldsymbol{\beta }%
_{i}=(\beta _{1i},\beta _{2i})^{\text{T}}$, ${\boldsymbol{z}}%
_{i}=(z_{i1}\ldots ,z_{i5})^{\text{T}}$ which are the 5 covariates described
above, and ${\boldsymbol{x}}_{i}=(1,x_{i})^{\text{T}}$, in which $x_{i}$ is
the binary variable for the treatment didanosine. All of the predictors are
centered and standardized before applying the regularization methods. Figure %
\ref{FIG:pathreal} displays the fusiongrams for the estimated treatment
coefficients, $(\widehat{\beta }_{21}(\lambda ),\ldots ,\widehat{\beta }%
_{2n}(\lambda ))$, by MCP. We obtain similar patterns by using SCAD. The
fusiongrams suggest the existence of heterogeneity in the treatment effects.
In particular, the modified BIC criterion selected the $\lambda $ value in
the region that gives two subgroups with different treatment effects.
Moreover, Figure \ref{FIG:density} (b) shows the kernel density plot of the
residuals by fitting the heterogeneous linear model (\ref{Mod1}). It shows a
uni-modal distribution.

\begin{table}[h]
\caption{The estimates (Est.), their standard errors (s.e.) and p-values
(P-value) for testing significance of $\boldsymbol{\protect\alpha }_{1}$ and
$\boldsymbol{\protect\alpha }_{2}$ by the MCP and SCAD methods, and those
values of $\boldsymbol{\protect\beta }=\boldsymbol{\protect\alpha }_{1}$ by
the OLS\ method. }
\begin{center}
\begin{tabular*}{0.95\textwidth}{l|c|@{\extracolsep{\fill}}cccc}
\hline
&  & $\alpha _{11}$(intercept) & $\alpha _{12}$(trt) & $\alpha _{21}$%
(intercept) & $\alpha _{22}$(trt) \\ \hline
MCP & Est. & $5.809$ & $-0.065$ & $5.735$ & $1.687$ \\
& s.e. & $0.016$ & $0.041$ & $0.055$ & $0.110$ \\
& p-value & $<0.001$ & $0.112$ & $<0.001$ & $<0.001$ \\ \hline
SCAD & Est. & $5.809$ & $-0.065$ & $5.737$ & $1.688$ \\
& s.e. & $0.016$ & $0.042$ & $0.055$ & $0.110$ \\
& p-value & $<0.001$ & $0.122$ & $<0.001$ & $<0.001$ \\ \hline
OLS & Est. & $5.787$ & $0.077$ & $-$ & $-$ \\
& s.e. & $0.019$ & $0.040$ & $-$ & $-$ \\
& p-value & $<0.001$ & $0.054$ & $-$ & $-$ \\ \hline
\end{tabular*}%
\end{center}
\label{TAB:aphareal}
\end{table}

Let $\widehat{\boldsymbol{\alpha }}_{1}=(\widehat{\alpha }_{11},\widehat{%
\alpha }_{12})^{\text{T}}$ and $\widehat{\boldsymbol{\alpha }}_{2}=(\widehat{%
\alpha }_{21},\widehat{\alpha }_{22})^{\text{T}}$ be the estimated
coefficients for $\boldsymbol{x}_{i}$ in the two identified groups $\widehat{%
\mathcal{G}}_{1}$ and $\widehat{\mathcal{G}}_{2}$, respectively, so that $%
\widehat{\boldsymbol{\beta }}_{i}=\widehat{\boldsymbol{\alpha }}_{1}$ for $%
i\in \widehat{\mathcal{G}}_{1}$ and $\widehat{\boldsymbol{\beta }}_{i}=%
\widehat{\boldsymbol{\alpha }}_{2}$ for $i\in \widehat{\mathcal{G}}_{2}$. In
Table \ref{TAB:aphareal}, we report the estimates (Est.) and their standard
errors (s.e.) of $\boldsymbol{\alpha }_{1}=(\alpha _{11},\alpha _{12})^{%
\text{T}}$ and $\boldsymbol{\alpha }_{2}=(\alpha _{21},\alpha _{22})^{\text{T%
}}$ by the MCP and SCAD methods, and those values of $\boldsymbol{\beta }=%
\boldsymbol{\alpha }_{1}$ by the OLS\ method. Note that the first and second
components represent the coefficients for intercept and the treatment
variable (trt), respectively. We also report the p-values (p-value) for
testing whether each coefficient is zero or not. By the OLS method, the
p-value for testing whether the coefficient of the therapy (didanosine), $%
\alpha _{12}$, is zero or not is greater than 0.05. This result indicates
that the effect of the therapy (didanosine) is not significantly different
from that of the standard treatment (zidovudine) by assuming that the
treatment effect is homogeneous. In comparison, by the proposed concave
fusion approach, the p-value for testing significance of the treatment
coefficient, $\alpha _{12}$, for one group is greater than 0.05, but the
p-value for the other group is less than 0.001. This result implies that
although the treatment has no statistically significant effect on one group
of patients, its effect on the other group is prominent.

Next, we employ the bootstrapping method given in Section \ref{SEC:bootstrap}
to test homogeneity of the treatment effect. We obtain the p-value $<0.01$.
Thus, the heterogeneity of the treatment effect is further confirmed by the
inference procedure. Lastly, we apply the method to model (\ref{Mod1}) with
all three therapies by letting ${\boldsymbol{x}}%
_{i}=(1,x_{i1},x_{i2},x_{i3})^{\text{T}}$, where $x_{i1}=$zidovudine and
didanosine (trt1), $x_{i2}=$zidovudine and zalcitabine (trt2), and $x_{i3}=$%
didanosine (trt3). {\color{black} Let $\widehat{\boldsymbol{\alpha }}_{1}=(\widehat{\alpha }%
_{11},\widehat{\alpha }_{12},\widehat{\alpha }_{13},\widehat{\alpha }_{14})^{%
\text{T}}$ and $\widehat{\boldsymbol{\alpha }}_{2}=(\widehat{\alpha }_{21},%
\widehat{\alpha }_{22},\widehat{\alpha }_{23},\widehat{\alpha }_{24})^{\text{%
T}}$ be the estimated coefficients for $\boldsymbol{x}_{i}$ in the two
identified groups $\widehat{\mathcal{G}}_{1}$ and $\widehat{\mathcal{G}}_{2}$%
, respectively. In Table \ref{TAB:aphareal2}, we report the estimates (Est.)
and their standard errors (s.e.) of $\boldsymbol{\alpha }_{1}=(\alpha
_{11},\alpha _{12},\alpha _{13},\alpha _{14})^{\text{T}}$ and $\boldsymbol{%
\alpha }_{2}=(\alpha _{21},\alpha _{22},\alpha _{23},\alpha _{24})^{\text{T}}
$ by the MCP and SCAD methods, and those values of $\boldsymbol{\beta }=%
\boldsymbol{\alpha }_{1}$ by the OLS method. We obtain two subgroups as
well by our proposed method. Moreover, the two methods, MCP and SCAD, have
similar results, and the effects of the three therapies are more significantly different from the standard therapy
in one group than they are in the other group.}
\begin{table}[h]
\caption{The estimates (Est.), their standard errors (s.e.) and p-values
(P-value) for testing significance of $\boldsymbol{\protect\alpha }_{1}$ and
$\boldsymbol{\protect\alpha }_{2}$ by MCP and SCAD, and those
values obtained by
OLS. }
\small
\begin{center}
\begin{tabular*}{0.95\textwidth}{l|c|@{\extracolsep{\fill}}cccccccc}
\hline
&  & $\alpha _{11}$ & $\alpha _{12}$ & $\alpha _{13}$ & $\alpha
_{14}$ & $\alpha _{21}$ & $\alpha _{22}$ & $\alpha _{23}$
& $\alpha _{24}$ \\ \hline
MCP & Est. & $3.985$ & $0.114$ & $0.104$ & $0.126$ & $5.525$ & $1.116$ & $%
1.200$ & $1.214$ \\
& s.e. & $0.086$ & $0.049$ & $0.051$ & $0.052$ & $0.074$ & $0.103$ & $0.106$
& $0.107$ \\
& p-value & $<0.001$ & $0.020$ & $0.041$ & $0.015$ & $<0.001$ & $<0.001$ & $%
<0.001$ & $<0.001$ \\ \hline
SCAD & Est. & $3.985$ & $0.114$ & $0.105$ & $0.127$ & $5.524$ & $1.116$ & $%
1.199$ & $1.214$ \\
& s.e. & $0.086$ & $0.047$ & $0.051$ & $0.052$ & $0.074$ & $0.102$ & $0.106$
& $0.107$ \\
& p-value & $<0.001$ & $0.015$ & $0.040$ & $0.015$ & $<0.001$ & $<0.001$ & $%
<0.001$ & $<0.001$ \\ \hline
OLS & Est. & $5.846$ & $0.069$ & $0.067$ & $0.065$ & $-$ & $-$ & $-$ & $-$
\\
& s.e. & $0.019$ & $0.024$ & $0.024$ & $0.024$ & $-$ & $-$ & $-$ & $-$ \\
& p-value & $<0.001$ & $0.004$ & $0.005$ & $0.007$ & $-$ & $-$ & $-$ & $-$
\\ \hline
\end{tabular*}%
\end{center}
\label{TAB:aphareal2}
\end{table}


\section{Discussion\label{SEC:Discussion}}

It will be of interest to extend the proposed method to a more general class
of regression problems including generalized linear and Cox regression
models. Moreover, it is also possible to relax the linearity assumption on
the covariates ${\boldsymbol{z}}_{i}$ by considering a nonparametric or
semiparametric functional form of ${\boldsymbol{z}}_{i}$.
For these more complicated models, the ADMM algorithm can still be employed
with some modifications. However, further work is needed to study the
theoretical properties. We refer to Section \ref{SEC:nonlinear} for detailed
discussions on these extensions. Also, we have assumed that the number of
treatment variables $p$ and the number of confounding variables $q$ are both
smaller than the sample size $n$, although we allow $p$ and $q$ to diverge
with $n$. For high-dimensional problems with $p>n$ or $q>n$, a sparsity
condition on the coefficients would be needed to ensure identifiability of
the model. Further studies are needed to develop computational algorithms
and theoretical results for the high-dimensional setting.




\section*{{\protect\normalsize Appendix\label{SEC:appendix}}}

\label{appendix}

\renewcommand{\thetheorem}{{\sc A.\arabic{theorem}}} \renewcommand{%
\theremark}{{\sc A.\arabic{remark}}}
\renewcommand{\theproposition}{{\sc
A.\arabic{proposition}}} \renewcommand{\thelemma}{{\sc
A.\arabic{lemma}}} \renewcommand{\thecorollary}{{\sc
A.\arabic{corollary}}} \renewcommand{\theequation}{A.\arabic{equation}} %
\renewcommand{\thesubsection}{{\it A.\arabic{subsection}}} %
\setcounter{equation}{0} \setcounter{lemma}{0} \setcounter{proposition}{0} %
\setcounter{theorem}{0} \setcounter{subsection}{0}\setcounter{corollary}{0}%
\setcounter{remark}{0}

In the Appendix, we give the computational details, the convergence property
of the ADMM algorithm and the technical proofs for Theorems \ref{THM:norm}-%
\ref{THM:normhomo}. 

\subsection{Computation}

\label{SEC:computation}

\subsubsection{ADMM with concave penalties}

We derive an ADMM algorithm for computing the solution (\ref{Min1}). The key
idea is to introduce a new set of parameters $\boldsymbol{\delta }_{ij}=%
\boldsymbol{\beta }_{i}-\boldsymbol{\beta }_{j}$. Then, we can reformulate
the problem of minimizing (\ref{EQ:objective}) as that of minimizing
\begin{gather}
L_{0}(\boldsymbol{\eta },\boldsymbol{\beta },\boldsymbol{\delta }\mathbf{)=}%
\frac{1}{2}\sum\nolimits_{i=1}^{n}(y_{i}-{\boldsymbol{z}}_{i}^{\text{T}}%
\boldsymbol{\eta }-{\boldsymbol{x}}_{i}^{\text{T}}\boldsymbol{\beta }_{i}%
\mathbf{)}^{2}+\sum\nolimits_{i<j}p_{\gamma }(\Vert \boldsymbol{\delta }%
_{ij}\Vert ,\lambda ),  \notag \\
\text{subject to }\boldsymbol{\beta }_{i}-\boldsymbol{\beta }_{j}-%
\boldsymbol{\delta }_{ij}=\mathbf{0},  \label{EQ:opt}
\end{gather}%
where {\color{black} $\boldsymbol{\delta }=\{\boldsymbol{\delta }_{ij}^{%
\text{T}},i<j\}^{\text{T}}$.} Let $\left\langle \mathbf{a,b}\right\rangle =%
\mathbf{a}^{\text{T}}\mathbf{b}$ be the inner product of two vectors $%
\mathbf{a}$ and $\mathbf{b}$ with the same dimension. The augmented
Lagrangian is
\begin{equation}
L(\boldsymbol{\eta },\boldsymbol{\beta },\boldsymbol{\delta },\boldsymbol{%
\upsilon })=L_{0}(\boldsymbol{\eta },\boldsymbol{\beta },\boldsymbol{\delta }%
)+\sum\nolimits_{i<j}\left\langle \boldsymbol{\upsilon }_{ij},\boldsymbol{%
\beta }_{i}-\boldsymbol{\beta }_{j}-\boldsymbol{\delta }_{ij}\right\rangle +%
\frac{\vartheta }{2}\sum\nolimits_{i<j}\Vert \boldsymbol{\beta }_{i}-%
\boldsymbol{\beta }_{j}-\boldsymbol{\delta }_{ij}\Vert ^{2},  \label{EQ:L}
\end{equation}%
where the dual variables {\color{black} $\boldsymbol{\upsilon }=\{%
\boldsymbol{\upsilon }_{ij}^{\text{T}},i<j\}^{\text{T}}$ }are Lagrange
multipliers and $\vartheta $ is a penalty parameter. We then compute the
estimates of $(\boldsymbol{\eta },\boldsymbol{\beta },\boldsymbol{\delta },%
\boldsymbol{\upsilon })$ through iterations using the ADMM.

For a given value of $\boldsymbol{\delta}^m$ and $\boldsymbol{\upsilon}^m$
at step $m$, the iteration goes as follows:
\begin{eqnarray}
(\boldsymbol{\eta }^{m+1},\boldsymbol{\beta }^{m+1}) &=&\mathop{\rm argmin}_{%
\boldsymbol{\eta },\boldsymbol{\beta }}L(\boldsymbol{\eta },\boldsymbol{%
\beta },\boldsymbol{\delta }^{m},\boldsymbol{\upsilon }^{m}),  \label{A1} \\
\boldsymbol{\delta }^{m+1} &=&\mathop{\rm argmin}_{\boldsymbol{\delta }}L(%
\boldsymbol{\eta }^{m+1},\boldsymbol{\beta }^{m+1},\boldsymbol{\delta },%
\boldsymbol{\upsilon }^{m}),  \label{A2} \\
\boldsymbol{\upsilon }_{ij}^{m+1} &=&\boldsymbol{\upsilon }%
_{ij}^{m}+\vartheta (\boldsymbol{\beta }_{i}^{m+1}-\boldsymbol{\beta }%
_{j}^{m+1}-\boldsymbol{\delta }_{ij}^{m+1}).  \label{A3}
\end{eqnarray}


In \eqref{A1}, the problem is equivalent to the minimization of the function
\begin{equation*}
f(\boldsymbol{\eta },\boldsymbol{\beta })=\frac{1}{2}\sum%
\nolimits_{i=1}^{n}(y_{i}-{\boldsymbol{z}}_{i}^{\text{T}}\boldsymbol{\eta }-{%
\boldsymbol{x}}_{i}^{\text{T}}\boldsymbol{\beta }_{i})^{2}+\frac{\vartheta }{%
2}\sum\nolimits_{i<j}\Vert \boldsymbol{\beta }_{i}-\boldsymbol{\beta }_{j}-%
\boldsymbol{\delta }_{ij}^{m}+\vartheta ^{-1}\boldsymbol{\upsilon }%
_{ij}^{m}\Vert ^{2}+C,
\end{equation*}%
where $C$ is a constant independent of $(\boldsymbol{\eta },\boldsymbol{%
\beta })$. Some algebra shows that we can write $f(\boldsymbol{\eta },%
\boldsymbol{\beta })$ as
\begin{equation}
f(\boldsymbol{\eta },\boldsymbol{\beta })=\frac{1}{2}\Vert \mathbf{Z}%
\boldsymbol{\eta }+\mathbf{X}\boldsymbol{\beta }-{\mathbf{y}}\Vert ^{2}+%
\frac{\vartheta }{2}\Vert \mathbf{A}\boldsymbol{\beta }-\boldsymbol{\delta }%
^{m}+\vartheta ^{-1}\boldsymbol{\upsilon }^{m}\Vert ^{2}+C,
\label{EQ:Lmubeta}
\end{equation}%
{\color{black} where $\mathbf{A=}D\otimes \mathbf{I}_{p}$. }Here $%
D=\{(e_{i}-e_{j}),i<j\}^{\text{T}}$ with $e_{i}$ being the $i$th unit $%
n\times 1$ vector whose $i$th element is 1 and the remaining ones are 0, $%
\mathbf{I}_{p}$ is a $p\times p$ identity matrix and $\otimes $ is the
Kronecker product.

Thus for given $\boldsymbol{\delta }^{m}$ and $\boldsymbol{\upsilon }^{m}$
at the $m$th step, the updates $\boldsymbol{\beta }^{m+1}$ and $\boldsymbol{%
\eta }^{m+1}$ are
\begin{eqnarray}
\boldsymbol{\beta }^{m+1} &=&(\mathbf{X}^{\text{T}}\mathbf{Q}_{Z}\mathbf{X}%
+\vartheta \mathbf{A}^{\text{T}}\mathbf{A)}^{-1}[\mathbf{X}^{\text{T}}%
\mathbf{Q}_{Z}\mathbf{y}+\vartheta \mathbf{A}^{\text{T}}(\boldsymbol{\delta }%
^{m}-\vartheta ^{-1}\boldsymbol{\upsilon }^{m})],  \notag \\
\boldsymbol{\eta }^{m+1} &=&(\mathbf{Z}^{\text{T}}\mathbf{Z)}^{-1}\mathbf{Z}%
^{\text{T}}(\mathbf{y-\mathbf{X}}\boldsymbol{\beta }^{m+1}),  \label{B2}
\end{eqnarray}%
where $\mathbf{Q}_{Z}=\mathbf{I}_{n}-\mathbf{\mathbf{Z}}(\mathbf{\mathbf{Z}}%
^{\text{T}}\mathbf{\mathbf{Z)}}^{-1}\mathbf{\mathbf{Z}}^{\text{T}}$. {%
\color{black} Since
\begin{equation*}
\mathbf{A}^{\text{T}}(\boldsymbol{\delta }^{m}-\vartheta ^{-1}\boldsymbol{%
\upsilon }^{m})=(D^{\text{T}}\otimes \mathbf{I}_{p})(\boldsymbol{\delta }%
^{m}-\vartheta ^{-1}\boldsymbol{\upsilon }^{m})=\text{vec}((\boldsymbol{%
\Delta }^{m}-\vartheta ^{-1}\boldsymbol{\Upsilon }^{m})D),
\end{equation*}%
where $\boldsymbol{\Delta }^{m}=\{\boldsymbol{\delta }_{ij}^{m},i<j\}_{p%
\times n(n-1)/2}$ and $\boldsymbol{\Upsilon }^{m}=\{\boldsymbol{\upsilon }%
_{ij}^{m},i<j\}_{p\times n(n-1)/2}$, then we have
\begin{equation}
\boldsymbol{\beta }^{m+1}=(\mathbf{X}^{\text{T}}\mathbf{Q}_{Z}\mathbf{X}%
+\vartheta \mathbf{A}^{\text{T}}\mathbf{A)}^{-1}[\mathbf{X}^{\text{T}}%
\mathbf{Q}_{Z}\mathbf{y}+\vartheta \text{vec}((\boldsymbol{\Delta }%
^{m}-\vartheta ^{-1}\boldsymbol{\Upsilon }^{m})D)].  \label{B1}
\end{equation}
} In \eqref{A2}, after discarding the terms independent of $\boldsymbol{%
\delta} $, we need to minimize
\begin{equation}
\frac{\vartheta }{2}\|{\boldsymbol{\zeta}}_{ij}^m-{\boldsymbol{\delta}}%
_{ij}\|^{2}+ p_{\gamma }(\|{\boldsymbol{\delta}}_{ij}\|,\lambda )
\label{EQ:etaij}
\end{equation}%
with respect to $\boldsymbol{\delta}_{ij}$, where $\boldsymbol{\zeta}_{ij}^m=%
\boldsymbol{\beta}_{i}^m-\boldsymbol{\beta}_{j}^m+\vartheta ^{-1}\boldsymbol{%
\upsilon}_{ij}^m$. This is a groupwise thresholding operator corresponding
to $p_{\gamma}$.

For the $L_{1}$ penalty, the solution is
\begin{equation}
\boldsymbol{\delta}_{ij}^{m+1}=S(\boldsymbol{\zeta}_{ij}^m,\lambda/%
\vartheta),  \label{EQ:etalasso}
\end{equation}
where $S({\boldsymbol{z}},t)=(1-t/\|\boldsymbol{z\|})_{+}\boldsymbol{z}$ is
the groupwise soft thresholding operator. Here $(x)_{+}=x$ if $x>0$ and $=0$%
, otherwise.

For the MCP with $\gamma >1/\vartheta $, the solution is
\begin{equation}
\boldsymbol{\delta}_{ij}^{m+1}=\left\{
\begin{array}{ll}
\frac{S(\boldsymbol{\zeta}_{ij}^m,\lambda /\vartheta )}{1-1/(\gamma
\vartheta )} & \text{ if }\|\boldsymbol{\zeta}_{ij}^m\|\leq \gamma \lambda,
\\
\boldsymbol{\zeta}_{ij} & \text{ if }\|\boldsymbol{\zeta}_{ij}^m\|>\gamma
\lambda.%
\end{array}
\right.  \label{EQ:etaMCP}
\end{equation}

For the SCAD penalty with $\gamma >1/\vartheta +1$, the solution is
\begin{equation}
\boldsymbol{\delta}_{ij}^{m+1}=\left\{
\begin{array}{ll}
S(\boldsymbol{\zeta}_{ij}^m,\lambda /\vartheta ) & \text{ if }\|\boldsymbol{%
\ \zeta}_{ij}^m\|\leq \lambda +\lambda /\vartheta, \\
\frac{S(\boldsymbol{\zeta}_{ij}^m,\gamma \lambda /((\gamma-1)\vartheta ))}{
1-1/((\gamma -1)\vartheta )} & \text{ if }\lambda +\lambda/\vartheta <\text{
}\|\boldsymbol{\zeta}_{ij}^m\|\leq \gamma \lambda, \\
\boldsymbol{\zeta}_{ij}^m & \text{ if }\|\boldsymbol{\zeta}
_{ij}^m\|>\gamma\lambda.%
\end{array}
\right.  \label{EQ:etaSCAD}
\end{equation}

Finally, the update of $\boldsymbol{\upsilon}_{ij}$ is given in \eqref{A3}.

We summarize the above analysis in Algorithm \ref{alg1}.
\begin{algorithm}[H]
\caption{ADMM for concave fusion}
\label{alg1}
\begin{algorithmic}[1]
\REQUIRE
Initialize $\bdelta^0$, $\bupsilon^0$. 
\FOR {$m=0, 1,2,\cdots$}
\STATE Compute $\bbeta^{m+1}$ using \eqref{B1}
\STATE Compute $\bata^{m+1}$ \eqref{B2}
\STATE  Compute $\bdelta^{m+1}$ \eqref{EQ:etalasso}, \eqref{EQ:etaMCP} or \eqref{EQ:etaSCAD}
\STATE Compute $\bupsilon^{m+1}$ using \eqref{A3}
 \IF {convergence criterion is met,}
 \STATE Stop and denote the last iteration by $(\hbata(\lambda), \hbbeta(\lambda))$,
 \ELSE
 \STATE $m = m+1$.
 \ENDIF
\ENDFOR
\ENSURE Output
\end{algorithmic}
\end{algorithm}

\begin{remark}
\label{Remark3} Our algorithm enables us to have $\widehat{\boldsymbol{%
\delta }}_{ij}=\mathbf{0}$ for a sufficiently large $\lambda$. We put
observations $i$ and $j$ in the group with the same treatment effect if $%
\widehat{\boldsymbol{\delta } }_{ij}=\mathbf{0}$. As a result, we have $%
\widehat{K}\ $estimated groups $\widehat{\mathcal{G}}_{1},\ldots ,\widehat{%
\mathcal{G}}_{\widehat{K}}$. The estimated treatment effect for the $k^{%
\text{th}}$ group is $\widehat{\boldsymbol{\alpha }}_{k}=|\widehat{\mathcal{G%
}}_{k}|^{-1}\sum \nolimits_{i\in \widehat{\mathcal{G}}_{k}}\widehat{%
\boldsymbol{\beta }}_{i},$ where $|\widehat{\mathcal{G}}_{k}|$ is the
cardinality of $\widehat{\mathcal{G}}_{k}$.
\end{remark}

\begin{remark}
\label{Invert1} {\color{black} In the algorithm, we require the
invertibility of $\mathbf{X}^{\text{T}}\mathbf{Q}_{Z}\mathbf{X}+\vartheta
\mathbf{A}^{\text{T}}\mathbf{A}$. It can be derived that $\mathbf{A}^{\text{T%
}}\mathbf{A=}$ $n\mathbf{I}_{np}-(1_{n}\otimes \mathbf{I}_{p})(1_{n}\otimes
\mathbf{I}_{p})^{\text{T}}$. For any nonzero vector $\mathbf{a}%
=(a_{ij},1\leq i\leq n,1\leq j\leq p)^{\text{T}}\in
\mathop{{\rm
I}\kern-.2em\hbox{\rm R}}\nolimits^{np}$, we have $\mathbf{a}^{\text{T}}(${$%
\vartheta \mathbf{A}^{\text{T}}\mathbf{A)}$}$\mathbf{a}\geq 0$ and $\mathbf{a%
}^{\text{T}}(${$\mathbf{X}^{\text{T}}\mathbf{Q}_{Z}\mathbf{X)}$}$\mathbf{a}%
\geq 0$. Note that $\mathbf{a}^{\text{T}}(${$\vartheta \mathbf{A}^{\text{T}}%
\mathbf{A)}$}$\mathbf{a}=0$ if and only if $a_{ij}=a_{j}$ for all $i $. When
$a_{ij}=a_{j}$ for all $i$, we have $\mathbf{a}^{\text{T}}(${$\mathbf{X}^{%
\text{T}}\mathbf{Q}_{Z}\mathbf{X)}$}$\mathbf{a}>0$ given that $\lambda
_{\min }(\sum\nolimits_{i=1}^{n}(\boldsymbol{x}_{i}^{\text{T}},\boldsymbol{z}%
_{i}^{\text{T}})^{\text{T}}(\boldsymbol{x}_{i}^{\text{T}},\boldsymbol{z}%
_{i}^{\text{T}}))>0$, which is a common assumption that the design matrix
needs to satisfy in linear regression. Therefore, $\mathbf{X}^{\text{T}}%
\mathbf{Q}_{Z}\mathbf{X}+\vartheta \mathbf{A}^{\text{T}}\mathbf{A}$ is
invertible.}
\end{remark}

\begin{remark}
\label{Remark2} We track the progress of the ADMM based on the primal
residual $\mathbf{r}^{m+1}=\mathbf{A}\boldsymbol{\beta }^{m+1}\mathbf{-}%
\boldsymbol{\delta }^{m+1}$. We stop the algorithm when $\mathbf{r}^{m+1}$
is close to zero such that $\left\Vert \mathbf{r}^{m+1}\right\Vert <a$ for
some small value $a$.
\end{remark}

\subsubsection{Initial value and computation of the solution path}

To start the ADMM algorithm described above, it is important to find a
reasonable initial value. For this purpose, we consider the ridge fusion
criterion given by
\begin{equation*}
L_{R}(\boldsymbol{\eta },\boldsymbol{\beta })=\frac{1}{2}\Vert {\mathbf{Z}}%
\boldsymbol{\eta }+{\mathbf{X}}\boldsymbol{\beta }-{\mathbf{y}}\Vert ^{2}+%
\frac{\lambda ^{\ast }}{2}\sum_{1\leq i<j\leq n}\Vert \boldsymbol{\beta }%
_{i}-\boldsymbol{\beta }_{j}\Vert ^{2},
\end{equation*}%
where $\lambda ^{\ast }$ is the tuning parameter having a small value. We
use $\lambda ^{\ast }=0.001$ in our analysis. Then $L_{R}(\boldsymbol{\eta },%
\boldsymbol{\beta })$ can be written as
\begin{equation*}
L_{R}(\boldsymbol{\eta },\boldsymbol{\beta })=\frac{1}{2}\Vert \mathbf{Z}%
\boldsymbol{\eta }+\mathbf{X}\boldsymbol{\beta }-{\mathbf{y}}\Vert ^{2}+%
\frac{\lambda ^{\ast }}{2}\Vert \mathbf{A}\boldsymbol{\beta }\Vert ^{2},
\end{equation*}%
where ${\boldsymbol{A}}$ is defined in \eqref{EQ:Lmubeta}. The solutions are
\begin{eqnarray*}
\boldsymbol{\beta }_{R}(\lambda ^{\ast }) &=&(\boldsymbol{\beta }_{R,1}^{%
\text{T}}(\lambda ^{\ast }),\ldots ,\boldsymbol{\beta }_{R,n}^{\text{T}%
}(\lambda ^{\ast }))^{\text{T}}=(\mathbf{X}^{\text{T}}\mathbf{Q}_{Z}\mathbf{X%
}+\lambda ^{\ast }\mathbf{A}^{\text{T}}\mathbf{A)}^{-1}\mathbf{X}^{\text{T}}%
\mathbf{Q}_{Z}\mathbf{y}, \\
\boldsymbol{\eta }_{R}(\lambda ^{\ast }) &=&(\mathbf{Z}^{\text{T}}\mathbf{Z)}%
^{-1}\mathbf{Z}^{\text{T}}(\mathbf{y-\mathbf{X}}\boldsymbol{\beta }%
_{R}(\lambda ^{\ast })),
\end{eqnarray*}%
where $\mathbf{Q}_{Z}$ is given in \eqref{B1}. Next, we assign the subjects
to $K^{\ast }$ groups by ranking the median values of $\boldsymbol{\beta }%
_{R,i}^{\text{T}}(\lambda ^{\ast })$. We let $K^{\ast }=\left\lfloor
n^{1/2}\right\rfloor $ to ensure that it is sufficiently large, where $%
\left\lfloor a\right\rfloor $ denotes the largest integer no greater than $a$%
. We then find the initial estimates $\boldsymbol{\eta }^{0}$ and $%
\boldsymbol{\beta }^{0}$ from least squares regression with $K^{\ast }$
groups. Let the initial estimates $\boldsymbol{\delta }_{ij}^{0}=\boldsymbol{%
\beta }_{i}^{0}-\boldsymbol{\beta }_{j}^{0}$ and $\boldsymbol{\upsilon }^{0}=%
\mathbf{0}$.

To compute the solution path of $\boldsymbol{\eta }$ and $\boldsymbol{\beta }
$ along the $\lambda $ values, we use the warm start and continuation
strategy to update the solutions. Let $[\lambda _{\min },\lambda _{\max }]$
be the interval on which we compute the solution path, where $0\leq \lambda
_{\min }<\lambda _{\max }<\infty $. Let $\lambda _{\min }=\lambda
_{0}<\lambda _{1}<\cdots <\lambda _{K}\equiv \lambda _{\max }$ be a grid of $%
\lambda $ values in $[\lambda _{\min },\lambda _{\max }]$. Compute $(%
\widehat{\boldsymbol{\eta }}(\lambda _{0}),\widehat{\boldsymbol{\beta }}%
(\lambda _{0}))$ using $(\boldsymbol{\eta }^{0},\boldsymbol{\beta }^{0})$ as
the initial value. Then compute $(\widehat{\boldsymbol{\eta }}(\lambda _{k}),%
\widehat{\boldsymbol{\beta }}(\lambda _{k}))$ using $(\widehat{\boldsymbol{%
\eta }}(\lambda _{k-1}),\widehat{\boldsymbol{\beta }}(\lambda _{k-1}))$ as
the initial value for each $k=1,\ldots ,K$.

Note that we start from the smallest $\lambda$ value in computing the
solution path. This is different from the coordinate descent algorithms for
computing the solution path in penalized regression problems %
\citep{Friedman:2007}, where the algorithms start at the $\lambda$ value
that forces all the coefficients to zero.

\subsubsection{Convergence of the algorithm}

We next derive the convergence properties of the ADMM algorithm.

\begin{proposition}
\label{PROP:primal} Let $\mathbf{r}^{m}=\mathbf{A}\boldsymbol{\beta }^{m}%
\mathbf{-}\boldsymbol{\delta }^{m}$ and $\mathbf{s}^{m+1}=\vartheta \mathbf{A%
}^{\text{T}}(\boldsymbol{\delta }^{m+1}-\boldsymbol{\delta }^{m})$ be the
primal residual and the dual residual in the ADMM described above,
respectively. It holds that $\lim_{m\rightarrow \infty }\|\mathbf{r}%
^{m}\|^{2}=0$ and $\lim_{m\rightarrow \infty }\|\mathbf{s}^{m}\|^{2}=0$ for
the MCP and SCAD penalties.
\end{proposition}

This proposition shows that the primal feasibility and the dual feasibility
are achieved by the algorithm.

\emph{Proof. } By the definition of $\boldsymbol{\delta }^{m+1}$, we have%
\begin{equation*}
L(\boldsymbol{\eta }^{m+1},\boldsymbol{\beta }^{m+1},\boldsymbol{\delta }%
^{m+1},\mathbf{\boldsymbol{\upsilon }}^{m})\mathbf{\leq }L(\boldsymbol{\eta }%
^{m+1},\boldsymbol{\beta }^{m+1}\mathbf{,\boldsymbol{\delta },\boldsymbol{%
\upsilon }}^{m}\mathbf{)}
\end{equation*}%
for any $\boldsymbol{\delta }$. Define
\begin{eqnarray*}
f^{m+1} &=&\inf_{\mathbf{A\boldsymbol{\beta }}^{m+1}\mathbf{-}\boldsymbol{%
\delta }=\mathbf{0}}\{\frac{1}{2}\left\Vert \mathbf{y-Z}\boldsymbol{\eta }%
^{m+1}\mathbf{-X\boldsymbol{\beta }}^{m+1}\right\Vert
^{2}+\sum\nolimits_{i<j}p_{\gamma }(|\boldsymbol{\delta }_{ij}|,\lambda )\}
\\
&=&\inf_{\mathbf{A\boldsymbol{\beta }}^{m+1}\mathbf{-}\boldsymbol{\delta }=%
\mathbf{0}}L(\boldsymbol{\eta }^{m+1},\boldsymbol{\beta }^{m+1}\mathbf{,%
\boldsymbol{\delta },\boldsymbol{\upsilon }}^{m}\mathbf{).}
\end{eqnarray*}%
Then
\begin{equation*}
L(\boldsymbol{\eta }^{m+1},\boldsymbol{\beta }^{m+1},\boldsymbol{\delta }%
^{m+1},\mathbf{\boldsymbol{\upsilon }}^{m})\mathbf{\leq }f^{m+1}.
\end{equation*}%
Let $t$ be an integer. Since $\mathbf{\boldsymbol{\upsilon }}^{m+1}=\mathbf{%
\boldsymbol{\upsilon }}^{m}+\vartheta (\mathbf{A\boldsymbol{\beta }}^{m+1}%
\mathbf{-}\boldsymbol{\delta }^{m+1})$, then we have%
\begin{equation*}
\mathbf{\boldsymbol{\upsilon }}^{m+t-1}=\mathbf{\boldsymbol{\upsilon }}%
^{m}+\vartheta \sum\nolimits_{i=1}^{t-1}(\mathbf{A\boldsymbol{\beta }}^{m+i}%
\mathbf{-}\boldsymbol{\delta }^{m+i}),
\end{equation*}%
and thus
\begin{eqnarray*}
&&L(\boldsymbol{\eta }^{m+t},\boldsymbol{\beta }^{m+t},\boldsymbol{\delta }%
^{m+t},\mathbf{\boldsymbol{\upsilon }}^{m+t-1}) \\
&=&\frac{1}{2}\left\Vert \mathbf{y-Z}\boldsymbol{\eta }^{m+t}\mathbf{-X%
\boldsymbol{\beta }}^{m+t}\right\Vert ^{2}+(\mathbf{\boldsymbol{\upsilon }}%
^{m+t-1})^{\text{T}}(\mathbf{A\boldsymbol{\beta }}^{m+t}\mathbf{-}%
\boldsymbol{\delta }^{m+t}\mathbf{\mathbf{)}} \\
&&+\frac{\vartheta }{2}\|\mathbf{A\boldsymbol{\beta }}^{m+t}\mathbf{-}%
\boldsymbol{\delta }^{m+t}\|^{2}+\sum\nolimits_{i<j}p_{\gamma }(|\boldsymbol{%
\delta }_{ij}^{m+t}|,\lambda ) \\
&=&\frac{1}{2}\left\Vert \mathbf{y-Z}\boldsymbol{\eta }^{m+t}\mathbf{-X%
\boldsymbol{\beta }}^{m+t}\right\Vert ^{2}+(\mathbf{\boldsymbol{\upsilon }}%
^{m})^{\text{T}}(\mathbf{A\boldsymbol{\beta }}^{m+t}\mathbf{-}\boldsymbol{%
\delta }^{m+t}\mathbf{\mathbf{)}} \\
&&\mathbf{\mathbf{+}}\vartheta \sum\nolimits_{i=1}^{t-1}(\mathbf{A%
\boldsymbol{\beta }}^{m+i}\mathbf{-}\boldsymbol{\delta }^{m+i}\mathbf{%
\mathbf{)}}^{\text{T}}(\mathbf{A\boldsymbol{\beta }}^{m+t}\mathbf{-}%
\boldsymbol{\delta }^{m+t}\mathbf{\mathbf{)}} \\
&&+\frac{\vartheta }{2}\|\mathbf{A\boldsymbol{\beta }}^{m+t}\mathbf{-}%
\boldsymbol{\delta }^{m+t}\|^{2}+p_{\gamma }(|\boldsymbol{\delta }%
_{ij}^{m+t}|,\lambda ) \\
&\mathbf{\leq }&f^{m+t}.
\end{eqnarray*}%
Since the objective function $L(\boldsymbol{\eta },\boldsymbol{\beta },%
\boldsymbol{\delta },\mathbf{\boldsymbol{\upsilon }})$ is differentiable
with respect to $(\boldsymbol{\eta },\boldsymbol{\beta }\mathbf{)}$ and is
convex with respect to $\boldsymbol{\delta }$, by applying the results in
Theorem 4.1 of \citep{tseng:2001}, the sequence $(\boldsymbol{\eta }^{m},%
\boldsymbol{\beta }^{m},\boldsymbol{\delta }^{m}\mathbf{)}$ has a limit
point, denoted by $(\boldsymbol{\eta }^{\ast },\boldsymbol{\beta }\mathbf{%
^{\ast },}\boldsymbol{\delta }^{\ast }\mathbf{)}$. Then we have
\begin{equation*}
f^{\ast }=\lim_{m\rightarrow \infty }f^{m+1}=\lim_{m\rightarrow \infty
}f^{m+t}=\inf_{\mathbf{A\boldsymbol{\beta }}^{\ast }\mathbf{-}\boldsymbol{%
\delta }=\mathbf{0}}\{\frac{1}{2}\left\Vert \mathbf{y-Z}\boldsymbol{\eta }%
^{\ast }\mathbf{-X\boldsymbol{\beta }}^{\ast }\right\Vert
^{2}+\sum\nolimits_{i<j}p_{\gamma }(|\boldsymbol{\delta }_{ij}|,\lambda )\},
\end{equation*}%
and for all $t\geq 0$
\begin{eqnarray*}
&&\lim_{m\rightarrow \infty }L(\mathbf{\boldsymbol{\mu }}^{m+t},\mathbf{%
\boldsymbol{\beta }}^{m+t}\mathbf{,\boldsymbol{\eta }}^{m+t}\mathbf{,%
\boldsymbol{\upsilon }}^{m+t-1}\mathbf{)} \\
&=&\frac{1}{2}\left\Vert \mathbf{y-Z}\boldsymbol{\eta }^{\ast }\mathbf{-X%
\boldsymbol{\beta }}^{\ast }\right\Vert ^{2}+\sum\nolimits_{i<j}p_{\gamma }(|%
\boldsymbol{\delta }_{ij}^{\ast }|,\lambda )+\lim_{m\rightarrow \infty }(%
\mathbf{\boldsymbol{\upsilon }}^{m})^{\text{T}}(\mathbf{A\boldsymbol{\beta }%
^{\ast }-}\boldsymbol{\delta }^{\ast }\mathbf{\mathbf{)+(}}t-\frac{1}{2}%
)\vartheta \|\mathbf{A\boldsymbol{\beta }^{\ast }-}\boldsymbol{\delta }%
^{\ast }\|^{2} \\
&\mathbf{\leq }&f^{\ast }.
\end{eqnarray*}%
Hence $\lim_{m\rightarrow \infty }\|\mathbf{r}^{m}\|^{2}=r^{\ast }\mathbf{=}%
\|\mathbf{A\boldsymbol{\beta }^{\ast }-}\boldsymbol{\delta }^{\ast }\|^{2}=0
$.

Since $\boldsymbol{\beta }^{m+1}$ minimizes $L(\boldsymbol{\eta }^{m},%
\boldsymbol{\beta },\boldsymbol{\delta }^{m},\mathbf{\boldsymbol{\upsilon }}%
^{m})$ by definition, we have that
\begin{equation*}
L(\boldsymbol{\eta }^{m},\boldsymbol{\beta },\boldsymbol{\delta }^{m},%
\mathbf{\boldsymbol{\upsilon }}^{m})\mathbf{/\partial \boldsymbol{\beta }=0},
\end{equation*}%
and moreover,%
\begin{eqnarray*}
&&L(\boldsymbol{\eta }^{m},\boldsymbol{\beta }^{m+1},\boldsymbol{\delta }%
^{m},\mathbf{\boldsymbol{\upsilon }}^{m})\mathbf{/\partial \boldsymbol{\beta
}} \\
&=&\mathbf{X}^{\text{T}}(\mathbf{Z}\boldsymbol{\eta }^{m}\mathbf{+X%
\boldsymbol{\beta }}^{m+1}\mathbf{-y)+A}^{\text{T}}\mathbf{\boldsymbol{%
\upsilon }}^{m}+\vartheta \mathbf{A}^{\text{T}}(\mathbf{A\mathbf{\boldsymbol{%
\beta }}}^{m+1}\mathbf{-}\boldsymbol{\delta }^{m}) \\
&=&\mathbf{X}^{\text{T}}(\mathbf{Z}\boldsymbol{\eta }^{m}\mathbf{+X%
\boldsymbol{\beta }}^{m+1}\mathbf{-y)+A}^{\text{T}}(\mathbf{\boldsymbol{%
\upsilon }}^{m}+\vartheta (\mathbf{A\mathbf{\mathbf{\boldsymbol{\beta }}}}%
^{m+1}\mathbf{-}\boldsymbol{\delta }^{m})) \\
&=&\mathbf{X}^{\text{T}}(\mathbf{Z}\boldsymbol{\eta }^{m}\mathbf{+X%
\boldsymbol{\beta }}^{m+1}\mathbf{-y)+A}^{\text{T}}(\mathbf{\boldsymbol{%
\upsilon }}^{m+1}-\vartheta (\mathbf{A\boldsymbol{\beta }}^{m+1}\mathbf{-}%
\boldsymbol{\delta }^{m+1})+\vartheta (\mathbf{A\boldsymbol{\beta }}^{m+1}%
\mathbf{-}\boldsymbol{\delta }^{m})) \\
&=&\mathbf{X}^{\text{T}}(\mathbf{Z}\boldsymbol{\eta }^{m}\mathbf{+X%
\boldsymbol{\beta }}^{m+1}\mathbf{-y)+A}^{\text{T}}\mathbf{\boldsymbol{%
\upsilon }}^{m+1}+\vartheta \mathbf{A}^{\text{T}}(\boldsymbol{\delta }^{m+1}-%
\boldsymbol{\delta }^{m}).
\end{eqnarray*}%
Therefore,
\begin{equation*}
\mathbf{s}^{m+1}=\vartheta \mathbf{A}^{\text{T}}(\boldsymbol{\delta }^{m+1}-%
\boldsymbol{\delta }^{m})=-(\mathbf{X}^{\text{T}}(\mathbf{Z}\boldsymbol{\eta
}^{m}\mathbf{+X\boldsymbol{\beta }}^{m+1}\mathbf{-y)+A}^{\text{T}}\mathbf{%
\boldsymbol{\upsilon }}^{m+1}).
\end{equation*}%
Since $\Vert \mathbf{A\boldsymbol{\beta }^{\ast }-}\boldsymbol{\delta }%
^{\ast }\Vert ^{2}=0$,
\begin{eqnarray*}
&&\lim_{m\rightarrow \infty }L(\boldsymbol{\eta }^{m},\boldsymbol{\beta }%
^{m+1},\boldsymbol{\delta }^{m},\mathbf{\boldsymbol{\upsilon }}^{m})\mathbf{%
/\partial \boldsymbol{\beta }} \\
&=&\lim_{m\rightarrow \infty }\mathbf{X}^{\text{T}}(\mathbf{Z}\boldsymbol{%
\eta }^{m}\mathbf{+X\boldsymbol{\beta }}^{m+1}\mathbf{-y)+A}^{\text{T}}%
\mathbf{\boldsymbol{\upsilon }}^{m+1} \\
&=&\mathbf{X}^{\text{T}}(\mathbf{Z}\boldsymbol{\eta }\mathbf{^{\ast }+X%
\boldsymbol{\beta }^{\ast }-y)+A}^{\text{T}}\mathbf{\boldsymbol{\upsilon }%
^{\ast }}=\mathbf{0}.
\end{eqnarray*}%
Therefore, $\lim_{m\rightarrow \infty }\mathbf{s}^{m+1}=\mathbf{0}$.

\subsection{Extension to nonlinear models\label{SEC:nonlinear}}

In this paper, we focus on the linear regression model (\ref{Mod1}) to
introduce our proposed method for exploring treatment heterogeneity. It is
worth mentioning that our method can be readily extended to semi-parametric
models by relaxing the linearity assumption on ${\boldsymbol{z}}_{i}$.
Considering the model:
\begin{equation*}
y_{i}=m({\boldsymbol{z}}_{i})+{\boldsymbol{x}}_{i}^{\text{T}}\boldsymbol{%
\beta }_{i}+\varepsilon _{i},i=1,\ldots ,n,
\end{equation*}%
where $m({\boldsymbol{\cdot }})$ is an unknown function of ${\boldsymbol{z}}%
_{i}$. This model has no constraint on the functional form of ${\boldsymbol{z%
}}_{i}$. Following the approach in \cite{Ma.Racine.Yang:2015}, we can
estimate $m({\boldsymbol{\cdot }})$ by tensor-product regression splines
weighted by categorical kernel functions, where the splines are used for the
continuous predictors and the categorical kernels are for the discrete
predictors, respectively. Then the objective function for obtaining the
estimates consists of two parts similar as given in (\ref{EQ:objective}). The first
part is a weighted least squares criterion as presented in equation (2) of
\cite{Ma.Racine.Yang:2015}, and the second part contains the same penalty
functions as given in (\ref{EQ:objective}). As a result, the proposed ADMM
algorithm proposed in Section \ref{SEC:computation} can be applied.
Moreover, we can also assume semi-parametric structures on $m({\boldsymbol{%
\cdot }})$ for further dimensionality reduction while retaining model
flexibility. For example, when ${\boldsymbol{z}}_{i}$ is a set of continuous
variables, we can assume the additive structure:
\begin{equation*}
m({\boldsymbol{z}}_{i})=m_{1}(z_{i1})+\cdots +m_{q}(z_{iq}),
\end{equation*}%
where $m_{k}(\cdot )$ for $k=1,...,q$ are unknown functions. Also the
partially linear additive structure is another popular semi-parametric
model, given as
\begin{equation*}
m({\boldsymbol{z}}_{i})=m_{1}(z_{i1})+\cdots +m_{q_{1}}(z_{iq_{1}})+{%
\boldsymbol{z}}_{i2}^{\text{T}}\boldsymbol{\eta ,}
\end{equation*}%
where ${\boldsymbol{z}}_{i}=({\boldsymbol{z}}_{i1}^{\text{T}},{\boldsymbol{z}%
}_{i2}^{\text{T}})^{\text{T}}$, ${\boldsymbol{z}}%
_{i1}=(z_{i1},...,z_{iq_{1}})^{\text{T}}$ and ${\boldsymbol{z}}%
_{i2}=(z_{i,q_{1}+1},...,z_{iq})^{\text{T}}$. If we use regression splines
to approximate the unknown functions, the same ADMM algorithm given in
Section \ref{SEC:computation} can be applied to obtain the parameter
estimators with the components in ${\boldsymbol{z}}_{i}$ replaced by their
spline basis functions. We refer to \cite{Ma.Song.Wang:2013} for the details
of using regression splines to estimate unknown functions in these settings.

It is also of interest to extend the proposed method to the case with
discrete responses. For a general scenario, one may fit a generalized linear
model:%
\begin{equation*}
E(y_{i}|{\boldsymbol{z}}_{i},{\boldsymbol{x}}_{i})=\mu _{i}=g^{-1}({%
\boldsymbol{z}}_{i}^{\text{T}}\boldsymbol{\eta }+{\boldsymbol{x}}_{i}^{\text{%
T}}\boldsymbol{\beta }_{i}),i=1,\ldots ,n,
\end{equation*}%
where $g$ is a known monotone link function. For obtaining the estimates of
the parameters, we can consider the negative quasi-likelihood function%
\begin{equation*}
Q(\mu ,y)=\int\nolimits_{\mu }^{y}\{(y-\xi )/V(\xi )\}d\xi ,
\end{equation*}%
where $V(\cdot )$ is the conditional variance of $y$ given $({\boldsymbol{z}}%
,{\boldsymbol{x}})$. Then the parameter estimates can be obtained by
minimizing
\begin{equation*}
\sum\nolimits_{i=1}^{n}Q(g^{-1}({\boldsymbol{z}}_{i}^{\text{T}}\boldsymbol{%
\eta }+{\boldsymbol{x}}_{i}^{\text{T}}\boldsymbol{\beta }_{i}),y_{i})+\sum%
\nolimits_{1\leq i<j\leq n}p(\Vert \boldsymbol{\beta }_{i}-\boldsymbol{\beta
}_{j}\Vert ,\lambda ).
\end{equation*}%
Because $Q(g^{-1}({\boldsymbol{z}}_{i}^{\text{T}}\boldsymbol{\eta }+{%
\boldsymbol{x}}_{i}^{\text{T}}\boldsymbol{\beta }_{i}),y_{i})$ is
differentiable with respect to ($\boldsymbol{\eta ,\beta }_{i}$), the ADMM
algorithm given in Section \ref{SEC:computation} can be straightforwardly
applied.

\subsection{Proof of Theorem \protect\ref{THM:norm}}

In this section we show the results in Theorem \ref{THM:norm}. For every $%
\mathbf{\boldsymbol{\beta }\in }\mathcal{M}_{\mathcal{G}}$, it can be
written as $\mathbf{\boldsymbol{\beta }=W\boldsymbol{\alpha }}$. Recall ${%
\mathbf{U}}=(\mathbf{Z},\mathbf{XW)}$. We have 
\begin{equation*}
\left(
\begin{array}{c}
\widehat{\mathbf{\boldsymbol{\eta }}}^{or} \\
\widehat{\boldsymbol{\alpha }}^{or}%
\end{array}%
\right) =\arg \min_{\mathbf{\boldsymbol{\eta }\in }R^{q},\boldsymbol{\alpha
\in }R^{Kp}}\frac{1}{2}\Vert \mathbf{y-Z\boldsymbol{\eta }-X\boldsymbol{%
\beta }\Vert }=\arg \min_{\mathbf{\boldsymbol{\eta }\in }R^{q},\boldsymbol{%
\alpha \in }R^{Kp}}\frac{1}{2}\Vert \mathbf{y-Z\boldsymbol{\eta }-XW%
\boldsymbol{\alpha }\Vert }^{2}.
\end{equation*}%
Thus
\begin{equation*}
\left(
\begin{array}{c}
\widehat{\mathbf{\boldsymbol{\eta }}}^{or} \\
\widehat{\boldsymbol{\alpha }}^{or}%
\end{array}%
\right) =[(\mathbf{Z},\mathbf{XW)}^{\text{T}}(\mathbf{Z},\mathbf{XW)]}^{-1}(%
\mathbf{Z},\mathbf{XW)}^{\text{T}}\mathbf{y}=({\mathbf{U}}^{\text{T}}{%
\mathbf{U}})^{-1}{\mathbf{U}}^{T}{\mathbf{y}}\text{. }
\end{equation*}%
Then
\begin{equation*}
\left(
\begin{array}{c}
\widehat{\mathbf{\boldsymbol{\eta }}}^{or}-\mathbf{\boldsymbol{\eta }}^{0}
\\
\widehat{\boldsymbol{\alpha }}^{or}-\boldsymbol{\alpha }^{0}%
\end{array}%
\right) =({\mathbf{U}}^{\text{T}}{\mathbf{U}})^{-1}{\mathbf{U}}^{\text{T}}%
\boldsymbol{\varepsilon }.
\end{equation*}%
Hence%
\begin{equation}
\left\Vert \left(
\begin{array}{c}
\widehat{\mathbf{\boldsymbol{\eta }}}^{or}-\mathbf{\boldsymbol{\eta }}^{0}
\\
\widehat{\boldsymbol{\alpha }}^{or}-\boldsymbol{\alpha }^{0}%
\end{array}%
\right) \right\Vert \leq \left\Vert ({\mathbf{U}}^{\text{T}}{\mathbf{U}}%
)^{-1}\right\Vert \left\Vert {\mathbf{U}}^{\text{T}}\boldsymbol{\varepsilon }%
\right\Vert .  \label{EQ:supnorm}
\end{equation}%
By Condition (C1), we have
\begin{equation}
\left\Vert ({\mathbf{U}}^{\text{T}}{\mathbf{U}})^{-1}\right\Vert \leq
C_{1}^{-1}\left\vert \mathcal{G}_{\min }\right\vert ^{-1}.  \label{EQ:zxsup}
\end{equation}%
Moreover
\begin{equation*}
P(\left\Vert {\mathbf{U}}^{\text{T}}\boldsymbol{\varepsilon }\right\Vert
_{\infty }>C\sqrt{n\log n})\leq P(\left\Vert \mathbf{Z}^{\text{T}}%
\boldsymbol{\varepsilon }\right\Vert _{\infty }>C\sqrt{n\log n}%
)+P(\left\Vert (\mathbf{XW)}^{\text{T}}\boldsymbol{\varepsilon }\right\Vert
_{\infty }>C\sqrt{n\log n}),
\end{equation*}%
for some constant $0<C<\infty $. Since $\mathbf{XW=}\left[ \mathbf{x}_{i}^{%
\text{T}}1\{i\in \mathcal{G}_{k}\}\right] _{i=1,k=1}^{n,K}$, we have
\begin{equation*}
\left\Vert (\mathbf{XW)}^{\text{T}}\boldsymbol{\varepsilon }\right\Vert
_{\infty }=\sup_{j,k}|\sum\nolimits_{i=1}^{n}x_{ij}\varepsilon _{i}1\{i\in
\mathcal{G}_{k}\}|
\end{equation*}%
and by union bound, Condition (C1) that $\sum\nolimits_{i=1}^{n}x_{ij}^{2}1%
\{i\in \mathcal{G}_{k\text{ }}\}=\left\vert \mathcal{G}_{k}\right\vert $ and
Condition (C3),
\begin{eqnarray*}
&&P\left( \left\Vert (\mathbf{XW)}^{\text{T}}\boldsymbol{\varepsilon }%
\right\Vert _{\infty }>C\sqrt{n\log n}\right) \\
&\leq &\sum\nolimits_{j=1,k=1}^{p,K}P\left(
|\sum\nolimits_{i=1}^{n}x_{ij}\varepsilon _{i}1\{i\in \mathcal{G}_{k}\}|>C%
\sqrt{n\log n}\right) \\
&\leq &\sum\nolimits_{j=1,k=1}^{p,K}P\left(
|\sum\nolimits_{i=1}^{n}x_{ij}\varepsilon _{i}1\{i\in \mathcal{G}_{k}\}|>%
\sqrt{|\mathcal{G}_{k}|}C\sqrt{\log n}\right) \\
&\leq &2Kp\exp (-c_{1}C^{2}\log n)=2Kpn^{-c_{1}C^{2}}.
\end{eqnarray*}%
By union bound, Condition (C1) that $\left\Vert \mathbf{Z}_{k}\right\Vert =%
\sqrt{n}$, where $\mathbf{Z}_{k}$ is the $k$th column of $\mathbf{Z}$, and
Condition (C3),
\begin{eqnarray*}
&&P\left( \left\Vert \mathbf{Z}^{\text{T}}\boldsymbol{\varepsilon }%
\right\Vert _{\infty }>C\sqrt{n\log n}\right) \\
&\leq &\sum\nolimits_{k=1}^{q}P\left( |\mathbf{Z}_{k}^{\text{T}}\boldsymbol{%
\varepsilon |}>\sqrt{n}C\sqrt{\log n}\right) \\
&\leq &2q\exp (-c_{1}C^{2}\log n)=2qn^{-c_{1}C^{2}}.
\end{eqnarray*}%
%
%
%
%
%
%
%
%
%
%
%
%
%
%
%
%
%
%
%
%
%
%
%
%
%
%
%
%
%
%
%
It follows that
\begin{equation*}
P(\left\Vert {\mathbf{U}}^{\text{T}}\boldsymbol{\varepsilon }\right\Vert
_{\infty }>C\sqrt{n\log n})\leq 2(Kp+q)n^{-c_{1}C^{2}}.
\end{equation*}%
Since $\left\Vert {\mathbf{U}}^{\text{T}}\boldsymbol{\varepsilon }%
\right\Vert \leq \sqrt{q+Kp}\left\Vert {\mathbf{U}}^{\text{T}}\boldsymbol{%
\varepsilon }\right\Vert _{\infty }$, then
\begin{equation}
P(\left\Vert {\mathbf{U}}^{\text{T}}\boldsymbol{\varepsilon }\right\Vert >C%
\sqrt{q+Kp}\sqrt{n\log n})\leq 2(Kp+q)n^{-c_{1}C^{2}}.  \label{EQ:zxeps}
\end{equation}%
Therefore, by (\ref{EQ:supnorm}), (\ref{EQ:zxsup}) and (\ref{EQ:zxeps}), we
have with probability at least $1-2(Kp+q)n^{-c_{1}C^{2}}$,
\begin{equation*}
\left\Vert \left(
\begin{array}{c}
\widehat{\mathbf{\boldsymbol{\eta }}}^{or}-\mathbf{\boldsymbol{\eta }}^{0}
\\
\widehat{\boldsymbol{\alpha }}^{or}-\boldsymbol{\alpha }^{0}%
\end{array}%
\right) \right\Vert \leq CC_{1}^{-1}\sqrt{q+Kp}\left\vert \mathcal{G}_{\min
}\right\vert ^{-1}\sqrt{n\log n}.
\end{equation*}%
The result (\ref{EQ:supnormmubeta}) in Theorem \ref{THM:norm} is proved by
letting $C=c_{1}^{-1/2}$. Moreover,%
\begin{eqnarray*}
\left\Vert \widehat{\mathbf{\boldsymbol{\beta }}}^{or}-\mathbf{\boldsymbol{%
\beta }}^{0}\right\Vert ^{2} &=&\sum\nolimits_{k=1}^{K}\sum\nolimits_{i\in
\mathcal{G}_{k}}\left\Vert \widehat{\boldsymbol{\alpha }}_{k}^{or}-%
\boldsymbol{\alpha }_{k}^{0}\right\Vert ^{2}\leq \left\vert \mathcal{G}%
_{\max }\right\vert \sum\nolimits_{k=1}^{K}\left\Vert \widehat{\boldsymbol{%
\alpha }}_{k}^{or}-\boldsymbol{\alpha }_{k}^{0}\right\Vert ^{2} \\
&=&\left\vert \mathcal{G}_{\max }\right\vert \left\Vert \widehat{\boldsymbol{%
\alpha }}^{or}-\boldsymbol{\alpha }^{0}\right\Vert ^{2}\leq \left\vert
\mathcal{G}_{\max }\right\vert \phi _{n}^{2},
\end{eqnarray*}%
and
\begin{equation*}
\sup_{i}\left\Vert \widehat{\mathbf{\boldsymbol{\beta }}}_{i}^{or}-\mathbf{%
\boldsymbol{\beta }}_{i}^{0}\right\Vert =\sup_{k}\left\Vert \widehat{%
\boldsymbol{\alpha }}_{k}^{or}-\boldsymbol{\alpha }_{k}^{0}\right\Vert \leq
\Vert \widehat{\boldsymbol{\alpha }}^{or}-\boldsymbol{\alpha }^{0}\Vert \leq
\phi _{n}.
\end{equation*}%
Let ${\mathbf{U}}=({\mathbf{U}}_{1},\ldots ,{\mathbf{U}}_{n})^{\text{T}}$,
and $\boldsymbol{\Xi }_{n}={\mathbf{U}}^{\text{T}}{\mathbf{U}}$. Then
\begin{equation*}
\mathbf{a}_{n}^{\text{T}}((\widehat{\boldsymbol{\eta }}^{or}-\boldsymbol{%
\eta }^{0})^{\text{T}},(\widehat{\boldsymbol{\alpha }}^{or}-\boldsymbol{%
\alpha }^{0})^{\text{T}})^{\text{T}}=\sum\nolimits_{i=1}^{n}\mathbf{a}_{n}^{%
\text{T}}\boldsymbol{\Xi }_{n}^{-1}{\mathbf{U}}_{i}\varepsilon _{i}.
\end{equation*}%
Hence
\begin{equation*}
E\{\mathbf{a}_{n}^{\text{T}}((\widehat{\boldsymbol{\eta }}^{or}-\boldsymbol{%
\eta }^{0})^{\text{T}},(\widehat{\boldsymbol{\alpha }}^{or}-\boldsymbol{%
\alpha }^{0})^{\text{T}})^{\text{T}}\}=0,
\end{equation*}%
and for any vector $\mathbf{a}_{n}\in \mathop{{\rm
I}\kern-.2em\hbox{\rm R}}\nolimits^{q+Kp}$ with $||\mathbf{a}_{n}||=1$,
\begin{eqnarray*}
&&\text{var}\{\mathbf{a}_{n}^{\text{T}}((\widehat{\boldsymbol{\eta }}^{or}-%
\boldsymbol{\eta }^{0})^{\text{T}},(\widehat{\boldsymbol{\alpha }}^{or}-%
\boldsymbol{\alpha }^{0})^{\text{T}})^{\text{T}}\} \\
&=&\sigma _{n}^{2}(\mathbf{a}_{n})=\sigma ^{2}\left[ \mathbf{a}_{n}^{\text{T}%
}({\mathbf{U}}^{\text{T}}{\mathbf{U}})^{-1}\mathbf{a}_{n}\right] \geq \sigma
^{2}\mathbf{a}_{n}^{\text{T}}\boldsymbol{\Xi }_{n}^{-1}\mathbf{a}_{n}.
\end{eqnarray*}%
Moreover, for any $\epsilon >0$,
\begin{eqnarray*}
&&\sum\nolimits_{i=1}^{n}E[(\mathbf{a}_{n}^{\text{T}}\boldsymbol{\Xi }%
_{n}^{-1}{\mathbf{U}}_{i}\varepsilon _{i})^{2}\cdot 1_{\{|\mathbf{a}_{n}^{%
\text{T}}\boldsymbol{\Xi }_{n}^{-1}{\mathbf{U}}_{i}\varepsilon
_{i}|>\epsilon \sigma _{n}(\mathbf{a}_{n})\}}] \\
&\leq &\sum\nolimits_{i=1}^{n}\{E(\mathbf{a}_{n}^{\text{T}}\boldsymbol{\Xi }%
_{n}^{-1}{\mathbf{U}}_{i}\varepsilon _{i})^{4}\}^{1/2}[P\{|\mathbf{a}_{n}^{%
\text{T}}\boldsymbol{\Xi }_{n}^{-1}{\mathbf{U}}_{i}\varepsilon
_{i}|>\epsilon \sigma _{n}(\mathbf{a}_{n})\}]^{1/2}.
\end{eqnarray*}%
Since $E(\varepsilon _{i}^{4})\leq c$ for some constant $c\in (0,\infty )$
by Condition (C2), then%
\begin{equation*}
\sum\nolimits_{i=1}^{n}\{E(\mathbf{a}_{n}^{\text{T}}\boldsymbol{\Xi }%
_{n}^{-1}{\mathbf{U}}_{i}\varepsilon _{i})^{4}\}^{1/2}\leq c^{1/2}\mathbf{a}%
_{n}^{\text{T}}\boldsymbol{\Xi }_{n}^{-1}\mathbf{a}_{n}.
\end{equation*}%
Moreover,
\begin{eqnarray*}
&&\max_{i}P\left\{ |\mathbf{a}_{n}^{\text{T}}\boldsymbol{\Xi }_{n}^{-1}{%
\mathbf{U}}_{i}\varepsilon _{i}|>\epsilon \sigma _{n}(\mathbf{a}_{n})\right\}
\\
&\leq &\max_{i}E(\mathbf{a}_{n}^{\text{T}}\boldsymbol{\Xi }_{n}^{-1}{\mathbf{%
U}}_{i}\varepsilon _{i})^{2}/\{\epsilon ^{2}\sigma _{n}^{2}(\mathbf{a}%
_{n})\}\leq c^{\prime }\epsilon ^{-2}(q+Kp)\mathbf{a}_{n}^{\text{T}}%
\boldsymbol{\Xi }_{n}^{-1}\boldsymbol{\Xi }_{n}^{-1}\mathbf{a}_{n}/(\mathbf{a%
}_{n}^{\text{T}}\boldsymbol{\Xi }_{n}^{-1}\mathbf{a}_{n})
\end{eqnarray*}%
for some constant $c^{\prime }\in (0,\infty )$. Therefore, by the above
results, we have
\begin{eqnarray*}
&&\sigma _{n}^{-2}(\mathbf{a}_{n})\sum\nolimits_{i=1}^{n}E[(\mathbf{a}_{n}^{%
\text{T}}\boldsymbol{\Xi }_{n}^{-1}{\mathbf{U}}_{i}\varepsilon
_{i})^{2}\cdot 1_{\{|\mathbf{a}_{n}^{\text{T}}\boldsymbol{\Xi }_{n}^{-1}{%
\mathbf{U}}_{i}\varepsilon _{i}|>\epsilon \sigma _{n}(\mathbf{a}_{n})\}}] \\
&\leq &\{\sigma ^{2}\mathbf{a}_{n}^{\text{T}}\boldsymbol{\Xi }_{n}^{-1}%
\mathbf{a}_{n}\}^{-1}c^{1/2}\mathbf{a}_{n}^{\text{T}}\boldsymbol{\Xi }%
_{n}^{-1}\mathbf{a}_{n}\{c^{\prime }\epsilon ^{-2}(q+Kp)\mathbf{a}_{n}^{%
\text{T}}\boldsymbol{\Xi }_{n}^{-1}\boldsymbol{\Xi }_{n}^{-1}\mathbf{a}_{n}/(%
\mathbf{a}_{n}^{\text{T}}\boldsymbol{\Xi }_{n}^{-1}\mathbf{a}_{n})\}^{1/2} \\
&\leq &c^{1/2}c^{\prime 1/2}C_{1}^{-1/2}\epsilon ^{-1}\sigma
^{-1}(q+Kp)^{1/2}\left\vert \mathcal{G}_{\min }\right\vert ^{-1/2}=o(1).
\end{eqnarray*}%
Then, the result (\ref{EQ:normal}) follows from Lindeberg--Feller Central
Limit Theorem.

\subsection{Proof of Theorem \protect\ref{THM:selection}}

In this section we show the results in Theorem \ref{THM:selection}. Define
\begin{eqnarray*}
L_{n}(\mathbf{\boldsymbol{\eta },\boldsymbol{\beta }}) &=&\frac{1}{2}\|%
\mathbf{y-Z\boldsymbol{\eta }-X\boldsymbol{\beta }}\|^{2},P_{n}(\mathbf{%
\boldsymbol{\beta }})=\lambda \sum_{i<j}\rho (\|\mathbf{\boldsymbol{\beta }}%
_{i}-\mathbf{\boldsymbol{\beta }}_{j}\|), \\
L_{n}^{\mathcal{G}}(\mathbf{\boldsymbol{\eta },\boldsymbol{\alpha } }) &=&%
\frac{1}{2}\|\mathbf{y-Z\boldsymbol{\eta }-XW\boldsymbol{\alpha } }%
\|^{2},P_{n}^{\mathcal{G}}(\boldsymbol{\alpha })=\lambda \sum_{k<k^{\prime
}}|\mathcal{G}_{k}\|\mathcal{G}_{k^{\prime }}|\rho (\|\boldsymbol{\alpha }%
_{k}-\boldsymbol{\alpha }_{k^{\prime }}\|),
\end{eqnarray*}%
and let
\begin{equation*}
Q_{n}(\mathbf{\boldsymbol{\eta },\boldsymbol{\beta }})=L_{n}(\mathbf{%
\boldsymbol{\eta },\boldsymbol{\beta }})+P_{n}(\mathbf{\boldsymbol{\beta }}%
),Q_{n}^{\mathcal{G}}(\mathbf{\boldsymbol{\eta },\boldsymbol{\alpha } }%
)=L_{n}^{\mathcal{G}}(\mathbf{\boldsymbol{\eta },\boldsymbol{\alpha }}%
)+P_{n}^{\mathcal{G}}(\mathbf{\boldsymbol{\alpha } }).
\end{equation*}%
Let $T:\mathcal{M}_{\mathcal{G}}\rightarrow R^{Kp}$ be the mapping that $T(%
\mathbf{\boldsymbol{\beta })}$ is the $Kp\times 1$ vector consisting of $K$
vectors with dimension $p$ and its $k^{\text{th}}$ vector component equals
to the common value of $\mathbf{\boldsymbol{\beta }}_{i}$ for $i\in \mathcal{%
G}_{k}$. Let $T^{\ast }:R^{np}\rightarrow R^{Kp}$ be the mapping that $%
T^{\ast }(\mathbf{\boldsymbol{\beta })=\{}|\mathcal{G}_{k}|^{-1}\sum%
\nolimits_{i\in \mathcal{G}_{k}}\mathbf{\boldsymbol{\beta }}_{i}^{\text{T}%
},k=1,\ldots ,K\}^{\text{T}}$. Clearly, when $\mathbf{\boldsymbol{\beta }\in
}\mathcal{M}_{\mathcal{G}}$, $T(\mathbf{\boldsymbol{\beta })=}T^{\ast }(%
\mathbf{\boldsymbol{\beta })}$. \

By calculation, for every $\mathbf{\boldsymbol{\beta }\in }\mathcal{M}_{%
\mathcal{G}}$, we have $P_{n}(\mathbf{\boldsymbol{\beta }})=P_{n}^{\mathcal{G%
}}(T(\mathbf{\boldsymbol{\beta }}))$ and for every $\boldsymbol{\alpha \in }%
R^{K}$, we have $P_{n}(T^{-1}(\boldsymbol{\alpha }))=P_{n}^{\mathcal{G}}(%
\boldsymbol{\alpha })$. Hence
\begin{equation}
Q_{n}(\mathbf{\boldsymbol{\eta },\boldsymbol{\beta }})=Q_{n}^{\mathcal{G}}(%
\mathbf{\boldsymbol{\eta },}T(\mathbf{\boldsymbol{\beta }})),Q_{n}^{\mathcal{%
G}}(\mathbf{\boldsymbol{\eta },\boldsymbol{\alpha }})=Q_{n}(\mathbf{%
\boldsymbol{\eta },}T^{-1}(\boldsymbol{\alpha })).  \label{EQ:qn}
\end{equation}%
Consider the neighborhood of $(\mathbf{\boldsymbol{\eta }}^{0}\mathbf{,%
\boldsymbol{\beta }}^{0})$:
\begin{equation*}
\Theta \mathcal{=\{}\mathbf{\boldsymbol{\eta }\in }R^{q},\mathbf{\boldsymbol{%
\beta }\mathbf{\in }}R^{Kp}\mathbf{:}\left\Vert \mathbf{\boldsymbol{\eta }}-%
\mathbf{\boldsymbol{\eta }}^{0}\right\Vert \leq \phi _{n},\sup_{i}\left\Vert
\mathbf{\boldsymbol{\beta }}_{i}-\mathbf{\boldsymbol{\beta }}%
_{i}^{0}\right\Vert \leq \phi _{n}\}.
\end{equation*}%
By Theorem \ref{THM:norm}, there exists an event $E_{1}$ in which
\begin{equation*}
\left\Vert \widehat{\mathbf{\boldsymbol{\eta }}}^{or}-\mathbf{\boldsymbol{%
\eta }}^{0}\right\Vert \leq \phi _{n},\sup_{i}\left\Vert \widehat{\mathbf{%
\boldsymbol{\beta }}}_{i}^{or}-\mathbf{\boldsymbol{\beta }}%
_{i}^{0}\right\Vert \leq \phi _{n}
\end{equation*}%
and $P(E_{1}^{C})\leq 2(q+Kp)n^{-1}$. Hence $(\widehat{\mathbf{\boldsymbol{%
\eta }}}^{or},\widehat{\mathbf{\boldsymbol{\beta }}}^{or})\in \Theta $ in $%
E_{1}$. For any $\mathbf{\boldsymbol{\beta }\in }R^{np}$, let $\mathbf{%
\boldsymbol{\beta }}^{\ast }=T^{-1}(T^{\ast }(\mathbf{\boldsymbol{\beta }))}$%
. We show that $(\widehat{\mathbf{\boldsymbol{\eta }}}^{or},\widehat{\mathbf{%
\boldsymbol{\beta }}}^{or})$ is a strictly local minimizer of the objective
function (\ref{EQ:objective}) with probability approaching $1$ through the
following two steps.

(i). In the event $E_{1}$, $Q_{n}(\mathbf{\boldsymbol{\eta },\boldsymbol{%
\beta }}^{\ast })>Q_{n}(\widehat{\mathbf{\boldsymbol{\eta }}}^{or}\mathbf{,}%
\widehat{\mathbf{\boldsymbol{\beta }}}^{or})$ for any $(\mathbf{\boldsymbol{%
\eta }}^{\text{T}}, \mathbf{\boldsymbol{\beta }}^{\text{T}})^{\text{T}}\in
\Theta $ and $((\mathbf{\boldsymbol{\eta }})^{\text{T}},(\mathbf{\boldsymbol{%
\beta }}^{\ast })^{\text{T}})^{\text{T}}\neq ((\widehat{\mathbf{\boldsymbol{%
\eta }}}^{or})^{\text{T}},(\widehat{\mathbf{\boldsymbol{\beta }}}^{or})^{%
\text{T}})^{\text{T}}$.

(ii). There is an event $E_{2}$ such that $P(E_{2}^{C})\leq 2n^{-1}$. In $%
E_{1}\cap E_{2}$, there is a neighborhood of $((\widehat{\mathbf{\boldsymbol{%
\eta }}}^{or})^{\text{T}},(\widehat{\mathbf{\boldsymbol{\beta }}}^{or})^{%
\text{T}})^{\text{T}}$, denoted by $\Theta _{n}$ such that $Q_{n}(\mathbf{%
\boldsymbol{\eta },\boldsymbol{\beta }})\geq Q_{n}(\mathbf{\boldsymbol{\eta }%
,\boldsymbol{\beta }}^{\ast })$ for any $((\mathbf{\boldsymbol{\eta }})^{%
\text{T}},(\mathbf{\boldsymbol{\beta }}^{\ast })^{\text{T}})^{\text{T}}\in
\Theta _{n}\cap \Theta $ for sufficiently large $n$.

Therefore, by the results in (i) and (ii), we have $Q_{n}(\mathbf{%
\boldsymbol{\eta },\boldsymbol{\beta }})>Q_{n}(\widehat{\mathbf{\boldsymbol{%
\eta }}}^{or}\mathbf{,}\widehat{\mathbf{\boldsymbol{\beta }}}^{or})$ for any
$(\mathbf{\boldsymbol{\eta }}^{\text{T}}, \mathbf{\boldsymbol{\beta }}^{%
\text{T}})^{\text{T}}\in \Theta _{n}\cap \Theta $ and $((\mathbf{\boldsymbol{%
\eta }})^{\text{T}},(\mathbf{\boldsymbol{\beta }})^{\text{T}})^{\text{T}%
}\neq ((\widehat{\mathbf{\boldsymbol{\eta }}}^{or})^{\text{T}},(\widehat{%
\mathbf{\boldsymbol{\beta }}}^{or})^{\text{T}})^{\text{T}}$ in $E_{1}\cap
E_{2}$, so that $((\widehat{\mathbf{\boldsymbol{\eta }}}^{or})^{\text{T}},(%
\widehat{\mathbf{\boldsymbol{\beta }}}^{or})^{\text{T}})^{\text{T}}$ is a
strict local minimizer of $Q_{n}(\mathbf{\boldsymbol{\eta },\boldsymbol{%
\beta }})$ (\ref{EQ:objective}) over the event $E_{1}\cap E_{2}$ with $%
P(E_{1}\cap E_{2})\geq 1-2(q+Kp+1)n^{-1}$ for sufficiently large $n$.

In the following we prove the result in (i). We first show $P_{n}^{\mathcal{G%
}}(T^{\ast }(\mathbf{\boldsymbol{\beta })})=C_{n}$ for any $\mathbf{%
\boldsymbol{\beta }\in }\Theta $, where $C_{n}$ is a constant which does not
depend on $\mathbf{\boldsymbol{\beta }}$. Let $T^{\ast }(\boldsymbol{\beta }%
)=\boldsymbol{\alpha } =(\boldsymbol{\alpha }_{1}^{\text{T}},\ldots ,%
\boldsymbol{\alpha }_{K}^{\text{T}})^{\text{T}}$. It suffices to show that $%
\|\boldsymbol{\alpha }_{k}-\boldsymbol{\alpha }_{k^{\prime }}\|>a\lambda $
for all $k$ and $k^{\prime }$. Then by Condition (C2), $\rho (\|\boldsymbol{%
\alpha }_{k}-\boldsymbol{\alpha }_{k^{\prime }}\|)$ is a constant, and as a
result $P_{n}^{\mathcal{G}}(T^{\ast }(\mathbf{\boldsymbol{\beta })})$ is a
constant. Since
\begin{equation*}
\|\boldsymbol{\alpha }_{k}-\boldsymbol{\alpha }_{k^{\prime }}\|\geq \|%
\boldsymbol{\alpha }_{k}^{0}-\boldsymbol{\alpha }_{k^{\prime
}}^{0}\|-2\sup_{k}\|\boldsymbol{\alpha }_{k}\boldsymbol{-\alpha }_{k}^{0}\|,
\end{equation*}%
and
\begin{eqnarray}
\sup_{k}\|\boldsymbol{\alpha }_{k}\boldsymbol{-\alpha }_{k}^{0}\|^{2}
&=&\sup_{k}\left\Vert |\mathcal{G}_{k}|^{-1}\sum\nolimits_{i\in \mathcal{G}%
_{k}}\mathbf{\boldsymbol{\beta }}_{i}-\boldsymbol{\alpha }%
_{k}^{0}\right\Vert ^{2}=\sup_{k}\left\Vert |\mathcal{G}_{k}|^{-1}\sum%
\nolimits_{i\in \mathcal{G}_{k}}(\mathbf{\boldsymbol{\beta }}_{i}-\mathbf{%
\boldsymbol{\beta }}_{i}^{0})\right\Vert ^{2}  \notag \\
&=&\sup_{k}|\mathcal{G}_{k}|^{-2}\left\Vert \sum\nolimits_{i\in \mathcal{G}%
_{k}}(\mathbf{\boldsymbol{\beta }}_{i}-\mathbf{\boldsymbol{\beta }}%
_{i}^{0})\right\Vert ^{2}\leq \sup_{k}|\mathcal{G}_{k}|^{-1}\sum\nolimits_{i%
\in \mathcal{G}_{k}}\left\Vert (\mathbf{\boldsymbol{\beta }}_{i}-\mathbf{%
\boldsymbol{\beta }}_{i}^{0})\right\Vert ^{2}  \notag \\
&\leq &\sup_{i}\left\Vert \mathbf{\boldsymbol{\beta }}_{i}-\mathbf{%
\boldsymbol{\beta }}_{i}^{0}\right\Vert ^{2}\leq \phi _{n}^{2},
\label{EQ:alpha}
\end{eqnarray}%
then for all $k$ and $k^{\prime }$
\begin{equation*}
\|\boldsymbol{\alpha }_{k}-\boldsymbol{\alpha }_{k^{\prime }}\|\geq \|%
\boldsymbol{\alpha }_{k}^{0}-\boldsymbol{\alpha }_{k^{\prime
}}^{0}\|-2\sup_{k}\|\boldsymbol{\alpha }_{k}\boldsymbol{-\alpha }%
_{k}^{0}\|\geq b_{n}-2\phi _{n}>a\lambda,
\end{equation*}%
where the last inequality follows from the assumption that $b_{n}>a\lambda
\gg \phi _{n}$. Therefore, we have $P_{n}^{\mathcal{G}}(T^{\ast }(\mathbf{%
\boldsymbol{\beta })})=C_{n}$, and hence $Q_{n}^{\mathcal{G}}(\mathbf{%
\boldsymbol{\eta },}T^{\ast }(\mathbf{\boldsymbol{\beta })})=L_{n}^{\mathcal{%
G}}(\mathbf{\boldsymbol{\eta },}T^{\ast }(\mathbf{\boldsymbol{\beta })}%
)+C_{n}$ for all $(\mathbf{\boldsymbol{\eta }}^{\text{T}}, \mathbf{%
\boldsymbol{\beta}}^{\text{T}})^{\text{T}}\in \Theta $. Since $((\widehat{%
\mathbf{\boldsymbol{\eta }}}^{or})^{\text{T}},(\widehat{\boldsymbol{\alpha }}%
^{or})^{\text{T}})^{\text{T}}$ is the unique global minimizer of $L_{n}^{%
\mathcal{G}}(\mathbf{\boldsymbol{\eta },\boldsymbol{\alpha } })$, then $%
L_{n}^{\mathcal{G}}(\mathbf{\boldsymbol{\eta },}T^{\ast }(\mathbf{%
\boldsymbol{\beta })})>L_{n}^{\mathcal{G}}(\widehat{\mathbf{\boldsymbol{\eta
}}}^{or}\mathbf{,}\widehat{\boldsymbol{\alpha }}^{or})$ for all $(\mathbf{%
\boldsymbol{\eta }}^{\text{T}},(T^{\ast }(\mathbf{\boldsymbol{\beta })})^{%
\text{T}})^{\text{T}}\neq ((\widehat{\mathbf{\boldsymbol{\eta }}}^{or})^{%
\text{T}},(\widehat{\boldsymbol{\alpha }}^{or})^{\text{T}})^{\text{T}}$ and
hence $Q_{n}^{\mathcal{G}}(\mathbf{\boldsymbol{\eta },}T^{\ast }(\mathbf{%
\boldsymbol{\beta })})>Q_{n}^{\mathcal{G}}(\widehat{\mathbf{\boldsymbol{\eta
}}}^{or}\mathbf{,}\widehat{\boldsymbol{\alpha }}^{or})$ for all $T^{\ast }(%
\mathbf{\boldsymbol{\beta })\neq }\widehat{\boldsymbol{\alpha }}^{or}$. By (%
\ref{EQ:qn}), we have $Q_{n}^{\mathcal{G}}(\widehat{\mathbf{\boldsymbol{\eta
}}}^{or},\widehat{\boldsymbol{\alpha }}^{or})=Q_{n}(\widehat{\mathbf{%
\boldsymbol{\eta }}}^{or}\mathbf{,}\widehat{\mathbf{\boldsymbol{\beta }}}%
^{or})$ and $Q_{n}^{\mathcal{G}}(\mathbf{\boldsymbol{\eta ,}}T^{\ast }(%
\mathbf{\boldsymbol{\beta })})=Q_{n}(\mathbf{\boldsymbol{\eta }}%
,T^{-1}(T^{\ast }(\mathbf{\boldsymbol{\beta })}))=Q_{n}(\mathbf{\boldsymbol{%
\eta ,\beta }}^{\ast })$. Therefore, $Q_{n}(\mathbf{\boldsymbol{\eta ,\beta }%
}^{\ast })>Q_{n}(\widehat{\mathbf{\boldsymbol{\eta }}}^{or}\mathbf{,}%
\widehat{\mathbf{\boldsymbol{\beta }}}^{or})$ for all $\mathbf{\boldsymbol{%
\beta }}^{\ast }\mathbf{\neq }\widehat{\mathbf{\boldsymbol{\beta}}}^{or}$,
and the result in (i) is proved.

Next we prove the result in (ii). For a positive sequence $t_{n}$, let $%
\Theta _{n}=\{\mathbf{\boldsymbol{\beta }}_{i}\mathbf{:}\sup_{i}\mathbf{%
\Vert \boldsymbol{\beta }}_{i}\mathbf{-}\widehat{\mathbf{\boldsymbol{\beta }}%
}_{i}^{or}\Vert \leq t_{n}\}$. For $(\mathbf{\boldsymbol{\eta }}^{\text{T}},%
\mathbf{\boldsymbol{\beta }}^{\text{T}})^{\text{T}}\in \Theta _{n}\cap
\Theta $, by Taylor's expansion, we have
\begin{equation*}
Q_{n}(\mathbf{\boldsymbol{\eta },\boldsymbol{\beta }})-Q_{n}(\mathbf{%
\boldsymbol{\eta },\boldsymbol{\beta }}^{\ast })=\Gamma _{1}+\Gamma _{2},
\end{equation*}%
where
\begin{eqnarray*}
\Gamma _{1} &=&-(\mathbf{y-Z\boldsymbol{\eta }}-\mathbf{X\boldsymbol{\beta }}%
^{m})^{\text{T}}\mathbf{X}(\mathbf{\boldsymbol{\beta }-\boldsymbol{\beta }}%
^{\ast }) \\
\Gamma _{2} &=&\sum_{i=1}^{n}\frac{\partial P_{n}(\mathbf{\boldsymbol{\beta }%
}^{m})}{\partial \mathbf{\boldsymbol{\beta }}_{i}^{\text{T}}}(\mathbf{%
\boldsymbol{\beta }}_{i}-\mathbf{\boldsymbol{\beta }}_{i}^{\ast }).
\end{eqnarray*}%
and $\mathbf{\boldsymbol{\beta }}^{m}=$ $\alpha \mathbf{\boldsymbol{\beta }}%
+(1-\alpha )\mathbf{\boldsymbol{\beta }}^{\ast }$ for some constant $\alpha
\in (0,1)$. Moreover,%
\begin{eqnarray}
\Gamma _{2} &=&\lambda \sum\nolimits_{\{j>i\}}\rho ^{\prime }(\Vert \mathbf{%
\boldsymbol{\beta }}_{i}^{m}-\mathbf{\boldsymbol{\beta }}_{j}^{m}\Vert
)\Vert \mathbf{\boldsymbol{\beta }}_{i}^{m}-\mathbf{\boldsymbol{\beta }}%
_{j}^{m}\Vert ^{-1}(\mathbf{\boldsymbol{\beta }}_{i}^{m}-\mathbf{\boldsymbol{%
\beta }}_{j}^{m})^{\text{T}}(\mathbf{\boldsymbol{\beta }}_{i}-\mathbf{%
\boldsymbol{\beta }}_{i}^{\ast })  \notag \\
&&+\lambda \sum\nolimits_{\{j<i\}}\rho ^{\prime }(\Vert \mathbf{\boldsymbol{%
\beta }}_{i}^{m}-\mathbf{\boldsymbol{\beta }}_{j}^{m}\Vert )\Vert \mathbf{%
\boldsymbol{\beta }}_{i}^{m}-\mathbf{\boldsymbol{\beta }}_{j}^{m}\Vert ^{-1}(%
\mathbf{\boldsymbol{\beta }}_{i}^{m}-\mathbf{\boldsymbol{\beta }}_{j}^{m})^{%
\text{T}}(\mathbf{\boldsymbol{\beta }}_{i}-\mathbf{\boldsymbol{\beta }}%
_{i}^{\ast })  \notag \\
&=&\lambda \sum\nolimits_{\{j>i\}}\rho ^{\prime }(\Vert \mathbf{\boldsymbol{%
\beta }}_{i}^{m}-\mathbf{\boldsymbol{\beta }}_{j}^{m}\Vert )\Vert \mathbf{%
\boldsymbol{\beta }}_{i}^{m}-\mathbf{\boldsymbol{\beta }}_{j}^{m}\Vert ^{-1}(%
\mathbf{\boldsymbol{\beta }}_{i}^{m}-\mathbf{\boldsymbol{\beta }}_{j}^{m})^{%
\text{T}}(\mathbf{\boldsymbol{\beta }}_{i}-\mathbf{\boldsymbol{\beta }}%
_{i}^{\ast })  \notag \\
&&+\lambda \sum\nolimits_{\{i<j\}}\rho ^{\prime }(\Vert \mathbf{\boldsymbol{%
\beta }}_{j}^{m}-\mathbf{\boldsymbol{\beta }}_{i}^{m}\Vert )\Vert \mathbf{%
\boldsymbol{\beta }}_{j}^{m}-\mathbf{\boldsymbol{\beta }}_{i}^{m}\Vert ^{-1}(%
\mathbf{\boldsymbol{\beta }}_{j}^{m}-\mathbf{\boldsymbol{\beta }}_{i}^{m})^{%
\text{T}}(\mathbf{\boldsymbol{\beta }}_{j}-\mathbf{\boldsymbol{\beta }}%
_{j}^{\ast })  \notag \\
&=&\lambda \sum\nolimits_{\{j>i\}}\rho ^{\prime }(\Vert \mathbf{\boldsymbol{%
\beta }}_{i}^{m}-\mathbf{\boldsymbol{\beta }}_{j}^{m}\Vert )\Vert \mathbf{%
\boldsymbol{\beta }}_{i}^{m}-\mathbf{\boldsymbol{\beta }}_{j}^{m}\Vert ^{-1}(%
\mathbf{\boldsymbol{\beta }}_{i}^{m}-\mathbf{\boldsymbol{\beta }}_{j}^{m})^{%
\text{T}}\{(\mathbf{\boldsymbol{\beta }}_{i}-\mathbf{\boldsymbol{\beta }}%
_{i}^{\ast })-(\mathbf{\boldsymbol{\beta }}_{j}-\mathbf{\boldsymbol{\beta }}%
_{j}^{\ast })\}.  \label{GAMMa2}
\end{eqnarray}%
When $i,j\in \mathcal{G}_{k}$, $\mathbf{\boldsymbol{\beta }}_{i}^{\ast }=%
\mathbf{\boldsymbol{\beta }}_{j}^{\ast }$, and $\mathbf{\boldsymbol{\beta }}%
_{i}^{m}-\mathbf{\boldsymbol{\beta }}_{j}^{m}=\alpha (\mathbf{\boldsymbol{%
\beta }}_{i}-\mathbf{\boldsymbol{\beta }}_{j})$. Thus,
\begin{eqnarray*}
\Gamma _{2} &=&\lambda \sum\limits_{k=1}^{K}\sum\limits_{\{i,j\in \mathcal{G}%
_{k},i<j\}}\rho ^{\prime }(\Vert \mathbf{\boldsymbol{\beta }}_{i}^{m}-%
\mathbf{\boldsymbol{\beta }}_{j}^{m}\Vert )\Vert \mathbf{\boldsymbol{\beta }}%
_{i}^{m}-\mathbf{\boldsymbol{\beta }}_{j}^{m}\Vert ^{-1}(\mathbf{\boldsymbol{%
\beta }}_{i}^{m}-\mathbf{\boldsymbol{\beta }}_{j}^{m})^{\text{T}}(\mathbf{%
\boldsymbol{\beta }}_{i}-\mathbf{\boldsymbol{\beta }}_{j}) \\
&&+\lambda \sum\limits_{k<k^{\prime }}\sum\limits_{\{i\in \mathcal{G}%
_{k},j^{\prime }\in \mathcal{G}_{k^{\prime }}\}}\rho ^{\prime }(\Vert
\mathbf{\boldsymbol{\beta }}_{i}^{m}-\mathbf{\boldsymbol{\beta }}%
_{j}^{m}\Vert )\Vert \mathbf{\boldsymbol{\beta }}_{i}^{m}-\mathbf{%
\boldsymbol{\beta }}_{j}^{m}\Vert ^{-1}(\mathbf{\boldsymbol{\beta }}_{i}^{m}-%
\mathbf{\boldsymbol{\beta }}_{j}^{m})^{\text{T}}\{(\mathbf{\boldsymbol{\beta
}}_{i}-\mathbf{\boldsymbol{\beta }}_{i}^{\ast })-(\mathbf{\boldsymbol{\beta }%
}_{j}-\mathbf{\boldsymbol{\beta }}_{j}^{\ast })\}.
\end{eqnarray*}%
Moreover,
\begin{equation}
\sup_{i}\Vert \mathbf{\boldsymbol{\beta }}_{i}^{\ast }\mathbf{-\boldsymbol{%
\beta }}_{i}^{0}\Vert ^{2}=\sup_{k}\Vert \boldsymbol{\alpha }_{k}\boldsymbol{%
-\alpha }_{k}^{0}\Vert ^{2}\leq \phi _{n}^{2},  \label{alphasup}
\end{equation}%
where the last inequality follows from (\ref{EQ:alpha}). Since $\mathbf{%
\boldsymbol{\beta }}_{i}^{m}$ is between $\mathbf{\boldsymbol{\beta }}_{i}$
and $\mathbf{\boldsymbol{\beta }}_{i}^{\ast }$,
\begin{equation}
\sup_{i}\Vert \mathbf{\boldsymbol{\beta }}_{i}^{m}\mathbf{-\boldsymbol{\beta
}}_{i}^{0}\Vert \leq \alpha \sup_{i}\Vert \mathbf{\boldsymbol{\beta }}_{i}%
\mathbf{-\boldsymbol{\beta }}_{i}^{0}\Vert +(1-\alpha )\sup_{i}\Vert \mathbf{%
\boldsymbol{\beta }}_{i}^{\ast }\mathbf{-\boldsymbol{\beta }}_{i}^{0}\Vert
\leq \alpha \phi _{n}+(1-\alpha )\phi _{n}=\phi _{n}.  \label{EQ:betam}
\end{equation}%
Hence for $k\neq k^{\prime }$, $i\in \mathcal{G}_{k},j^{\prime }\in \mathcal{%
G}_{k^{\prime }}$,
\begin{equation*}
\Vert \mathbf{\boldsymbol{\beta }}_{i}^{m}-\mathbf{\boldsymbol{\beta }}%
_{j}^{m}\Vert \geq \min_{i\in \mathcal{G}_{k},j^{\prime }\in \mathcal{G}%
_{k^{\prime }}}\Vert \mathbf{\boldsymbol{\beta }}_{i}^{0}-\mathbf{%
\boldsymbol{\beta }}_{j}^{0}\Vert -2\max_{i}\Vert \mathbf{\boldsymbol{\beta }%
}_{i}^{m}-\mathbf{\boldsymbol{\beta }}_{i}^{0}\Vert \geq b_{n}-2\phi
_{n}>a\lambda ,
\end{equation*}%
and thus $\rho ^{\prime }(\Vert \mathbf{\boldsymbol{\beta }}_{i}^{m}-\mathbf{%
\boldsymbol{\beta }}_{j}^{m}\Vert )=0$. Therefore,
\begin{eqnarray*}
\Gamma _{2} &=&\lambda \sum\limits_{k=1}^{K}\sum\limits_{\{i,j\in \mathcal{G}%
_{k},i<j\}}\rho ^{\prime }(\Vert \mathbf{\boldsymbol{\beta }}_{i}^{m}-%
\mathbf{\boldsymbol{\beta }}_{j}^{m}\Vert )\Vert \mathbf{\boldsymbol{\beta }}%
_{i}^{m}-\mathbf{\boldsymbol{\beta }}_{j}^{m}\Vert ^{-1}(\mathbf{\boldsymbol{%
\beta }}_{i}^{m}-\mathbf{\boldsymbol{\beta }}_{j}^{m})^{\text{T}}(\mathbf{%
\boldsymbol{\beta }}_{i}-\mathbf{\boldsymbol{\beta }}_{j}) \\
&=&\lambda \sum\limits_{k=1}^{K}\sum\limits_{\{i,j\in \mathcal{G}%
_{k},i<j\}}\rho ^{\prime }(\Vert \mathbf{\boldsymbol{\beta }}_{i}^{m}-%
\mathbf{\boldsymbol{\beta }}_{j}^{m}\Vert )\Vert \mathbf{\boldsymbol{\beta }}%
_{i}-\mathbf{\boldsymbol{\beta }}_{j}\Vert ,
\end{eqnarray*}%
where the last step follows from $\mathbf{\boldsymbol{\beta }}_{i}^{m}-%
\mathbf{\boldsymbol{\beta }}_{j}^{m}=\alpha (\mathbf{\boldsymbol{\beta }}%
_{i}-\mathbf{\boldsymbol{\beta }}_{j})$. Furthermore, by the same reasoning
as (\ref{EQ:alpha}), we have%
\begin{equation*}
\sup_{i}\Vert \mathbf{\boldsymbol{\beta }}_{i}^{\ast }\mathbf{-}\widehat{%
\mathbf{\boldsymbol{\beta }}}_{i}^{or}\Vert =\sup_{k}\Vert \boldsymbol{%
\alpha }_{k}\mathbf{-}\widehat{\boldsymbol{\alpha }}_{k}^{or}\Vert ^{2}\leq
\sup_{i}\Vert \mathbf{\boldsymbol{\beta }-}\widehat{\mathbf{\boldsymbol{%
\beta }}}_{i}^{or}\Vert .
\end{equation*}%
Then
\begin{eqnarray*}
\sup_{i}\Vert \mathbf{\boldsymbol{\beta }}_{i}^{m}-\mathbf{\boldsymbol{\beta
}}_{j}^{m}\Vert &\leq &2\sup_{i}\Vert \mathbf{\boldsymbol{\beta }}_{i}^{m}-%
\mathbf{\boldsymbol{\beta }}_{i}^{\ast }\Vert \leq 2\sup_{i}\Vert \mathbf{%
\boldsymbol{\beta }}_{i}-\mathbf{\boldsymbol{\beta }}_{i}^{\ast }\Vert \\
&\leq &2(\sup_{i}\Vert \mathbf{\boldsymbol{\beta }}_{i}-\widehat{\mathbf{%
\boldsymbol{\beta }}}_{i}^{or}\Vert +\sup_{i}\Vert \mathbf{\boldsymbol{\beta
}}_{i}^{\ast }-\widehat{\mathbf{\boldsymbol{\beta }}}_{i}^{or}\Vert )\leq
4\sup_{i}\Vert \mathbf{\boldsymbol{\beta }}_{i}-\widehat{\mathbf{\boldsymbol{%
\beta }}}_{i}^{or}\Vert \leq 4t_{n}.
\end{eqnarray*}%
Hence $\rho ^{\prime }(\Vert \mathbf{\boldsymbol{\beta }}_{i}^{m}-\mathbf{%
\boldsymbol{\beta }}_{j}^{m}\Vert )\geq \rho ^{\prime }(4t_{n})$ by
concavity of $\rho (\cdot )$. As a result,
\begin{equation}
\Gamma _{2}\geq \sum\nolimits_{k=1}^{K}\sum\nolimits_{\{i,j\in \mathcal{G}%
_{k},i<j\}}\lambda \rho ^{\prime }(4t_{n})\Vert \mathbf{\boldsymbol{\beta }}%
_{i}-\mathbf{\boldsymbol{\beta }}_{j}\Vert .  \label{EQ:GAMMA2}
\end{equation}%
Let
\begin{equation*}
\mathbf{Q=(Q}_{1}^{\text{T}},\ldots ,\mathbf{Q}_{n}^{\text{T}})^{\text{T}}=[(%
\mathbf{y-Z\boldsymbol{\eta }}-\mathbf{X\boldsymbol{\beta }}^{m})^{\text{T}}%
\mathbf{X]}^{\text{T}}.
\end{equation*}%
Then
\begin{eqnarray}
\Gamma _{1} &=&-\mathbf{Q}^{\text{T}}(\mathbf{\boldsymbol{\beta }-%
\boldsymbol{\beta }}^{\ast
})=-\sum\nolimits_{k=1}^{K}\sum\nolimits_{\{i,j\in \mathcal{G}_{k}\}}\frac{%
\mathbf{Q}_{i}^{\text{T}}(\mathbf{\boldsymbol{\beta }}_{i}-\mathbf{%
\boldsymbol{\beta }}_{j})}{|\mathcal{G}_{k}|}  \notag \\
&=&-\sum\nolimits_{k=1}^{K}\sum\nolimits_{\{i,j\in \mathcal{G}_{k}\}}\frac{%
\mathbf{Q}_{i}^{\text{T}}(\mathbf{\boldsymbol{\beta }}_{i}-\mathbf{%
\boldsymbol{\beta }}_{j})}{2|\mathcal{G}_{k}|}-\sum\nolimits_{k=1}^{K}\sum%
\nolimits_{\{i,j\in \mathcal{G}_{k}\}}\frac{\mathbf{Q}_{i}^{\text{T}}(%
\mathbf{\boldsymbol{\beta }}_{i}-\mathbf{\boldsymbol{\beta }}_{j})}{2|%
\mathcal{G}_{k}|}  \notag \\
&=&-\sum\nolimits_{k=1}^{K}\sum\nolimits_{\{i,j\in \mathcal{G}_{k}\}}\frac{(%
\mathbf{Q}_{j}-\mathbf{Q}_{i})^{\text{T}}(\mathbf{\boldsymbol{\beta }}_{j}-%
\mathbf{\boldsymbol{\beta }}_{i})}{2|\mathcal{G}_{k}|}  \notag \\
&=&-\sum\nolimits_{k=1}^{K}\sum\nolimits_{\{i,j\in \mathcal{G}_{k},i<j\}}%
\frac{(\mathbf{Q}_{j}-\mathbf{Q}_{i})^{\text{T}}(\mathbf{\boldsymbol{\beta }}%
_{j}-\mathbf{\boldsymbol{\beta }}_{i})}{|\mathcal{G}_{k}|}.
\label{EQ:GAMMA11}
\end{eqnarray}%
Moreover,
\begin{equation*}
\mathbf{Q}_{i}=(y_{i}-\boldsymbol{z}_{i}^{\text{T}}\boldsymbol{\eta }-%
\boldsymbol{x}_{i}^{\text{T}}\boldsymbol{\beta }_{i}^{m})\boldsymbol{x}%
_{i}=(\varepsilon _{i}+\mathbf{z}_{i}^{\text{T}}(\boldsymbol{\eta }^{0}-%
\boldsymbol{\eta })+\boldsymbol{x}_{i}^{\text{T}}(\boldsymbol{\beta }%
_{i}^{0}-\boldsymbol{\beta }_{i}^{m}))\boldsymbol{x}_{i},
\end{equation*}%
and then%
\begin{equation*}
\sup_{i}\Vert \mathbf{Q}_{i}\Vert \leq \sup_{i}\left\{ \Vert \boldsymbol{x}%
_{i}\Vert (\Vert \boldsymbol{\varepsilon }\Vert _{\infty }+\Vert \boldsymbol{%
z}_{i}\Vert \Vert \boldsymbol{\eta }^{0}-\boldsymbol{\eta }\mathbf{\Vert +}%
\Vert \boldsymbol{x}_{i}\Vert \Vert \boldsymbol{\beta }_{i}^{0}-\boldsymbol{%
\beta }_{i}^{m}\mathbf{\Vert )}\right\}
\end{equation*}%
By Condition (C1) that $\sup_{i}\Vert \boldsymbol{x}_{i}\Vert \leq C_{2}%
\sqrt{p}$ and $\sup_{i}\Vert \boldsymbol{z}_{i}\Vert \leq C_{3}\sqrt{q}$, (%
\ref{EQ:betam}) that $\sup_{i}\Vert \boldsymbol{\beta }_{i}^{0}-\boldsymbol{%
\beta }_{i}^{m}\mathbf{\Vert }\leq \phi _{n}$ and $\Vert \boldsymbol{\eta }%
^{0}-\boldsymbol{\eta }\Vert \leq \phi _{n}$, we have
\begin{equation*}
\sup_{i}\Vert \mathbf{Q}_{i}\Vert \leq C_{2}\sqrt{p}(\Vert \boldsymbol{%
\varepsilon }\Vert _{\infty }+C_{3}\sqrt{q}\phi _{n}+C_{2}\sqrt{p}\phi _{n}).
\end{equation*}%
By Condition (C3)
\begin{equation*}
P(\Vert \boldsymbol{\varepsilon }\Vert _{\infty }>\sqrt{2c_{1}^{-1}}\sqrt{%
\log n})\leq \sum\nolimits_{i=1}^{n}P(|\varepsilon _{i}\mathbf{|}>\sqrt{%
2c_{1}^{-1}}\sqrt{\log n})\leq 2n^{-1}.
\end{equation*}%
Thus there is an event $E_{2}$ such that $P(E_{2}^{C})\leq 2n^{-1}$, and
over the event $E_{2}$,
\begin{equation*}
\sup_{i}\Vert \mathbf{Q}_{i}\Vert \leq C_{2}\sqrt{p}(\sqrt{2c_{1}^{-1}}\sqrt{%
\log n}+C_{3}\sqrt{q}\phi _{n}+C_{2}\sqrt{p}\phi _{n}).
\end{equation*}%
Then
\begin{eqnarray}
&&|\frac{(\mathbf{Q}_{j}-\mathbf{Q}_{i})^{\text{T}}(\mathbf{\boldsymbol{%
\beta }}_{j}-\mathbf{\boldsymbol{\beta }}_{i})}{|\mathcal{G}_{k}|}|  \notag
\\
&\leq &|\mathcal{G}_{\min }|^{-1}\Vert \mathbf{Q}_{j}-\mathbf{Q}_{i}\Vert
\Vert \mathbf{\boldsymbol{\beta }}_{i}-\mathbf{\boldsymbol{\beta }}_{j}\Vert
\leq |\mathcal{G}_{\min }|^{-1}2\sup_{i}\Vert \mathbf{Q}_{i}\Vert \Vert
\mathbf{\boldsymbol{\beta }}_{i}-\mathbf{\boldsymbol{\beta }}_{j}\Vert
\notag \\
&\leq &2C_{2}|\mathcal{G}_{\min }|^{-1}\sqrt{p}(\sqrt{2c_{1}^{-1}}\sqrt{\log
n}+C_{3}\sqrt{q}\phi _{n}+C_{2}\sqrt{p}\phi _{n})\Vert \mathbf{\boldsymbol{%
\beta }}_{i}-\mathbf{\boldsymbol{\beta }}_{j}\Vert .  \label{EQ:GAMMA}
\end{eqnarray}%
Therefore, by (\ref{EQ:GAMMA2}), (\ref{EQ:GAMMA11}) and (\ref{EQ:GAMMA}), we
have
\begin{eqnarray*}
&&Q_{n}(\mathbf{\boldsymbol{\eta },\boldsymbol{\beta }})-Q_{n}(\mathbf{%
\boldsymbol{\eta },\boldsymbol{\beta }}^{\ast }) \\
&\geq &\sum\limits_{k=1}^{K}\sum\limits_{\{i,j\in \mathcal{G}%
_{k},i<j\}}\{\lambda \rho ^{\prime }(4t_{n})-2C_{2}|\mathcal{G}_{\min }|^{-1}%
\sqrt{p}(\sqrt{2c_{1}^{-1}}\sqrt{\log n}+C_{3}\sqrt{q}\phi _{n}+C_{2}\sqrt{p}%
\phi _{n})\}\Vert \mathbf{\boldsymbol{\beta }}_{i}-\mathbf{\boldsymbol{\beta
}}_{j}\Vert .
\end{eqnarray*}%
Let $t_{n}=o(1)$, then $\rho ^{\prime }(4t_{n})\rightarrow 1$. Since $%
\lambda \gg \phi _{n}$, $p=o(n)$, and $|\mathcal{G}_{\min }|^{-1}p=o(1)$,
then $\lambda \gg |\mathcal{G}_{\min }|^{-1}\sqrt{p}\sqrt{\log n}$, $\lambda
\gg |\mathcal{G}_{\min }|^{-1}\sqrt{pq}$ and $\lambda \gg |\mathcal{G}_{\min
}|^{-1}p\phi _{n}$. Therefore, $Q_{n}(\mathbf{\boldsymbol{\eta },\boldsymbol{%
\beta }})-Q_{n}(\mathbf{\boldsymbol{\eta },\boldsymbol{\beta }}^{\ast })\geq
0$ for sufficiently large $n$, so that the result in (ii) is proved.

\subsection{Proof of Theorem \protect\ref{THM:normhomo}}

In this section we show the results in Theorem \ref{THM:normhomo}. The
proofs of (\ref{EQ:supnormbetahomo}) and (\ref{EQ:normhomo}) follow the same
arguments as the proof of Theorem \ref{THM:norm} by letting $\widetilde{%
\mathbf{X}}=\mathbf{x}$ and $|\mathcal{G}_{\min }|=n$, and thus they are
omitted. Next, we will show (\ref{EQ:penhomo}). It follows similar
procedures as the proof of Theorem \ref{THM:selection} with the details
given below. Define $\mathcal{M}=\{\boldsymbol{\beta }\in
\mathop{{\rm
I}\kern-.2em\hbox{\rm R}}\nolimits^{np}:\boldsymbol{\beta }_{1}=\cdots =%
\boldsymbol{\beta }_{n}\}$. For each $\boldsymbol{\beta }\in \mathcal{M}$,
we have $\boldsymbol{\beta }_{i}=\boldsymbol{\alpha }$ for all $i$. Let $T:%
\mathcal{M}\rightarrow R^{p}$ be the mapping that $T(\boldsymbol{\beta }%
\mathbf{)}$ is the $p\times 1$ vector equal to the common vector $%
\boldsymbol{\alpha }$. Let $T^{\ast }:R^{np}\rightarrow R^{p}$ be the
mapping that $T^{\ast }(\boldsymbol{\beta }\mathbf{)=\{}n^{-1}\sum%
\nolimits_{i=1}^{n}\boldsymbol{\beta }_{i}$. Clearly, when $\mathbf{%
\boldsymbol{\beta }\in }\mathcal{M}$, $T(\mathbf{\boldsymbol{\beta })=}%
T^{\ast }(\mathbf{\boldsymbol{\beta })}$. Consider the neighborhood of $(%
\mathbf{\boldsymbol{\eta }}^{0}\mathbf{,\boldsymbol{\beta }}^{0})$:
\begin{equation*}
\Theta \mathcal{=\{}\boldsymbol{\eta }\mathbf{\in }R^{q},\boldsymbol{\beta }%
\mathbf{\mathbf{\in }}R^{p}\mathbf{:}\left\Vert \boldsymbol{\eta }-%
\boldsymbol{\eta }^{0}\right\Vert \leq \phi _{n},\sup_{i}\left\Vert
\boldsymbol{\beta }_{i}-\boldsymbol{\beta }_{i}^{0}\right\Vert \leq \phi
_{n}\},
\end{equation*}%
where $\phi _{n}=c_{1}^{-1/2}C_{1}^{-1}\sqrt{q+p}\sqrt{n^{-1}\log n}$. By
the result in (\ref{EQ:supnormbetahomo}), there exists an event $E_{1}$ such
that on the event $E_{1}$,%
\begin{equation*}
\left\Vert \widehat{\mathbf{\boldsymbol{\eta }}}^{or}-\mathbf{\boldsymbol{%
\eta }}^{0}\right\Vert \leq \phi _{n},\sup_{i}\left\Vert \widehat{\mathbf{%
\boldsymbol{\beta }}}_{i}^{or}-\mathbf{\boldsymbol{\beta }}%
_{i}^{0}\right\Vert \leq \phi _{n},
\end{equation*}%
and $P(E_{1}^{C})\leq 2(q+p)n^{-1}$. Hence $(\widehat{\mathbf{\boldsymbol{%
\eta }}}^{or},\widehat{\mathbf{\boldsymbol{\beta }}}^{or})\in \Theta $ on
the event $E_{1}$. For any $\boldsymbol{\beta }\mathbf{\in }R^{np}$, let $%
\boldsymbol{\beta }^{\ast }=T^{-1}(T^{\ast }(\boldsymbol{\beta }\mathbf{))}$%
. We show that $(\widehat{\mathbf{\boldsymbol{\eta }}}^{or},\widehat{\mathbf{%
\boldsymbol{\beta }}}^{or})$ is a strictly local minimizer of the objective
function (\ref{EQ:objective}) with probability approaching $1$ through the
following two steps.

(i). On the event $E_{1}$, $Q_{n}(\mathbf{\boldsymbol{\eta },\boldsymbol{%
\beta }}^{\ast })>Q_{n}(\widehat{\mathbf{\boldsymbol{\eta }}}^{or}\mathbf{,}%
\widehat{\mathbf{\boldsymbol{\beta }}}^{or})$ for any $(\mathbf{\boldsymbol{%
\eta }}^{\text{T}},\mathbf{\boldsymbol{\beta }}^{\text{T}})^{\text{T}}\in
\Theta $ and $((\mathbf{\boldsymbol{\eta }})^{\text{T}},(\mathbf{\boldsymbol{%
\beta }}^{\ast })^{\text{T}})^{\text{T}}\neq ((\widehat{\mathbf{\boldsymbol{%
\eta }}}^{or})^{\text{T}},(\widehat{\mathbf{\boldsymbol{\beta }}}^{or})^{%
\text{T}})^{\text{T}}$.

(ii). There is an event $E_{2}$ such that $P(E_{2}^{C})\leq 2n^{-1}$. On $%
E_{1}\cap E_{2}$, there is a neighborhood of $((\widehat{\mathbf{\boldsymbol{%
\eta }}}^{or})^{\text{T}},(\widehat{\mathbf{\boldsymbol{\beta }}}^{or})^{%
\text{T}})^{\text{T}}$, denoted by $\Theta _{n}$ such that $Q_{n}(\mathbf{%
\boldsymbol{\eta },\boldsymbol{\beta }})\geq Q_{n}(\mathbf{\boldsymbol{\eta }%
,\boldsymbol{\beta }}^{\ast })$ for any $((\mathbf{\boldsymbol{\eta }})^{%
\text{T}},(\mathbf{\boldsymbol{\beta }}^{\ast })^{\text{T}})^{\text{T}}\in
\Theta _{n}\cap \Theta $ for sufficiently large $n$.

Therefore, by the results in (i) and (ii), we have $Q_{n}(\mathbf{%
\boldsymbol{\eta },\boldsymbol{\beta }})>Q_{n}(\widehat{\mathbf{\boldsymbol{%
\eta }}}^{or}\mathbf{,}\widehat{\mathbf{\boldsymbol{\beta }}}^{or})$ for any
$(\mathbf{\boldsymbol{\eta }}^{\text{T}},\mathbf{\boldsymbol{\beta }}^{\text{%
T}})^{\text{T}}\in \Theta _{n}\cap \Theta $ and $((\mathbf{\boldsymbol{\eta }%
})^{\text{T}},(\mathbf{\boldsymbol{\beta }})^{\text{T}})^{\text{T}}\neq ((%
\widehat{\mathbf{\boldsymbol{\eta }}}^{or})^{\text{T}},(\widehat{\mathbf{%
\boldsymbol{\beta }}}^{or})^{\text{T}})^{\text{T}}$ in $E_{1}\cap E_{2}$, so
that $((\widehat{\mathbf{\boldsymbol{\eta }}}^{or})^{\text{T}},(\widehat{%
\mathbf{\boldsymbol{\beta }}}^{or})^{\text{T}})^{\text{T}}$ is a strict
local minimizer of $Q_{n}(\mathbf{\boldsymbol{\eta },\boldsymbol{\beta }})$ (%
\ref{EQ:objective}) on the event $E_{1}\cap E_{2}$ with $P(E_{1}\cap
E_{2})\geq 1-2(q+p+1)n^{-1}$ for sufficiently large $n$.

By the definition of $((\widehat{\mathbf{\boldsymbol{\eta }}}^{or})^{\text{T}%
},(\widehat{\mathbf{\boldsymbol{\beta }}}^{or})^{\text{T}})^{\text{T}}$, we
have $\frac{1}{2}\Vert \mathbf{y}-\mathbf{Z\boldsymbol{\eta }}-\mathbf{X%
\boldsymbol{\beta }}^{\ast }\Vert ^{2}>\frac{1}{2}\Vert \mathbf{y}-\mathbf{Z}%
\widehat{\mathbf{\boldsymbol{\eta }}}^{or}-\mathbf{X}\widehat{\mathbf{%
\boldsymbol{\beta }}}^{or}\Vert ^{2}$ for any $((\mathbf{\boldsymbol{\eta }}%
)^{\text{T}},(\mathbf{\boldsymbol{\beta }})^{\text{T}})^{\text{T}}\in \Theta
$ and $((\mathbf{\boldsymbol{\eta }})^{\text{T}},(\mathbf{\boldsymbol{\beta }%
}^{\ast })^{\text{T}})^{\text{T}}\neq ((\widehat{\mathbf{\boldsymbol{\eta }}}%
^{or})^{\text{T}},(\widehat{\mathbf{\boldsymbol{\beta }}}^{or})^{\text{T}})^{%
\text{T}}$. Moreover, since $p_{\gamma }(\Vert \widehat{\boldsymbol{\beta }}%
_{i}^{or}-\widehat{\boldsymbol{\beta }}_{j}^{or}\Vert ,\lambda )=p_{\gamma
}(\Vert \boldsymbol{\beta }_{i}^{\ast }-\boldsymbol{\beta }_{j}^{\ast }\Vert
,\lambda )=0$ for $1\leq i,j\leq n$, we have $Q_{n}(\mathbf{\boldsymbol{\eta
},\boldsymbol{\beta }}^{\ast })=\frac{1}{2}\Vert \mathbf{y}-\mathbf{Z%
\boldsymbol{\eta }}-\mathbf{X\boldsymbol{\beta }}^{\ast }\Vert ^{2}$ and $%
Q_{n}(\widehat{\mathbf{\boldsymbol{\eta }}}^{or}\mathbf{,}\widehat{\mathbf{%
\boldsymbol{\beta }}}^{or})=\frac{1}{2}\Vert \mathbf{y}-\mathbf{Z}\widehat{%
\mathbf{\boldsymbol{\eta }}}^{or}-\mathbf{X}\widehat{\mathbf{\boldsymbol{%
\beta }}}^{or}\Vert ^{2}$. Therefore, $Q_{n}(\mathbf{\boldsymbol{\eta },%
\boldsymbol{\beta }}^{\ast })>Q_{n}(\widehat{\mathbf{\boldsymbol{\eta }}}%
^{or}\mathbf{,}\widehat{\mathbf{\boldsymbol{\beta }}}^{or})$.

Next we prove the result in (ii). For a positive sequence $t_{n}$, let $%
\Theta _{n}=\{\mathbf{\boldsymbol{\beta }}_{i}\mathbf{:}\sup_{i}\mathbf{%
\Vert \boldsymbol{\beta }}_{i}\mathbf{-}\widehat{\mathbf{\boldsymbol{\beta }}%
}_{i}^{or}\Vert \leq t_{n}\}$. For $(\mathbf{\boldsymbol{\eta }}^{\text{T}},%
\mathbf{\boldsymbol{\beta }}^{\text{T}})^{\text{T}}\in \Theta _{n}\cap
\Theta $, by Taylor's expansion, we have
\begin{equation*}
Q_{n}(\mathbf{\boldsymbol{\eta },\boldsymbol{\beta }})-Q_{n}(\mathbf{%
\boldsymbol{\eta },\boldsymbol{\beta }}^{\ast })=\Gamma _{1}+\Gamma _{2},
\end{equation*}%
where
\begin{eqnarray*}
\Gamma _{1} &=&-(\mathbf{y-Z\boldsymbol{\eta }}-\mathbf{X\boldsymbol{\beta }}%
^{m})^{\text{T}}\mathbf{X}(\mathbf{\boldsymbol{\beta }-\boldsymbol{\beta }}%
^{\ast }) \\
\Gamma _{2} &=&\sum_{i=1}^{n}\frac{\partial P_{n}(\mathbf{\boldsymbol{\beta }%
}^{m})}{\partial \mathbf{\boldsymbol{\beta }}_{i}^{\text{T}}}(\mathbf{%
\boldsymbol{\beta }}_{i}-\mathbf{\boldsymbol{\beta }}_{i}^{\ast }).
\end{eqnarray*}%
$P_{n}(\mathbf{\boldsymbol{\beta }})=\lambda \sum_{i<j}\rho (\Vert \mathbf{%
\boldsymbol{\beta }}_{i}-\mathbf{\boldsymbol{\beta }}_{j}\Vert )$, and $%
\mathbf{\boldsymbol{\beta }}^{m}=$ $a\mathbf{\boldsymbol{\beta }}+(1-a)%
\mathbf{\boldsymbol{\beta }}^{\ast }$ for some constant $a\in (0,1)$.
Moreover, by (\ref{GAMMa2}),%
\begin{eqnarray*}
\Gamma _{2} &=&\lambda \sum\nolimits_{\{j>i\}}\rho ^{\prime }(\Vert \mathbf{%
\boldsymbol{\beta }}_{i}^{m}-\mathbf{\boldsymbol{\beta }}_{j}^{m}\Vert
)\Vert \mathbf{\boldsymbol{\beta }}_{i}^{m}-\mathbf{\boldsymbol{\beta }}%
_{j}^{m}\Vert ^{-1}(\mathbf{\boldsymbol{\beta }}_{i}^{m}-\mathbf{\boldsymbol{%
\beta }}_{j}^{m})^{\text{T}}\{(\mathbf{\boldsymbol{\beta }}_{i}-\mathbf{%
\boldsymbol{\beta }}_{i}^{\ast })-(\mathbf{\boldsymbol{\beta }}_{j}-\mathbf{%
\boldsymbol{\beta }}_{j}^{\ast })\} \\
&=&\lambda \sum\nolimits_{\{j>i\}}\rho ^{\prime }(\Vert \mathbf{\boldsymbol{%
\beta }}_{i}^{m}-\mathbf{\boldsymbol{\beta }}_{j}^{m}\Vert )\Vert \mathbf{%
\boldsymbol{\beta }}_{i}-\mathbf{\boldsymbol{\beta }}_{j}\Vert ,
\end{eqnarray*}%
where the second equality holds due to the fact that $\mathbf{\boldsymbol{%
\beta }}_{i}^{\ast }=\mathbf{\boldsymbol{\beta }}_{j}^{\ast }$ and $\mathbf{%
\boldsymbol{\beta }}_{i}^{m}-\mathbf{\boldsymbol{\beta }}_{j}^{m}=a(\mathbf{%
\boldsymbol{\beta }}_{i}-\mathbf{\boldsymbol{\beta }}_{j})$. Let $T^{\ast }(%
\boldsymbol{\beta }\mathbf{)}=\boldsymbol{\alpha }$. Then, following the
same argument as (\ref{alphasup}), we have
\begin{equation}
\sup_{i}\Vert \mathbf{\boldsymbol{\beta }}_{i}^{\ast }\mathbf{-\boldsymbol{%
\beta }}_{i}^{0}\Vert ^{2}=\Vert \boldsymbol{\alpha }-\boldsymbol{\alpha }%
^{0}\Vert ^{2}\leq \sup_{i}\Vert \mathbf{\boldsymbol{\beta }}_{i}\mathbf{-%
\boldsymbol{\beta }}_{i}^{0}\Vert ^{2}.  \notag
\end{equation}%
Then
\begin{eqnarray*}
\sup_{i}\Vert \mathbf{\boldsymbol{\beta }}_{i}^{m}-\mathbf{\boldsymbol{\beta
}}_{j}^{m}\Vert &\leq &2\sup_{i}\Vert \mathbf{\boldsymbol{\beta }}_{i}^{m}-%
\mathbf{\boldsymbol{\beta }}_{i}^{\ast }\Vert \leq 2\sup_{i}\Vert \mathbf{%
\boldsymbol{\beta }}_{i}-\mathbf{\boldsymbol{\beta }}_{i}^{\ast }\Vert \\
&\leq &2(\sup_{i}\Vert \mathbf{\boldsymbol{\beta }}_{i}-\widehat{\mathbf{%
\boldsymbol{\beta }}}_{i}^{or}\Vert +\sup_{i}\Vert \mathbf{\boldsymbol{\beta
}}_{i}^{\ast }-\widehat{\mathbf{\boldsymbol{\beta }}}_{i}^{or}\Vert )\leq
4\sup_{i}\Vert \mathbf{\boldsymbol{\beta }}_{i}-\widehat{\mathbf{\boldsymbol{%
\beta }}}_{i}^{or}\Vert \leq 4t_{n}.
\end{eqnarray*}%
Hence $\rho ^{\prime }(\Vert \mathbf{\boldsymbol{\beta }}_{i}^{m}-\mathbf{%
\boldsymbol{\beta }}_{j}^{m}\Vert )\geq \rho ^{\prime }(4t_{n})$ by
concavity of $\rho (\cdot )$. As a result,
\begin{equation}
\Gamma _{2}\geq \sum\nolimits_{\{i<j\}}\lambda \rho ^{\prime }(4t_{n})\Vert
\mathbf{\boldsymbol{\beta }}_{i}-\mathbf{\boldsymbol{\beta }}_{j}\Vert .
\label{EQ:GAMMA2homo}
\end{equation}%
Let
\begin{equation*}
\mathbf{Q=(Q}_{1}^{\text{T}},\ldots ,\mathbf{Q}_{n}^{\text{T}})^{\text{T}}=[(%
\mathbf{y-Z\boldsymbol{\eta }}-\mathbf{X\boldsymbol{\beta }}^{m})^{\text{T}}%
\mathbf{X]}^{\text{T}}.
\end{equation*}%
By the same reasoning as the proof for (\ref{EQ:GAMMA11}), we have
\begin{equation}
\Gamma _{1}=-\mathbf{Q}^{\text{T}}(\mathbf{\boldsymbol{\beta }-\boldsymbol{%
\beta }}^{\ast })=-n^{-1}\sum\nolimits_{\{i<j\}}(\mathbf{Q}_{j}-\mathbf{Q}%
_{i})^{\text{T}}(\mathbf{\boldsymbol{\beta }}_{j}-\mathbf{\boldsymbol{\beta }%
}_{i}).  \label{EQ:GAMMA11homo}
\end{equation}%
By the same argument as the proof for (\ref{EQ:GAMMA}), we have that there
is an event $E_{2}$ such that $P(E_{2}^{C})\leq 2n^{-1}$, and on the event $%
E_{2}$,%
\begin{eqnarray}
&&n^{-1}|(\mathbf{Q}_{j}-\mathbf{Q}_{i})^{\text{T}}(\mathbf{\boldsymbol{%
\beta }}_{j}-\mathbf{\boldsymbol{\beta }}_{i})|  \notag \\
&\leq &2C_{2}n^{-1}\sqrt{p}(\sqrt{2c_{1}^{-1}}\sqrt{\log n}+C_{3}\sqrt{q}%
\phi _{n}+C_{2}\sqrt{p}\phi _{n})\Vert \mathbf{\boldsymbol{\beta }}_{i}-%
\mathbf{\boldsymbol{\beta }}_{j}\Vert .  \label{EQ:GAMMAhomo}
\end{eqnarray}%
Therefore, by (\ref{EQ:GAMMA2homo}), (\ref{EQ:GAMMA11homo}) and (\ref%
{EQ:GAMMAhomo}), we have
\begin{eqnarray*}
&&Q_{n}(\mathbf{\boldsymbol{\eta },\boldsymbol{\beta }})-Q_{n}(\mathbf{%
\boldsymbol{\eta },\boldsymbol{\beta }}^{\ast }) \\
&\geq &\sum\nolimits_{\{i<j\}}\{\lambda \rho ^{\prime }(4t_{n})-2C_{2}n^{-1}%
\sqrt{p}(\sqrt{2c_{1}^{-1}}\sqrt{\log n}+C_{3}\sqrt{q}\phi _{n}+C_{2}\sqrt{p}%
\phi _{n})\}\Vert \mathbf{\boldsymbol{\beta }}_{i}-\mathbf{\boldsymbol{\beta
}}_{j}\Vert .
\end{eqnarray*}%
Let $t_{n}=o(1)$, then $\rho ^{\prime }(4t_{n})\rightarrow 1$. Since $%
\lambda \gg \phi _{n}$, $p=o(n)$, and $n^{-1}p=o(1)$, then $\lambda \gg
n^{-1}\sqrt{p}\sqrt{\log n}$, $\lambda \gg n^{-1}\sqrt{pq}$ and $\lambda \gg
n^{-1}p\phi _{n}$. Therefore, $Q_{n}(\mathbf{\boldsymbol{\eta },\boldsymbol{%
\beta }})-Q_{n}(\mathbf{\boldsymbol{\eta },\boldsymbol{\beta }}^{\ast })\geq
0$ for sufficiently large $n$, so that the result in (ii) is proved.

\end{document}